%% file: Paper_-_Varenna_Proceedings_-_Master_-_Submit.tex
\documentclass[varenna]{cimento}
\usepackage[dvips]{graphicx}
\usepackage{subfigure}
\usepackage{amsmath}
\usepackage{amssymb}
\usepackage{bm}
\graphicspath{{./}}
\def\Eq{eq.~}
\def\Eqs{eqs.~}
\def\Fig{fig.~}

\def\Ref{ref.~}
\def\Refs{refs.~}
\def\Sec{sect.~}

\def\be{\begin{equation}}
\def\ee{\end{equation}}
\def\bea{\begin{eqnarray}}
\def\eea{\end{eqnarray}}

\def\ie{\textit{i.e.}~}
\def\eg{\textit{e.g.}~}
\def\etc{\textit{etc.}}
\def\diff{\mathrm{d}}
\newcommand{\bra}[1]{\left\langle #1 \, \right|}
\newcommand{\ket}[1]{\left| #1 \right\rangle}

\newcommand{\refeqn}[1]{(\ref{#1})}
\begin{document}

\title{Mobile and remote inertial sensing\\with atom interferometers}


\author{B. Barrett, P.-A. Gominet, E. Cantin, L. Antoni-Micollier, A. Bertoldi, B. Battelier \atque P. Bouyer}
\institute{Laboratoire Photonique Num\'{e}rique et Nanosciences, Institut d'Optique d'Aquitaine and Universit\'{e} de Bordeaux, rue Fran\c{c}ois Mitterrand, 33400 Talence, France}

\author{J. Lautier, \atque A. Landragin}
\institute{LNE-SYRTE, Observatoire de Paris, CNRS and UPMC, 61 avenue de l'Observatoire, F-75014 Paris, France}

\shortauthor{B. Barrett, P.-A. Gominet, E. Cantin, L. Antoni-Micollier, etc.}


\maketitle

\begin{abstract}
The past three decades have shown dramatic progress in the ability to manipulate and coherently control the motion of atoms. This exquisite control offers the prospect of a new generation of inertial sensors with unprecedented sensitivity and accuracy, which will be important for both fundamental and applied science. In this article, we review some of our recent results regarding the application of atom interferometry to inertial measurements using compact, mobile sensors. This includes some of the first interferometer measurements with cold $^{39}$K atoms, which is a major step toward achieving a transportable, dual-species interferometer with rubidium and potassium for equivalence principle tests. We also discuss future applications of this technology, such as remote sensing of geophysical effects, gravitational wave detection, and precise tests of the weak equivalence principle in Space.
\end{abstract}


\input{Varenna-Introduction}
\input{Varenna-TheoreticalBackground}
\input{Varenna-SensitivityFunction}

\input{Varenna-InertialSensors}
\input{Varenna-MobileSensors}
\input{Varenna-Geophysics}
\input{Varenna-Space}
\input{Varenna-Conclusion}

\acknowledgments

This work is supported by the French national agencies CNES (Centre National d'Etudes Spatiales), l'Agence Nationale pour la Recherche (MiniAtom: ANR-09-BLAN-0026), the D\'{e}l\'egation G\'{e}n\'{e}rale de l'Armement, the European Space Agency, IFRAF (Institut Francilien de Recherche sur les Atomes Froids), action sp\'{e}cifique GRAM (Gravitation, Relativit\'{e}, Astronomie et M\'{e}trologie) and RTRA ``Triangle de la Physique''. We would like to thank our partners of the MiniAtom collaboration: IXBLUE, KLOE, THALES and III-V Lab, and for the ICE project: ONERA. B. Barrett, P.-A. Gominet and L. Antoni-Micollier also thank CNES for financial support. P. Bouyer thanks Conseil R\'{e}gional d'Aquitaine for the Excellence Chair. Finally, the ICE team thanks T. Rvachov of the Massachusetts Institute of Technology for his assistance with the ``Cicero Word Generator'' experimental control software.

\bibliographystyle{varenna}
\bibliography{Bib-VarennaProceedings}
\end{document}

%% file: Varenna-Introduction.tex
\section{Introduction}
\label{sec:Introduction}

In 1923, Louis de Broglie generalized the wave-particle duality of photons to material particles \cite{deBroglie-ComptRend-1923} with his famous expression, $\lambda_{\rm dB} = h/p$, relating the momentum of the particle, $p$, to its wavelength. Shortly afterwards, the first matter-wave diffraction experiments were carried out with electrons \cite{Davisson-PR-1927}, and later with a beam of He atoms \cite{Estermann-ZPhys-1930}. Although these experiments were instrumental to the field of matter-wave interference, they also revealed two major challenges. First, due to the relatively high temperature of most accessible particles, typical de Broglie wavelengths were much less than a nanometer (thousands of times smaller than that of visible light)---making the wave-like behavior of particles difficult to observe. For a long time, only low-mass particles such as neutrons or electrons could be coaxed to behave like waves since their small mass resulted in a relatively large de Broglie wavelength. Second, there is no natural mirror or beam-splitter for matter waves because solid matter usually scatters or absorbs atoms. Initially, diffraction from the surface of solids, and later from micro-fabricated gratings, was used as the first type of atom optic. After the development of the laser in the 1960's, it became possible to use the electric dipole interaction with near-resonance light to diffract atoms from ``light gratings''.

In parallel, the coherent manipulation of internal atomic states with resonant radio frequency (rf) waves was demonstrated in experiments by Rabi \cite{Rabi-PR-1938}. Later, pioneering work by Ramsey \cite{Ramsey-PR-1949} lead to long-lived coherent superpositions of quantum states. The techniques developed by Ramsey would later be used to develop the first atomic clocks, which were the first matter-wave sensors to find industrial applications.

From the late 1970's until the mid-90's, a particular focus was placed on laser-cooling and trapping neutral atoms \cite{Ashkin-PRL-1978, Phillips-PRL-1985, Raab-PRL-1987, Anderson-Science-1995, Davis-PRL-1995} which eventually led to two nobel prizes in physics \cite{Chu-NobelLecture-1997, Cornell-NobelLecture-2001}. Heavy neutral atoms such as sodium and cesium were slowed to velocities of a few millimeters per second (corresponding to temperatures of a few hundred nano-Kelvin), thus making it possible to directly observe the wave-like nature of matter. 

The concept of an atom interferometer was initially patented in 1973 by Altschuler and Franz \cite{Altschuler-Patent-1973}. By the late 1980s, multiple proposals had emerged regarding the experimental realization of different types of atom interferometers \cite{Chebotayev-JOSAB-1985, Clauser-PhysicaB-1988, Borde-PhysLettA-1989, Pritchard-Patent-1989}. The first demonstration of cold-atom-based interferometers using stimulated Raman transitions was carried out by Chu and co-workers \cite{Kasevich-PRL-1991, Kasevich-ApplPhysB-1992, Weiss-PRL-1993}. Since then, the field of atom interferometry has evolved quickly. Although the state-labeled, Raman-transition-based interferometer remains the most developed and commonly used type, a significant effort has been directed toward the exploration and development of new types of interferometers \cite{Cahn-PRL-1997, Berman-Book-1997, Strekalov-PRA-2002, Gupta-PRL-2002, Weel-PRA(R)-2003, Battesti-PRL-2004, Gerlich-NaturePhys-2007, Wu-PRL-2007, Beattie-PRA(R)-2009, Clade-PRL-2009, Tonyushkin-PRA(R)-2009, Alberti-NJP-2010, Su-PRA-2010, Barrett-Advances-2011, Barrett-PRA-2011, Andia-PRA(R)-2013, Jamieson-arXiv-2014}.\footnote{For a more complete history and review of atom interferometry experiments see, for example, \Ref \cite{Cronin-RevModPhys-2009}.}

Atom interferometry is nowadays one of the most promising candidates for ultra-precise and ultra-accurate measurements of inertial forces and fundamental constants \cite{Peters-Metrologia-2001, Wicht-PhysScr-2002, Fixler-Science-2007, Lamporesi-PRL-2008, Bouchendira-PRL-2011}. The realization of a Bose-Einstein condensation (BEC) from a dilute gas of trapped atoms \cite{Anderson-Science-1995, Davis-PRL-1995} has produced the matter-wave analog of a laser in optics \cite{Mewes-PRL-1997, Anderson-Science-1998, Hagley-Science-1999, Bloch-PRL-1999}. Similar to the revolution brought about by lasers in optical interferometry, BEC-based interferometry is expected to bring the field to an unprecedented level of accuracy \cite{Bouyer-PRA-1997}.

Lastly, there remained a very promising application for the future: atomic inertial sensors. Such devices are inherently sensitive to, for example, the acceleration due to gravity, or the acceleration or rotation undergone by the interferometer when placed in a non-inertial reference frame. Apart from industrial applications, which include navigation and mineral prospecting, their ability to detect minuscule changes in inertial fields can be utilized for testing fundamental physics, such as the detection of gravitational waves or geophysical effects. Inertial sensors based on ultra-cold atoms are only expected to reach their full potential in space-based applications, where a micro-gravity environment will allow the interrogation time, and therefore the sensitivity, to increase by orders of magnitude compared to ground-based sensors.

The remainder of the article is organized as follows. In \Sec \ref{sec:Theory}, we review, briefly, the basic principles of an interferometer based on matter waves and give some theoretical background for calculating interferometer phase shifts. Section \ref{sec:SensitivityFunction} provides a detailed description of the interferometer sensitivity function, and the important role it plays in measuring phase shifts in the presence of noise. In \Sec \ref{sec:InertialSensors}, we discuss various types of lab-based inertial sensors. This is followed by \Sec \ref{sec:MobileSensors} with a description of mobile sensors and recent experimental results. Section \ref{sec:Geophysics} reviews some applications of remote atomic sensors to geophysics and gravitational wave detection. Finally, in \Sec \ref{sec:Space}, we outline the advantages of space-based atom interferometry experiments, and describe two proposals for precise tests of the weak equivalence principle. We conclude the article in \Sec \ref{sec:Conclusion}.

%% file: Varenna-TheoreticalBackground.tex
\section{Theoretical background}
\label{sec:Theory}

\subsection{Principles of a matter-wave interferometer}

\begin{figure}[!t]
  \centering
  \subfigure[]{\includegraphics[width=0.45\textwidth]{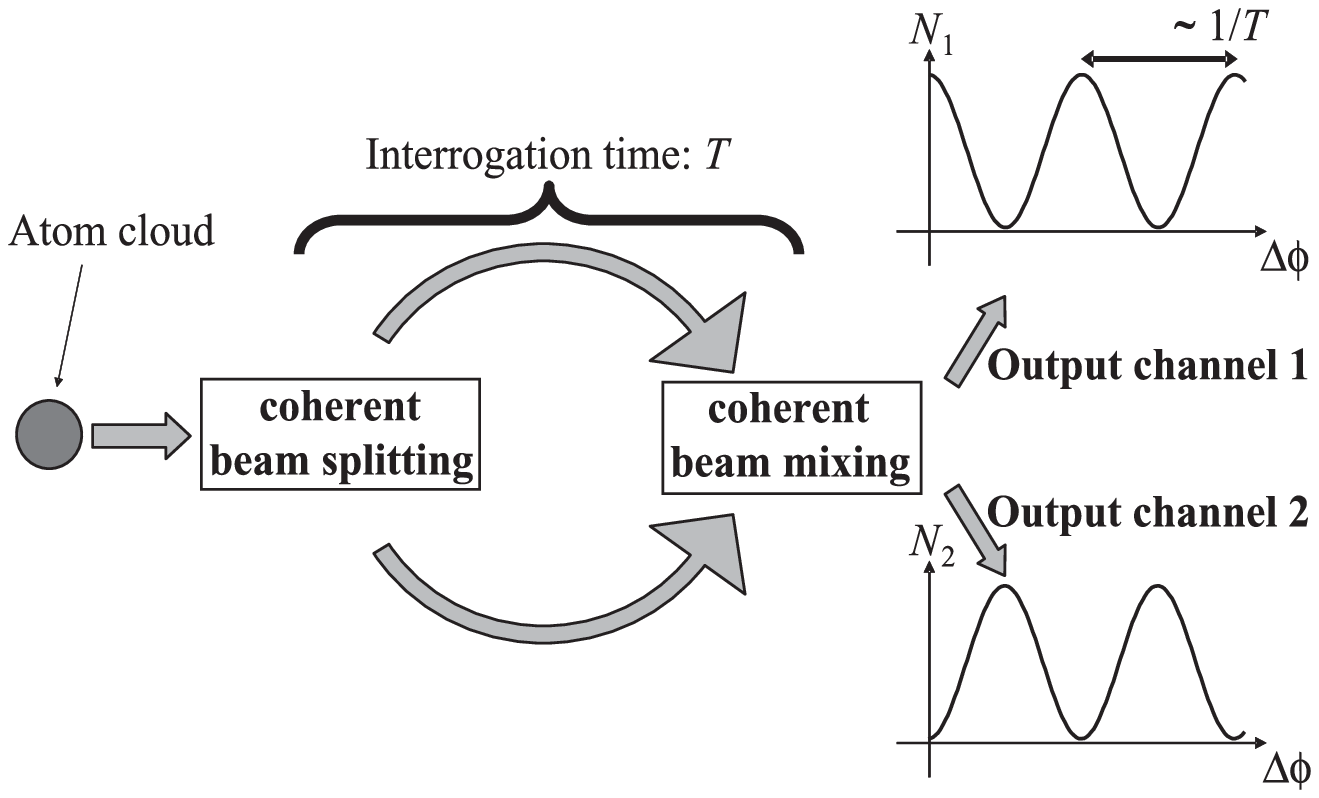}}
  \hspace{0.5cm}
  \subfigure[]{\includegraphics[width=0.45\textwidth]{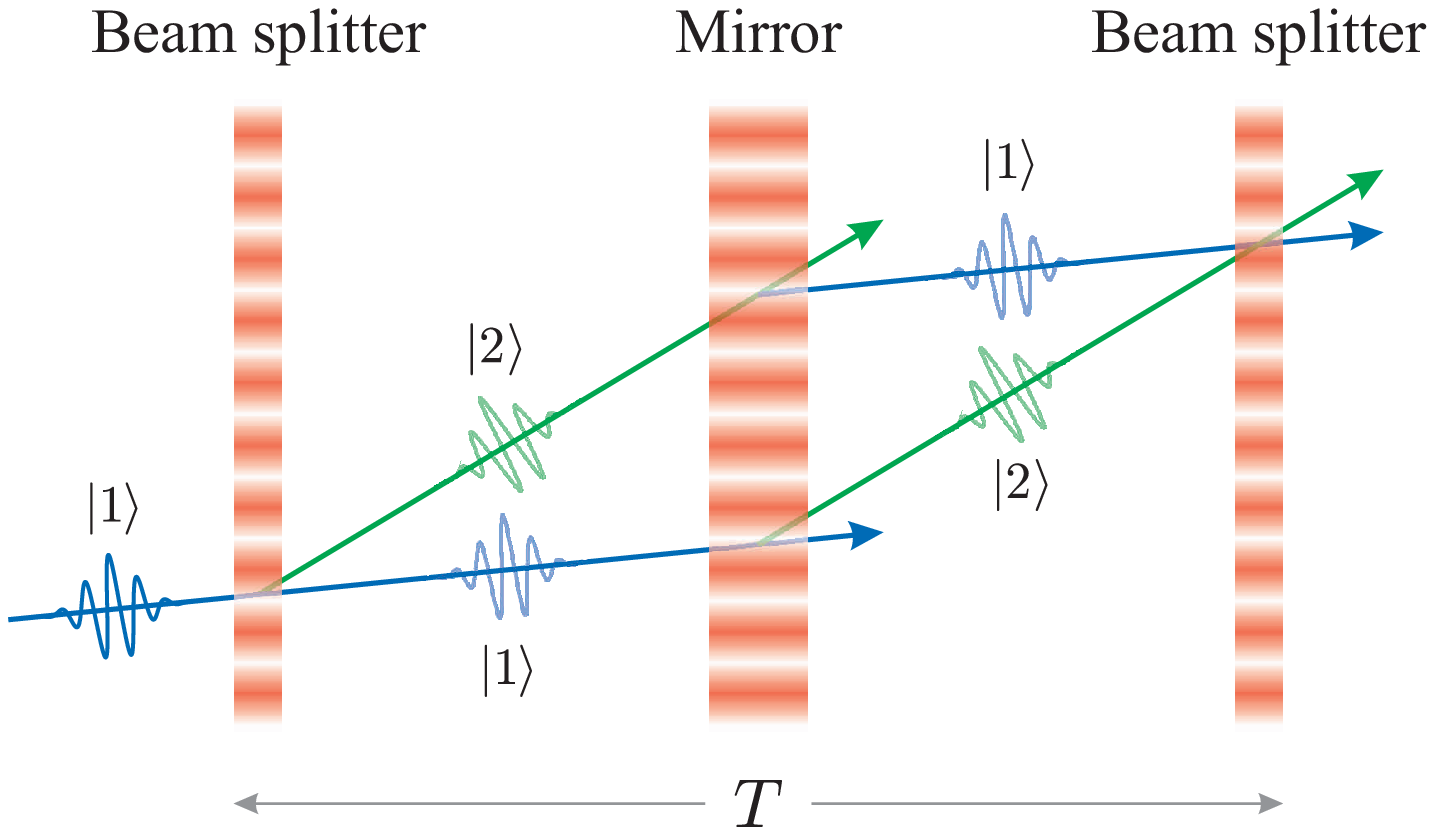}}
  \caption{(Colour online) (a) Principle of an atom interferometer. An initial atomic wavepacket is split into two parts by a coherent beam-splitting process. The wavepackets then propagate freely along the two different paths for an interrogation time $T$, during which the two wavepackets can accumulate different phases. After this time, the wavepackets are coherently mixed and interference causes the number of atoms at each output port, $N_1$ and $N_2$, to oscillate sinusoidally with respect to this phase difference, $\Delta \phi$. (b) The basic Mach-Zehnder configuration of an atom interferometer. An atom initially in a quantum state $\ket{1}$ is coherently split into a superposition of states $\ket{1}$ and $\ket{2}$. A mirror is placed at the center to close the two atomic trajectories. Interference between the two paths occurs at the second beam-splitter.}
  \label{fig:AIPrinciple}
\end{figure}

Generally, atom interferometry is performed by applying a sequence of coherent beam-splitting processes separated by an interrogation time $T$, to an ensemble of particles. This is followed by detection of the particles in each of the two output channels, as is illustrated in \Fig \ref{fig:AIPrinciple}(a). The interpretation in terms of matter waves follows from the analogy with optical interferometry. The incoming matter wave is separated into two different paths by the first beam-splitter, and the accumulation of phase along the two paths leads to interference at the last beam-splitter. This produces complementary probability amplitudes in the two channels, where the detection probability oscillates sinusoidally as a function of the total phase difference, $\Delta \phi$. In general, the sensitivity of the interferometer is proportional to the enclosed area between the two interfering pathways.

A well-known configuration of an atom interferometer is designed after the optical Mach-Zehnder interferometer: two splitting processes with a mirror placed at the center to close the two paths [see \Fig \ref{fig:AIPrinciple}(b)]. Usually, a matter-wave diffraction process replaces the mirrors and the beam-splitters and, when compared with optical diffraction, these processes can be separated either in space or in time. During the interferometer sequence, the atom resides in two different internal states while following the spatially-separated paths. In comparison, interferometers using diffraction gratings (which can be comprised of either light or matter) utilize atoms that have been separated spatially, but reside in the same internal state. This is the case, for example, with single-state Talbot-Lau interferometers \cite{Cahn-PRL-1997, Gupta-PRL-2002, Barrett-Advances-2011, Su-PRA-2010}, which have also been demonstrated with heavy molecules \cite{Gerlich-NaturePhys-2007}.

Light-pulse interferometers work on the principle that, when an atom absorbs or emits a photon, momentum must be conserved between the atom and the light field. Consequently, when an atom absorbs (emits) a photon of momentum $\hbar \bm{k}$, it will receive a momentum impulse of $\hbar \bm{k}$ ($-\hbar \bm{k}$). When a resonant traveling wave is used to excite the atom, the internal state of the atom becomes correlated with its momentum: an atom in its ground state $\ket{1}$ with momentum $\bm{p}$ (labeled $\ket{1,\bm{p}}$) is coupled to an excited state $\ket{2}$ of momentum $\bm{p}+\hbar \bm{k}$ (labeled $\ket{2,\bm{p}+\hbar\bm{k}}$).

The most developed type of light-pulse atom interferometer is that which utilizes two-photon velocity-selective Raman transitions to manipulate the atom between separate long-lived ground states. With the Raman method, two laser beams of frequency $\omega_1$ and $\omega_2$ are tuned to be nearly resonant with an optical transition. Their frequency difference $\omega_1 - \omega_2$ is chosen to be resonant with a microwave transition between two hyperfine ground states. Under appropriate conditions, the atomic population oscillates between these two states as a function of the interaction time with the lasers, $\tau$. The ``Rabi'' frequency associated with this oscillation, $\Omega_{\rm eff}$, is proportional to the product of the two single-photon Rabi frequencies of the each Raman beam, and inversely proportional to the optical detuning from a common hyperfine excited state. Thus, pulses of the Raman lasers can be tuned to coherently split (with a pulse area $\Omega_{\rm eff} \tau = \pi/2$) or reflect (with a pulse area $\Omega_{\rm eff} \tau = \pi$) the atomic wavepackets.

When the Raman beams are counter-propagating (\ie when the wave vector $\bm{k}_2 \approx -\bm{k}_1$), a momentum exchange of approximately twice the single photon momentum accompanies these transitions: $\hbar (\bm{k_1} - \bm{k}_2) \approx 2\hbar\bm{k}_1$. This results in a strong sensitivity to the Doppler frequency associated with the motion of the atom.\footnote{In contrast, when the beams are aligned to be co-propagating (\ie $\bm{k}_2 \approx \bm{k}_1$), these transitions have a negligible effect on the atomic momentum and the transition frequency is essentially insensitive to the Doppler shift of moving atoms.}

\begin{figure}[!t]
  \centering
  \includegraphics[width=0.7\textwidth]{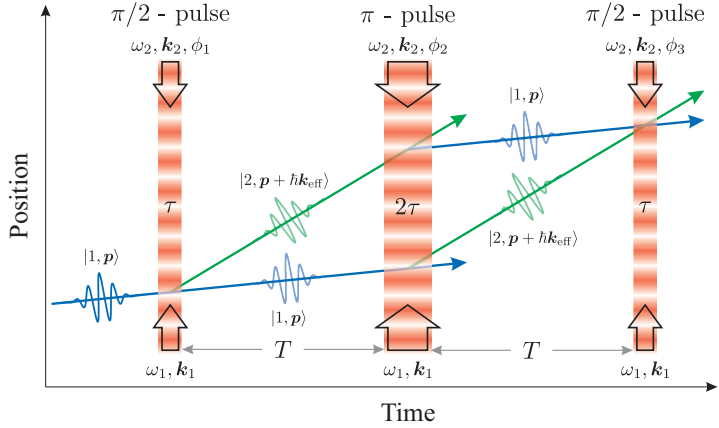}
  \caption{(Colour online) Three-pulse atom interferometer based on stimulated Raman transitions. Here, $\bm{k}_{\rm eff} = \bm{k}_1 + \bm{k}_2$ is the effective wave vector for the two-photon transition, and the pulse duration $\tau$ is defined by $\Omega_{\rm eff} \tau = \pi/2$.}
  \label{fig:MachZehnder-Raman}
\end{figure}

Henceforth, we shall consider only the most commonly used interferometer configuration, which is the so-called ``three-pulse'' or ``Mach-Zehnder'' configuration formed from a $\pi/2 - \pi - \pi/2$ pulse sequence to coherently divide, reflect and finally recombine atomic wavepackets.\footnote{Other possible configurations include that of the Ramsey-Bord\'e interferometer: $\pi/2-\pi/2-\pi/2-\pi/2$, or those utilizing Bloch-oscillation pulses or large momentum transfer pulses to increase the interferometer sensitivity.} This pulse sequence is illustrated in \Fig \ref{fig:MachZehnder-Raman}. Here, the first $\pi/2$-pulse excites an atom initially in the $\ket{1,\bm{p}}$ state into a coherent superposition of ground states $\ket{1,\bm{p}}$ and $\ket{2,\bm{p}+\hbar\bm{k}_{\rm eff}}$, where $\bm{k}_{\rm eff}$ is the difference between the two Raman wave vectors. In a time $T$, the two parts of the wavepacket drift apart by a distance $\hbar \bm{k}_{\rm eff} T/M$. Each partial wavepacket is redirected by a $\pi$-pulse which induces the transitions $\ket{1,\bm{p}} \to \ket{2,\bm{p}+\hbar\bm{k}}$ and $\ket{2,\bm{p}+\hbar\bm{k}_{\rm eff}} \to \ket{1,\bm{p}}$. After another interval $T$ the two partial wavepackets overlap again. A final $\pi/2$-pulse causes the two wavepackets to recombine and interfere. The interference is detected, for example, by measuring the total number of atoms in the internal state $\ket{2}$ at any point after the Raman pulse sequence.
%
%
This allows one to easily access the interferometer transition probability, which oscillates sinusoidally with the interferometer phase $\Phi$:
\be
  P(\Phi) = \frac{N_1}{N_1+N_2} = P_0 - \frac{C}{2} \cos(\Phi).
\ee
Here, $N_1$ ($N_2$) represents the number of atoms detected in the state $\ket{1}$ ($\ket{2}$), the offset of the probability is usually $P_0 \sim 0.5$, and $C$ is the contrast of the fringe pattern. In comparison, the detection scheme for single-state interferometers \cite{Berman-Book-1997, Cahn-PRL-1997, Gupta-PRL-2002, Su-PRA-2010, Barrett-Advances-2011, Barrett-PRA-2011, Mok-PRA-2013} requires a near-resonant traveling wave laser to coherently backscatter off of the atomic density grating formed at $t = 2T$. Here, the interferometer phase can be detected by heterodyning the backscattered beam with an optical local oscillator.

Another positive feature of this type of interferometer is that the linewidth of stimulated Raman transitions can be adjusted to tune the spread of transverse velocities addressed by the pulse. This relaxes the ``velocity collimation'' requirements and can increase the number of atoms that contribute to the interferometer signal. In contrast, Bragg scattering from standing waves is efficient only for narrow velocity spreads, where the width is much less than the photon recoil velocity.

\subsection{Phase shifts from the classical action}

In this section, we give a brief review of the Feynman path integral approach to computing the interferometer phase shift from the classical action. Then, in \Sec \ref{sec:Theory-3pulse}, we apply this formalism to the specific example of the three-pulse interferometer in the presence of a constant acceleration. Both of these sections are largely based on \Ref \cite{Storey-JPhysIIFrance-1994}.

According to the principle of least action, the actual path, $z(t)$, taken by a classical particle is the one for which the action $S$ is extremal. The action is defined as
\be
  S = \int_{t_a}^{t_b} \diff t \mathcal{L}[z(t),\dot{z}(t)],
\ee
where $\mathcal{L}(z,\dot{z})$ is the Lagrangian of the system. The action corresponding to this path is called the classical action, $S_{\rm cl}$, and it can be shown to depend on only the initial and final points $\{ z_a t_a, z_b t_b \}$ in spacetime: $S_{\rm cl}(z_b t_b, z_a t_a)$.

Given the initial state of a quantum system at time $t_a$, the state at a later time $t_b$ is determined through the evolution operator $U$
\be
  \ket{\Psi(t_b)} = U(t_b, t_a) \ket{\Psi(t_a)}.
\ee
The projection of this state on the position basis gives the wavefunction at time $t_b$
\be
  \label{Psi(zb,tb)}
  \Psi(z_b, t_b) = \int \diff z_a K(z_b t_b, z_a t_a) \Psi(z_a, t_a),
\ee
where $K$ is called the quantum propagator, and is defined as \cite{Storey-JPhysIIFrance-1994}
\be
  K(z_b t_b, z_a t_a) \equiv \bra{z_b} U(t_b, t_a) \ket{z_a}.
\ee
The quantity $|K(z_b t_b, z_a t_a)|^2$ gives the probability of finding the particle at the spacetime position $z_b t_b$, provided it started from the point $z_a t_a$. As demonstrated by Feynman \cite{Feynman-RevModPhys-1948}, the quantum propagator can be expressed equivalently as a sum over all possible paths, $\mathcal{P}$, connecting point $z_a t_a$ to $z_b t_b$.
\be
  K(z_b t_b, z_a t_a) \propto \sum_{\mathcal P} e^{i S_{\mathcal P}/\hbar} = \int_a^b \diff z(t) e^{i S_{\mathcal P}/\hbar}.
\ee
Since the action is extremal for the classical path, the phase factors $e^{i S_{\mathcal P}/\hbar}$ associated with neighboring paths tend to interfere constructively. For other paths, $S_{\mathcal P}$ generally varies rapidly compared to $S_{\rm cl}$, thus, they interfere destructively and don't contribute to $K(z_b t_b, z_a t_a)$.

In the general case of a system that can be described by a Lagrangian that is, at most, quadratic in $z(t)$ and $\dot{z}(t)$, the quantum propagator can be expressed in the simple form \cite{Storey-JPhysIIFrance-1994}
\be
  \label{K}
  K(z_b t_b, z_a t_a) = F(t_b, t_a) e^{i S_{\rm cl}(z_b t_b, z_a t_a)/\hbar},
\ee
where $F(t_b, t_a)$ is a function that depends on only the initial and final times. Inserting this result into \Eq \refeqn{Psi(zb,tb)} for the final wavefunction gives
\be
  \label{Psi(zb,tb)-2}
  \Psi(z_b, t_b) = F(t_b, t_a) \int \diff z_a e^{i S_{\rm cl}(z_b t_b, z_a t_a)/\hbar} \Psi(z_a, t_a).
\ee
In the integral over $z_a$, the neighborhood of the positions where the phase of $e^{i S_{\rm cl}(z_b t_b, z_a, t_a)/\hbar}$ cancels the phase of $\phi(z_a, t_a)$ will be the most dominant. Equation \refeqn{Psi(zb,tb)-2} has a simple interpretation: the phase of the final wavefunction, $\varphi_b$, is determined by the classical action, $S_{\rm cl}(z_b t_b, z_a t_a)$, and the phase of the wavefunction at the initial point, $\varphi_a$. In the case of an atom interferometer, the phase shift introduced between two arms is then simply the difference in classical action between the two closed paths.

\subsection{Application to the three-pulse interferometer}
\label{sec:Theory-3pulse}

In this section, we will apply the formalism of the previous section to compute the phase shift of the three-pulse Mach-Zehnder atom interferometer (shown in \Fig \ref{fig:MachZehnder-Raman}) in the presence of the acceleration due to gravity, $g$. This type of interferometer, which has the Raman lasers oriented along the vertical direction, was first demonstrated by Kasevich \& Chu \cite{Kasevich-PRL-1991, Kasevich-ApplPhysB-1992} and later developed for precise measurements of $g$ in an atomic fountain \cite{Peters-Nature-1999, Peters-Metrologia-2001}.

\begin{figure}[!t]
  \centering
  \includegraphics[width=0.4\textwidth]{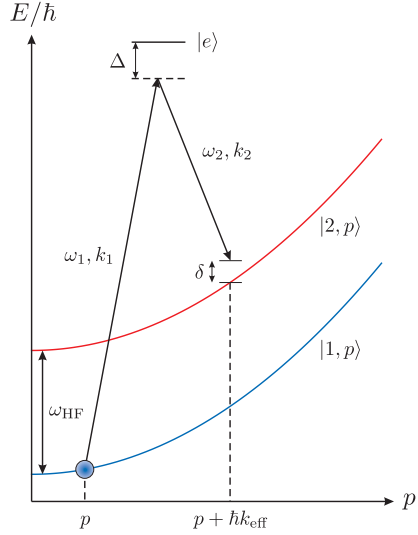}
  \caption{(Colour online) Raman transition energy levels. Atoms initially in the state $\ket{1,p}$ are transferred to $\ket{2,p+\hbar k_{\rm eff}}$ via a two-photon transition from counter-propagating Raman beams. The hyperfine ground states are separated by $\hbar \omega_{\rm HF}$ in energy, the one-photon detuning of the Raman beams from the common excited state $\ket{e}$ is $\Delta$, and $\delta$ is the two-photon detuning given by \Eq \refeqn{delta}.}
  \label{fig:RamanTransition}
\end{figure}

To evaluate the phase of the wavefunction after the interferometer pulse sequence (which governs the probability of detecting the atoms in either of the two ground states), we first describe the physics of two-photon Raman transitions. Figure \ref{fig:RamanTransition} illustrates the energy levels of the atom as a function of momentum, $p$. Two counter-propagating plane waves, with frequencies $\omega_1$ and $\omega_2$, wave vectors $\bm{k}_1$ and $\bm{k}_2$, and phase difference $\phi$, induce a transition between ground states $\ket{1}$ and $\ket{2}$ via off-resonant coupling from a common excited state $\ket{e}$. In the process, the atom scatters one photon from each beam for a total momentum transfer of $\hbar \bm{k}_{\rm eff} = \hbar(\bm{k}_1 + \bm{k}_2)$. The two-photon detuning $\delta$, which characterizes the resonance condition for the Raman transition, is given by
\be
  \label{delta}
  \delta = \omega_{\rm eff} - \omega_{\rm HF} - \frac{\bm{k}_{\rm eff} \cdot \bm{p}}{M} - \frac{\hbar k_{\rm eff}^2}{2M},
\ee
where $\omega_{\rm eff} \equiv \omega_1 - \omega_2$ is the frequency difference between Raman lasers, $\omega_{\rm HF}$ is the hyperfine splitting between the two ground states, and $M$ is the mass of the atom. \footnote{We have ignored shifts in the atomic energy levels due to the AC Stark effect in \Eq \refeqn{delta}.} The last two terms in \Eq \refeqn{delta} are the Doppler frequency and two-photon recoil frequency, respectively.

Under certain conditions, the Schr\"{o}dinger equation associated with the interaction with the Raman beams can be written as
\be
  \label{Schrodinger}
  \frac{\diff}{\diff t} \ket{\Psi(t)} =
  \frac{\diff}{\diff t} \left(
  \begin{array}{cc}
    c_1(t) \\ c_2(t)
  \end{array}
  \right) =
  i \left(
  \begin{array}{cc}
    |\chi_{\rm eff}| & \chi_{\rm eff} e^{i(\delta t + \phi)} \\
    \chi_{\rm eff}^* e^{-i(\delta t + \phi)} & |\chi_{\rm eff}|  \\
  \end{array}
  \right) \left(
  \begin{array}{cc}
    c_1(t_0) \\ c_2(t_0)
  \end{array}
  \right).
\ee
Here, the wave function is defined as a time-dependent superposition between the two states:
\be
  \label{Psi(t)}
  \ket{\Psi(t)} \equiv c_1(t) e^{-i (p^2/2M\hbar) t} \ket{1,p} + c_2(t) e^{-i [(p+\hbar k_{\rm eff})^2/2M\hbar + \omega_{\rm HF}] t} \ket{2,p+\hbar k_{\rm eff}},
\ee
and $\chi_{\rm eff} \equiv \Omega_{\rm eff}/2 = \Omega_1^* \Omega_2/\Delta$ is half of the effective Rabi frequency. To arrive at \Eq \refeqn{Schrodinger}, we have made a number of assumptions. First, the two Raman frequencies, $\omega_1$ and $\omega_2$, are shifted far from the excited state such that their one-photon detuning is much larger than the transition linewidth: $|\Delta| \gg \Gamma$. This allows us to ignore spontaneous emission effects and to eliminate the evolution of the excited state. Second, we assume the light intensity is constant, and that the two Rabi frequencies, $\Omega_1$ and $\Omega_2$, associated with each single-photon transition are equal: $\Omega_1 = \Omega_2 \equiv \Omega$. Third, we assume that $|\delta| \ll |\Omega|$ and ignore terms of order $\delta/\Omega$.

The solution to \Eq \refeqn{Schrodinger} can be shown to be \cite{Moler-PRA-1992}
\be
  \left( \begin{array}{c}
    c_1(t) \\
    c_2(t) \\
  \end{array} \right)
  = U_{\chi,\,\phi}(t, t_0)
  \left( \begin{array}{c}
    c_1(t_0) \\
    c_2(t_0) \\
  \end{array} \right),
\ee
where $U_{\chi,\,\phi}(t, t_0)$ is the evolution matrix from time $t_0 \to t$ given by
\be
  \label{U_chi_phi}
  U_{\chi,\,\phi}(t, t_0) = e^{i \chi_{\rm eff}(t-t_0)}
  \left( \begin{array}{cc}
    \cos\chi_{\rm eff}(t-t_0) &
    i e^{i(\delta t_0 + \phi)} \sin\chi_{\rm eff}(t-t_0) \\
    i e^{-i(\delta t_0 + \phi)} \sin\chi_{\rm eff}(t-t_0) &
    \cos\chi_{\rm eff}(t-t_0) \\
  \end{array} \right).
\ee
This expression describes the time-dependence of the atomic state amplitudes during a Raman pulse with phase difference $\phi$ and effective Rabi frequency $\Omega_{\rm eff} = 2\chi_{\rm eff}$.

\begin{figure}[!t]
  \centering
  \includegraphics[width=0.6\textwidth]{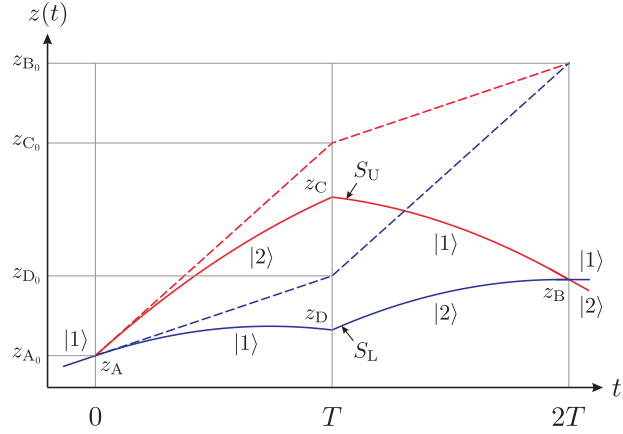}
  \caption{(Colour online) Center-of-mass trajectories taken by the atoms in a Mach-Zehnder interferometer with gravity (dashed lines) and without gravity (solid lines).}
  \label{fig:MachZehnder-Gravity}
\end{figure}

The total phase shift of the interferometer is equal to the difference in phase accumulated between the upper and lower pathways shown in \Fig \ref{fig:MachZehnder-Gravity}. It can be divided into three terms:
\be
  \Phi_{\rm total} = \Phi_{\rm propagation} + \Phi_{\rm light} + \Phi_{\rm separation},
\ee
the phase shift from the free propagation of the atom, $\Phi_{\rm propagation}$, the phase shift from the atom-laser interaction during the Raman transitions, $\Phi_{\rm light}$, and the phase shift originating from a difference in initial position of the interfering wave packets, $\Phi_{\rm separation}$. This last term is zero in this case, because the two wave packets are initially overlapped. However, it is non-zero when considering higher-order potentials \cite{Peters-Metrologia-2001}, such as that produced by a gravity gradient (which varies as $z^2$).

First, we examine the phase due to the propagation of the atoms along the two arms of the interferometer. To do this, we use the relation
\be
  \Phi_{\rm propagation} = (S_{\rm L} - S_{\rm U})/\hbar,
\ee
where $S_{\rm U}$ and $S_{\rm L}$ are the classical actions evaluated along the upper and lower atomic trajectories, respectively, as shown in \Fig \ref{fig:MachZehnder-Gravity}. For an atom of mass $M$ in free-fall with Earth's gravitational field, the classical action takes the form \cite{Storey-JPhysIIFrance-1994}
\begin{align}
\begin{split}
  S_{\rm cl}(z_b t_b, z_a t_a)
  & = \int_{t_a}^{t_b} \diff t \left[\frac{1}{2} M v(t)^2 - M g z(t)\right] \\
  & = \frac{M}{2} \frac{(z_b - z_a)^2}{(t_b - t_a)} - \frac{M g}{2} (z_b + z_a)(t_b - t_a) - \frac{M g^2}{24} (t_b - t_a)^3.
\end{split}
\end{align}
Computing the difference in action between the two classical paths we find
\be
  \label{SU-SL}
  S_{\rm L} - S_{\rm U} = \frac{M}{T}(z_{\rm D} - z_{\rm C})(z_{\rm C} + z_{\rm D} - z_{\rm A} - z_{\rm B} - gT^2).
\ee
%
%
However, from the equations of motion, it is straightforward to show that the vertices along the parabolic trajectories are related to the corresponding points along the straight-line paths in the absence of gravity (see \Fig \ref{fig:MachZehnder-Gravity}):
$$
  z_{\rm A} = z_{\rm A_0}, \;\;\;
  z_{\rm C} = z_{\rm C_0} - \frac{1}{2} g T^2, \;\;\;
  z_{\rm D} = z_{\rm D_0} - \frac{1}{2} g T^2, \;\;\;
  z_{\rm B} = z_{\rm B_0} - 2 g T^2.
$$
Evaluating the last term in \Eq \refeqn{SU-SL}, we find
\be
  \label{parallelogram}
  z_{\rm C} + z_{\rm D} - z_{\rm A} - z_{\rm B} - gT^2 =
  z_{\rm C_0} + z_{\rm D_0} - z_{\rm A_0} - z_{\rm B_0} = 0,
\ee
since the straight-line trajectories enclose a parallelogram. Hence, the phase shift due to the propagation of the wavefunction vanishes: $\Phi_{\rm propagation} = 0$. The interferometer phase is then completely determined by the contribution from the interaction with the Raman beams, $\Phi_{\rm light}$, which we now discuss.\footnote{There is zero contribution to $\Phi_{\rm total}$ from the evolution of the internal atomic energies because the atom spends the same amount of time in each state.}

Similar to $\Phi_{\rm propagation}$, the laser phase can be written as the difference between the phase accumulated along the upper and lower pathways:
\be
  \Phi_{\rm light} = \varphi_{\rm L}^{\rm light} - \varphi_{\rm U}^{\rm light}.
\ee
Table \ref{tab:RamanTransitionPhases} summarizes the laser phase contributions to the wave function that result from Raman transitions \cite{Peters-Metrologia-2001}. As a result of the $\pi/2 - \pi -\pi/2$ sequence, along the upper path the atomic state changes from $\ket{1} \to \ket{2} \to \ket{1} \to \ket{2}$, giving
\bea
  \varphi_{\rm U}^{\rm light}
  & = & \big[ k_{\rm eff} (z_{\rm A_0}) - \omega_{\rm eff} (0) - \phi_1 \big]
  - \left[ k_{\rm eff} \left(z_{\rm C_0} - \frac{1}{2} g T^2\right) - \omega_{\rm eff} (T) - \phi_2 \right] \nonumber \\
  &   & + \; \big[ k_{\rm eff} \left(z_{\rm B_0} - 2 g T^2\right) - \omega_{\rm eff} (2T) - \phi_3 \big] \nonumber \\
  & = & k_{\rm eff} \left(z_{\rm A_0} + z_{\rm B_0} - z_{\rm C_0} - \frac{3}{2} g T^2\right) - \omega_{\rm eff} T - (\phi_1 - \phi_2 + \phi_3). \nonumber
\eea
Here, $\phi_1$, $\phi_2$ and $\phi_3$ are the Raman phases during the first, second and third pulses, respectively. Similarly, along the lower path we have $\ket{1} \to \ket{1} \to \ket{2} \to \ket{2}$, thus the laser phase is
$$
  \varphi_{\rm L}^{\rm light}
  = \left[ k_{\rm eff} \left( z_{\rm D_0} - \frac{1}{2} g T^2 \right) - \omega_{\rm eff} T - \phi_2 \right].
$$
Finally, using relation \refeqn{parallelogram}, we find the interferometer phase to be
\be
  \label{Phitotal}
  \Phi_{\rm total} = \Phi_{\rm light} = k_{\rm eff} g T^2 + (\phi_1 - 2\phi_2 + \phi_3).
\ee
Since the phase scales as $g T^2$, this relation portrays the intrinsically high sensitivity of the interferometer to gravitational acceleration. One can generalize this result to show that the interferometer is sensitive to a variety of inertial effects arising from different forces.

\begin{table}[!tb]
  \centering
  \caption{Phase contributions to the wave function for different Raman transitions.}
  \begin{tabular}{ccc}
    \hline
     Internal State & Momentum & Phase shift \\
    \hline
      $1 \to 2$ & $p \to p + \hbar k_{\rm eff}$ & $+[k_{\rm eff} z(t) - \omega_{\rm eff} t - \phi]$ \\
      $2 \to 1$ & $p + \hbar k_{\rm eff} \to p$ & $-[k_{\rm eff} z(t) - \omega_{\rm eff} t - \phi]$ \\
      $1 \to 1$ & $p \to p$ & 0 \\
      $2 \to 2$ & $p + \hbar k_{\rm eff} \to p + \hbar k_{\rm eff}$ & 0 \\
    \hline
  \end{tabular}
  \label{tab:RamanTransitionPhases}
\end{table}

In summary, inertial forces manifest themselves in the interferometer by changing the relative phase of the matter waves with respect to the phase of the driving light field. The physical manifestation of the phase shift is a change in the probability of finding the atoms in, for example, the state $\ket{1}$, after the interferometer pulse sequence. A complete relativistic treatment of wave packet phase shifts in the case of an acceleration, an acceleration with a spatial gradient, or a rotation can be realized with the ABCD$\xi$ formalism \cite{Borde-CRAcadSciParis-2001, Borde-Metrologia-2002, Antoine-PhysLettA-2003}, which is a generalization of ABCD matrices for light optics.

%% file: Varenna-SensitivityFunction.tex
\section{The sensitivity function}
\label{sec:SensitivityFunction}

In this section, we provide a detailed analysis of the sensitivity function, $g(t)$, which characterizes how the atomic transition probability, and therefore the measured interferometer phase, behaves in the presence of fluctuations in the phase difference $\phi$ between Raman beams. Developed previously for use with atomic clocks \cite{Dick-Proc19PTTI-1987}, it is an extremely useful tool that can be applied, for example, to evaluate the response of the interferometer to laser phase noise \cite{Cheinet-IEEETransInstrumMeas-2008}, or to correct the interferometer phase for unwanted vibrations in the Raman beam optics \cite{Geiger-NatureComm-2011}.

Suppose there is a small, instantaneous phase jump of $\delta\phi$ at time $t$ during the Raman pulse sequence. This changes the transition probability $P(\Phi)$ by a corresponding amount $\delta P$. The sensitivity function is a unitless quantity defined as
\be
  \label{g(t)}
  g(t) = 2 \lim_{\delta\phi \to 0} \frac{\delta P(\delta\phi, t)}{\delta\phi}.
\ee
The utility of this function can be demonstrated by considering the case of an arbitrary, time-dependent phase noise, $\phi(t)$, in the Raman lasers. The change in interferometer phase, $\delta \Phi$, induced by this noise is
\be
  \label{deltaPhi}
  \delta\Phi = \int g(t) \diff \phi(t) = \int g(t) \frac{\diff \phi(t)}{\diff t} \diff t.
\ee
Thus, for a sinusoidally modulated phase given by $\phi(t) = A_{\phi} \cos(\omega_{\phi} t + \theta)$, we find $\delta\Phi = A_{\phi} \omega_{\phi} \mbox{Im}[G(\omega_{\phi})] \cos \theta$, where $G(\omega)$ is the Fourier transform of the sensitivity function:
\be
  \label{G}
  G(\omega) = \int e^{-i\omega t} g(t) \diff t.
\ee
If we then average over a random distribution of the modulation phase $\theta$, the root-mean-squared (rms) value of the interferometer phase can be shown to be $\delta\Phi_{\rm rms} = |A_{\phi} \omega_{\phi} G(\omega_{\phi})|$. From this relation, we can deduce the weight function that transforms sinusoidal laser phase noise into interferometer phase noise (the so-called transfer function):
\be
  \label{H}
  H(\omega) = \omega G(\omega).
\ee
Using the transfer function, we can tackle the more general case of broad-spectral phase noise [with power spectral density given by $S_{\phi}(\omega)$], and compute the rms standard deviation of the interferometric phase noise, $\sigma_{\Phi}^{\rm rms}$, using the following relation
\be
  \label{sigmaPhirms}
  (\sigma_{\Phi}^{\rm rms})^2 = \int_{0}^{\infty} |H(\omega)|^2 S_{\phi}(\omega) \diff \omega.
\ee
At this point, we need to know the exact form of the sensitivity function, $g(t)$, to determine the response of a given atom interferometer. For this purpose, we will use the three-pulse Mach-Zehnder configuration as an example. More specifically, we consider a pulse sequence $\tau_R$ -- $T$ -- $2\tau_R$ -- $T$ -- $\tau_R$, where $\tau_R$ is the duration of the beam-splitting pulse (with a pulse area $\Omega_{\rm eff} \tau_R = \pi/2$), $T$ is a period of free evolution, $2\tau_R$ is the duration of the reflection pulse, and so on. This pulse sequence results in the well known transition probability
\be
  \label{P}
  P(\Phi) = \frac{1}{2}(1 - \cos \Phi),
\ee
where $\Phi = \phi_1 - 2 \phi_2 + \phi_3$ is the total phase of the interferometer,\footnote{We have assumed that the Raman beams are oriented horizontally such that the interferometer is insensitive to gravity.} and $\phi_j$ is the Raman phase difference at the time of the $j^{\rm th}$ pulse (taken at the center of the atomic wavepacket). Usually, the interferometer is operated at $\Phi = \pi/2$, where the transition probability is $1/2$ and the sensitivity to phase fluctuations is maximized.

It is straightforward to compute $g(t)$ if the phase jump $\delta\phi$ occurs between Raman pulses. For instance, if the phase jump occurs between the first and second pulses, we use \Eq \refeqn{P} with $\phi_1 = \phi$, $\phi_2 = \phi + \delta\phi$, and $\phi_3 = \phi + \delta\phi + \pi/2$ to obtain $P(\delta\phi) = \big(1 - \cos(\pi/2 - \delta\phi)\big)/2$. For small $\delta\phi$, it follows that
\be
  \delta P
  = \frac{\partial P}{\partial(\delta\phi)} \delta\phi
  = -\frac{1}{2} \sin(\pi/2 - \delta\phi) \delta\phi,
\ee
and from \Eq \refeqn{g(t)} we find $g(t) = -1$. Similarly, it can be shown that $g(t) = +1$ if the phase jump occurs between the second and third pulses.

In general, however, $g(t)$ depends on the evolution of the atomic states resulting from the interaction with the Raman beams. The quantum mechanical nature of the atom plays a crucial role on the sensitivity function, particularly when a phase jump occurs \emph{during} any of the laser pulses. To determine how $g(t)$ behaves during these times, we must evaluate the time-dependent state amplitudes, $c_1(t)$ and $c_1(t)$, of the atomic wave function [see \Eq \refeqn{Psi(t)}]. To do this, we solve the Schr\"{o}dinger equation under the same conditions mentioned in \Sec \ref{sec:Theory-3pulse}, and use the evolution operator, $U_{\chi,\,\phi}(t, t_0)$, given by \Eq \refeqn{U_chi_phi}. This operator describes the evolution of the atomic state amplitudes from time $t_0$ to $t$ during (i) a Raman pulse with phase $\phi$ if $\chi_{\rm eff} > 0$, or (ii) during a period of free evolution if $\chi_{\rm eff} = 0$. A product of these matrices in the appropriate order simulates the $\pi/2 - \pi - \pi/2$ Raman pulse sequence, and can be used to compute the final state population at the output of the interferometer. Choosing the initial wave function as $\ket{\Psi(0)} = \ket{1,p}$ such that $c_1(0) = 1$ and $c_2(0) = 0$, the transition probability is given by $P = |c_2(t_f)|^2$. Here, $t_f = 2T + 4\tau_R$ and $c_2(t_f)$ is calculated from
\begin{align}
\begin{split}
  \left( \begin{array}{c}
    c_1(t_f) \\
    c_2(t_f) \\
  \end{array} \right)
  & =
  U_{\chi,\,\phi_3}(2T+4\tau_R,2T+3\tau_R)
  U_{0,\,\phi_2}(2T+3\tau_R,T+3\tau_R) \\
  & \times
  U_{\chi,\,\phi_2}(T+3\tau_R,T+\tau_R)
  U_{0,\,\phi_1}(T+\tau_R,\tau_R)
  U_{\chi,\,\phi_1}(\tau_R,0)
  \left( \begin{array}{c}
    1 \\
    0 \\
  \end{array} \right).
\end{split}
\end{align}
Equation \refeqn{P} can also be validated using this expression.

\begin{figure}[!t]
  \centering
  \includegraphics[width=0.6\textwidth]{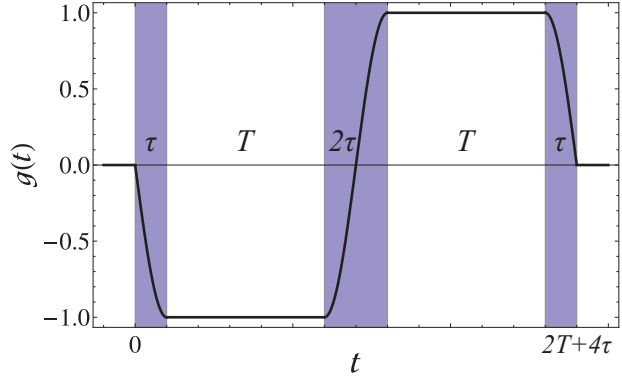}
  \caption{(Colour online) Plot of the sensitivity function, $g(t)$, for the three-pulse interferometer given by \Eq \refeqn{g(t)-3pulse}. The pulse duration, $\tau$, satisfies $\Omega_{\rm eff} \tau = \pi/2$.}
  \label{fig:g(t)-3pulse}
\end{figure}

To simulate a phase jump during a Raman pulse, we replace the matrix associated with the $j^{\rm th}$ pulse, $U_{\chi,\,\phi_j}(T_j+\tau_j,T_j)$, with the product of two matrices: $U_{\chi,\,\phi_j+\delta\phi}(T_j+\tau_j,T_j+t)U_{\chi,\,\phi_j}(T_j+t,T_j)$. Here, the first matrix on the right evolves the wave function from time $T_j$ to $T_j + t$ with a Raman phase $\phi_j$. At this time there is a phase jump, and the second matrix carries the wave function from $T_j + t$ to $T_j + \tau_j$ with a phase $\phi_j + \delta\phi$. The times $T_j$ and $\tau_j$ represent the onset time and duration of the $j^{\rm th}$ pulse, respectively. Carrying out this procedure, the resulting sensitivity function can be shown to be
\be
  \label{g(t)-3pulse}
  g(t) = \left\{
  \begin{array}{cc}
    -\sin(\Omega_{\rm eff} t)               & 0 < t \le \tau_R, \\
    -1                                      & \tau_R < t \le T + \tau_R, \\
    -\sin\big(\Omega_{\rm eff}(t - T)\big)  & T + \tau_R < t \le T + 3\tau_R, \\
    1                                       & T + 3\tau_R < t \le 2T + 3\tau_R, \\
    -\sin\big(\Omega_{\rm eff}(t - 2T)\big) & 2T + 3\tau_R < t \le 2T + 4\tau_R, \\
    0                                       & \mbox{otherwise}.
  \end{array}
  \right.
\ee
This function is illustrated in \Fig \ref{fig:g(t)-3pulse}.

\subsection{Interferometer response to laser phase noise}

\begin{figure}[!t]
  \centering
  \includegraphics[width=0.6\textwidth]{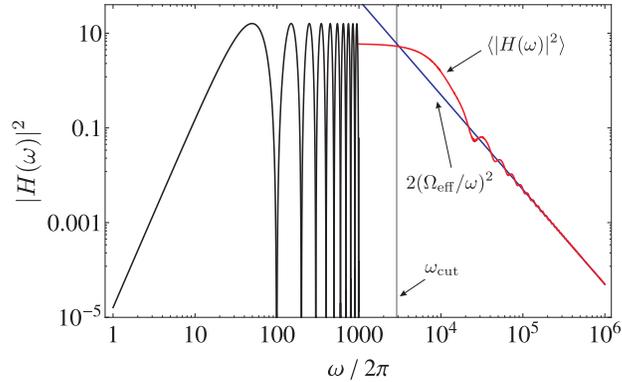}
  \caption{(Colour online) Plot of the phase noise weight function, $|H(\omega)|^2$, for the three-pulse interferometer with $T = 10$ ms, $\tau_R = 50$ $\mu$s and $\Omega_{\rm eff} = \pi/2\tau_R = 2\pi \times 5$ kHz. Here, the black curve indicates $|H(\omega)|^2$, which has been terminated at 1 kHz. The red curve shows the average of $|H(\omega)|^2$ over one period of oscillation, $(T+\tau_R)^{-1}$. In blue is the scale factor for the weight function at large frequencies, $2(\Omega_{\rm eff}/\omega)^2$. The gray vertical line shows the effective cut-off frequency of $|H(\omega)|^2$ at $\omega_{\rm cut} = \Omega_{\rm eff}/\sqrt{3} \sim 2\pi \times 2.9$ kHz.}
  \label{fig:H2}
\end{figure}

It is interesting to understand how this interferometer responds to phase noise at a given frequency. Recall that the standard deviation of interferometer phase noise, $\sigma_{\Phi}^{\rm rms}$, is composed of a sum over the laser phase noise harmonics, $S_{\phi}(\omega)$, weighted by $|H(\omega)|^2$ [see \Eq \refeqn{sigmaPhirms}]. Thus, to investigate the interferometer response to phase noise, we first compute the transfer function, $H(\omega)$, using \Eqs \refeqn{G}, \refeqn{H} and \refeqn{g(t)-3pulse}:
\be
  \label{H-3pulse}
  H(\omega) = \frac{2 i \omega \Omega_{\rm eff}}{\omega^2 - \Omega_{\rm eff}^2}
  \left[ \sin\big(\omega(T+2\tau_R)\big) + 2\frac{\Omega_{\rm eff}}{\omega} \sin\left(\frac{\omega T}{2}\right) \sin\left(\frac{\omega(T+2\tau_R)}{2}\right) \right].
\ee
An example of the weight function, $|H(\omega)|^2$, is displayed in \Fig \ref{fig:H2}, which has two important features. First, for frequencies much less than the Rabi frequency ($\omega \ll \Omega_{\rm eff}$), the transfer function can be approximated by
\be
  H(\omega) \approx -4i \sin^2\left(\frac{\omega (T+\tau_R)}{2}\right),
\ee
which originates from the second term in \Eq \refeqn{H-3pulse}. In this regime, the weight function oscillates periodically, with zeroes at integer multiples of the fundamental harmonic: $f_0 = (T+\tau_R)^{-1}$. Thus, the interferometer is relatively insensitive to phase noise at frequencies much less than $f_0$, since the weight function scales as $|H(\omega)|^2 \sim \omega^4(T+\tau_R)^4 \ll 1$. Second, for frequencies $\omega \gg \Omega_{\rm eff}$, the transfer function is dominated by the first term in \Eq \refeqn{H-3pulse}:
\be
  H(\omega) \approx 2i \frac{\Omega_{\rm eff}}{\omega} \sin\big(\omega (T+2\tau_R)\big).
\ee
This expression indicates that there is a natural low-pass filtering of the higher harmonics due to the finite duration Raman pulses. As a result, the sensitivity of the interferometer to high-frequency phase noise scales as $(\Omega_{\rm eff}/\omega)^2$, with an effective cut-off frequency at $\omega_{\rm cut} = \Omega_{\rm eff}/\sqrt{3}$. These features have been confirmed experimentally in \Ref \cite{Cheinet-IEEETransInstrumMeas-2008}.

In order to correctly evaluate the sensitivity of the interferometer to phase noise, it is necessary to take into account the fact that interferometer measurements are pulsed cyclically at a rate $f_c = 1/T_c$. A natural tool to characterize the sensitivity is the Allan variance of the atom interferometric phase fluctuations:
\be
  \sigma_{\Phi}^2(\tau_{\rm avg}) = \frac{1}{2} \lim_{n \to \infty}
  \left[ \frac{1}{n} \sum_{k=1}^n \big( \left< \delta\Phi_{k+1} \right> - \left< \delta\Phi_{k} \right> \big)^2 \right],
\ee
where $\left< \delta\Phi_{k} \right>$ is the mean value of $\delta\Phi$ over the measurement interval $[t_k,t_{k+1}=t_k + \tau_{\rm avg}]$ of duration $\tau_{\rm avg}$, which is an integer multiple of the cycle time: $\tau_{\rm avg} = mT_c$. For large enough averaging times $\tau_{\rm avg}$, where the fluctuations between successive averages are not correlated, the Allan variance can be shown to be
\be
  \sigma_{\Phi}^2(\tau_{\rm avg}) = \frac{1}{\tau_{\rm avg}} \sum_{n=1}^{\infty} |H(2\pi n f_c)|^2 S_{\phi}(2\pi n f_c).
\ee
This expression indicates that the sensitivity of the interferometer is limited by an aliasing phenomenon similar to the Dick effect for atomic clocks \cite{Dick-Proc19PTTI-1987}. Only the phase noise at harmonics of the cycling frequency $f_c$ contributes to the Allan variance, and they are weighted by the square of the transfer function at these frequencies.

\subsection{Sensitivity to mirror vibrations}

The sensitivity function can also be used to investigate the response of the interferometer to motion of the retro-reflection mirror that acts as the inertial reference frame for absolute measurements of inertial effects. In this case, the phase noise can be expressed as $\phi(t) = \bm{k}_{\rm eff} \cdot \bm{r}(t)$, where $\bm{r}(t)$ represents the time-dependent position of the mirror. Using \Eq \refeqn{deltaPhi}, the change in the interferometer phase due to mirror motion is
\be
  \label{deltaPhi_v}
  \delta\Phi_v = \int_{-\infty}^{\infty} g(t) \bm{k}_{\rm eff} \cdot \bm{v}(t) \diff t,
\ee
where $\bm{v}(t) = \dot{\bm{r}}(t)$ is the velocity of the mirror. Using the chain rule, \Eq \refeqn{deltaPhi_v} can be converted to a more useful form:
\be
  \label{deltaPhi_a}
  \delta\Phi_a = -\bm{k}_{\rm eff} \cdot \Big[ f(t) \bm{v}(t) \Big]_{-\infty}^{\infty} + \bm{k}_{\rm eff} \cdot \int_{-\infty}^{\infty} f(t) \bm{a}(t) \diff t.
\ee
Here, $\bm{a}(t) = \dot{\bm{v}}(t)$ is the acceleration noise of the mirror and $f(t)$ is called the response function of the interferometer, which is defined as
\be
  f(t) = -\int_{0}^{t} g(t') \diff t'.
\ee
In what follows, we will illustrate how $f(t)$ characterizes the sensitivity of the three-pulse interferometer to mirror vibrations. Integrating the sensitivity function given by \Eq \refeqn{g(t)-3pulse}, we find
\be
  \label{f(t)-3pulse}
  f(t) = \left\{
  \begin{array}{cc}
    \frac{1}{\Omega_{\rm eff}} \big( 1 - \cos\Omega_{\rm eff} t \big) &
    0 < t \le \tau_R, \\

    t + \frac{1}{\Omega_{\rm eff}} - \tau_R &
    \tau_R < t \le T + \tau_R, \\

    T + \frac{1}{\Omega_{\rm eff}} \big(1 - \cos \Omega_{\rm eff}(t - T) \big) &
    T + \tau_R < t \le T + 3\tau_R, \\

    2T + 3\tau_R + \frac{1}{\Omega_{\rm eff}} - t &
    T + 3\tau_R < t \le 2T + 3\tau_R, \\

    \frac{1}{\Omega_{\rm eff}} \big( 1 - \cos \Omega_{\rm eff}(t-2T) \big) &
    2T + 3\tau_R < t \le 2T + 4\tau_R, \\

    0
    & \mbox{otherwise}.
  \end{array}
  \right.
\ee
%
\begin{figure}[!t]
  \centering
  \includegraphics[width=0.6\textwidth]{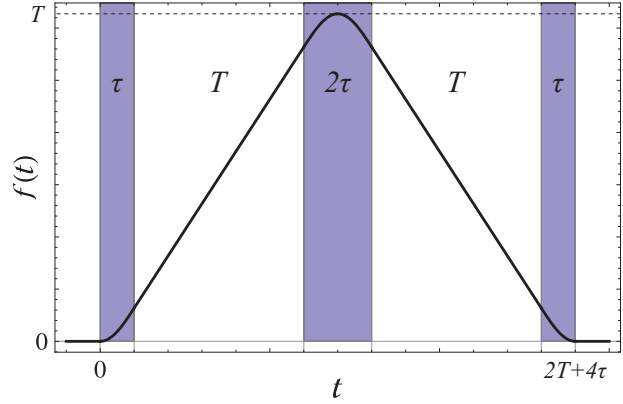}
  \caption{(Colour online) Plot of the response function, $f(t)$, given by \Eq \refeqn{f(t)-3pulse}. Again, the pulse duration satisfies $\Omega_{\rm eff} \tau = \pi/2$.}
  \label{fig:f(t)-3pulse}
\end{figure}
%
The response function for the three-pulse interferometer is a triangle-shaped function with units of time, as shown in \Fig \ref{fig:f(t)-3pulse}. Since it is equal to zero outside of the interval $t \in [0, 2T+4\tau_R]$, the first term in \Eq \refeqn{deltaPhi_a} vanishes and the phase variation of the interferometer due to the acceleration noise of the mirror is
\be
  \label{deltaPhi_a2}
  \delta\Phi_a = \bm{k}_{\rm eff} \cdot \int_{-\infty}^{\infty} f(t) \bm{a}(t) \diff t.
\ee
At its heart, this expression is a generalization of the phase shift $\bm{k}_{\rm eff}\cdot\bm{a}T^2$ produced by atoms moving in a non-inertial reference frame with a constant acceleration.\footnote{Equation \refeqn{deltaPhi_a2} can be evaluated with a constant acceleration to obtain: $\delta\Phi_a = \bm{k}_{\rm eff} \cdot \bm{a} (T+2\tau_R)(T+4\tau_R/\pi)$, which reduces to the well-known result $\bm{k}_{\rm eff} \cdot \bm{a} T^2$ in the limit of short Raman pulses: $\tau_R \ll T$.} Here, $f(t)$ acts as a weight function that determines how strongly the mirror acceleration at time $t$ contributes to the interferometer phase shift. The phase contributions are smallest near $t = 0$ and $2T + 4\tau_R$, where the wavepacket separation is a minimum. Similarly, the weight is strongest near the mid-point, $t = T + 2\tau_R$, where the separation between the interfering states is a maximum.

\begin{figure}[!t]
  \centering
  \subfigure[]{\includegraphics[width=0.45\textwidth]{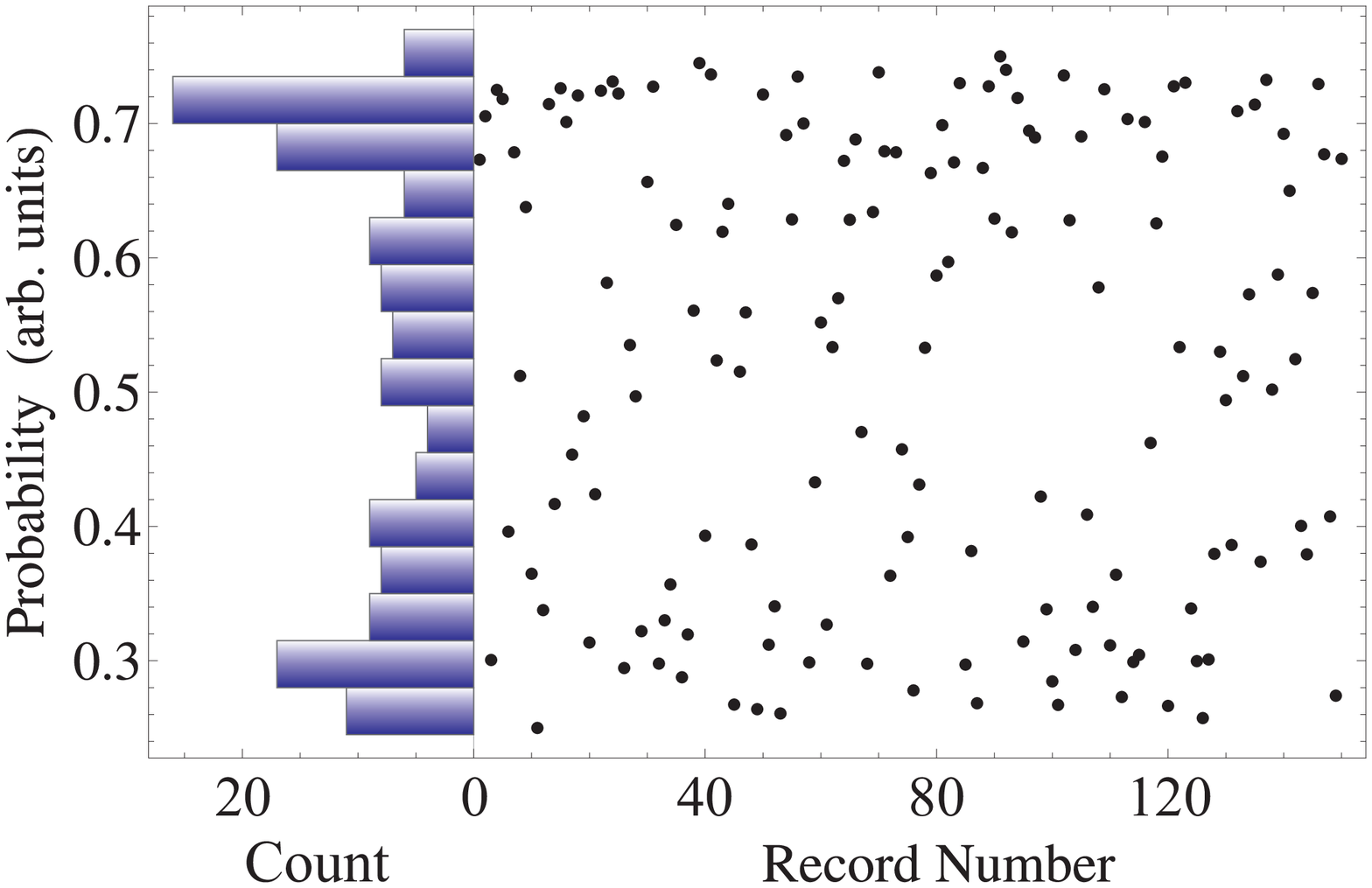}}
  \hspace{0.5cm}
  \subfigure[]{\includegraphics[width=0.45\textwidth]{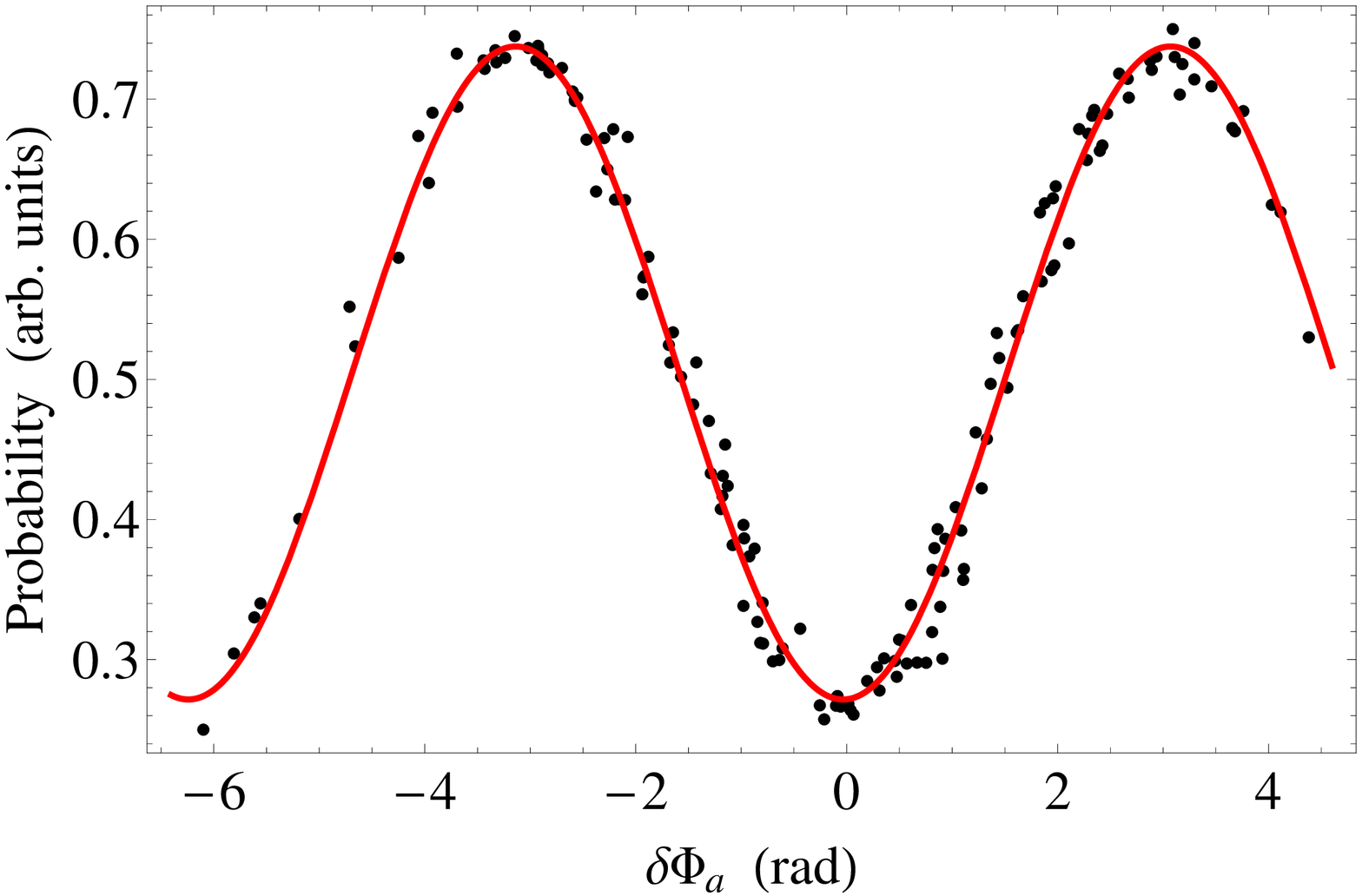}}
  \caption{(Colour online) (a) Transition probability of a three-pulse $^{87}$Rb interferometer with total interrogation time $2T = 6$ ms, measured in the presence of strong Raman mirror vibrations (standard deviation of acceleration noise: $\sigma_a \sim 1.4 \times 10^{-3}$ $g$). The data are plotted chronologically. On the left is a histogram of the measured probabilities, which resembles the probability distribution of $\cos^{-1}(\phi)$. The double-peaked structure indicates that the interferometer is operating normally, but the Raman phase is randomized by the mirror vibrations. (b) Same data as shown in (a) plotted as a function of the acceleration-induced phase, $\delta\Phi_a$, given by \Eq \refeqn{deltaPhi_a2}. The time-dependent acceleration, $a(t)$, was recorded for each shot of the experiment with a mechanical accelerometer (Colibrys SF3600A) attached to the Raman mirror. These data clearly show that the interferometer fringes can be reconstructed with a high degree of accuracy even in noisy environments.}
  \label{fig:RbFringes-Noise}
\end{figure}

Equation \refeqn{deltaPhi_a2} suggests that, if the acceleration of the retro-reflecting mirror is measured during the interferometer pulse sequence, one can correct for changes in the mirror position that induce parasitic phase shifts on the atoms. This principle is illustrated in \Fig \ref{fig:RbFringes-Noise}, where the initially randomized signal from a Mach-Zehnder interferometer in a noisy environment is recovered using the aforementioned analysis. Here, the fringes are effectively ``scanned'' by vibrations on the retro-reflecting Raman mirror. This has also been demonstrated with a mobile matter-wave interferometer in a micro-gravity environment during parabolic flights onboard a zero-g aircraft \cite{Geiger-NatureComm-2011}. We give a detailed description of this experiment and recent results in \Sec \ref{sec:ICE}.

%% file: Varenna-InertialSensors.tex
\section{Inertial sensors based on atom interferometry}
\label{sec:InertialSensors}

In general, an inertial sensor is a device that can detect changes in momentum, for example, a change in direction caused by rotation, or a change in velocity caused by the presence of a force. High precision inertial sensors have found scientific applications in the areas of general relativity, geophysics and geology, as well as industrial applications, such as the non-invasive detection of massive objects, or oil and mineral prospecting.

In the years following the first demonstration of an atom interferometer, many theoretical and experimental studies were carried out to investigate these new kinds of inertial sensors \cite{Berman-Book-1997}. To date, ground-based experiments using atomic gravimeters (measuring acceleration) \cite{Peters-Nature-1999, Peters-Metrologia-2001, Bertoldi-EuroPhysJD-2006, Bodart-ApplPhysLett-2010, Bidel-ApplPhysLett-2013}, gravity gradiometers (measuring acceleration gradients) \cite{Snadden-PRL-1998, McGuirk-PRA-2002, Bidel-ApplPhysLett-2013} and gyroscopes (measuring rotations) \cite{Gustavson-PRL-1997, Gustavson-ClassQuantumGrav-2000, Canuel-PRL-2006} have been realized and proved to be competitive with existing optical or artifact-based devices.

\begin{figure}[!t]
  \centering
  \includegraphics[width=0.99\textwidth]{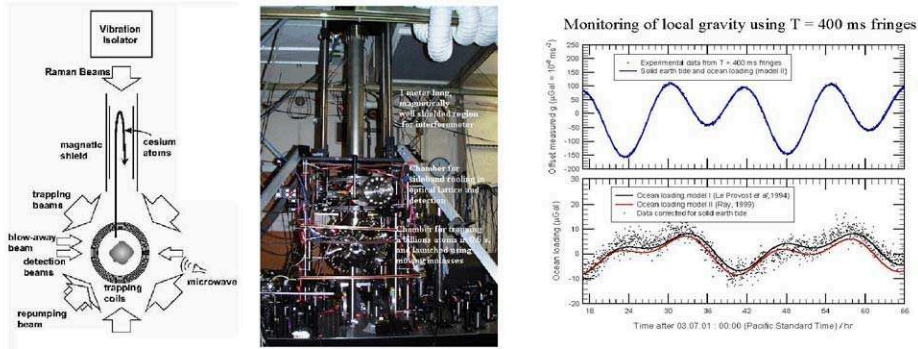}
  \caption{(Colour online) The atomic-fountain-based gravimeter developed by the Chu group at Stanford during the 1990's \cite{Kasevich-ApplPhysB-1992, Peters-Nature-1999, Peters-Metrologia-2001}. On the right is a two-day recording of the variation of gravity. The high accuracy enables ocean loading effects to be observed. Photo courtesy of S. Chu and M. Kasevich.}
  \label{fig:ChuExperiment}
\end{figure}

In this section, we present a brief summary of different inertial sensors based on atom interferometry that were designed as proof-of-principle experiments for use only in the laboratory. A classic example of such an experiment is the gravimeter developed at Stanford in the early 1990s shown in \Fig \ref{fig:ChuExperiment}. Later, in \Sec \ref{sec:MobileSensors}, we focus on projects designed for ``field'' use and give detailed descriptions of some mobile sensors developed by our research groups.

\subsection{Accelerometers and gravimeters}
\label{sec:Gravimeters}

\begin{figure}[!t]
  \centering
  \includegraphics[width=0.6\textwidth]{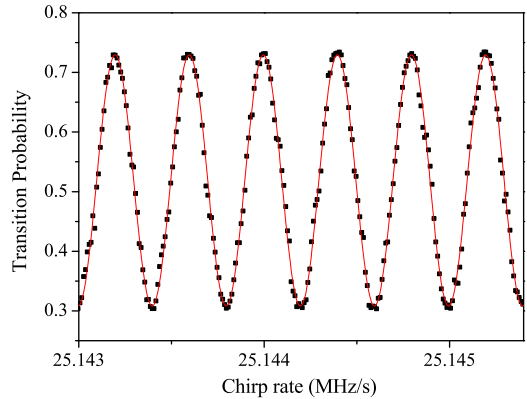}
  \caption{(Colour online) Transition probability as a function of the Raman beam chirp rate, $\alpha$, for $T = 50$ ms. Taken from \Ref \cite{Cheinet-ApplPhysB-2006}.}
  \label{fig:Fringes-ChirpRate}
\end{figure}

If the three light pulses of the interferometer sequence are separated only in time, and not in space, the interferometer is in an accelerometer (or gravimeter) configuration. For a uniform acceleration $\bm{a}$, in the atom's frame the frequency of the Raman lasers changes linearly with time at a rate of $-\bm{k}_{\rm eff} \cdot \bm{a}$. The resulting phase shift that arises from the interaction between the light and the atoms can be shown to be (see \Sec \ref{sec:Theory})
\be
  \label{Phi_a}
  \Phi_a = \bm{k}_{\rm eff} \cdot \bm{a} T^2 + (\phi_1 - \phi_2 + \phi_3).
\ee
Similarly, if the Raman beams are oriented along the vertical, the gravitationally induced chirp on the Raman frequency is $-\bm{k}_{\rm eff} \cdot \bm{g}$. In this case, the usual procedure to measure $g$ is to chirp the frequency of the Raman beams during the pulse sequence, such that the Doppler frequency of the atoms is canceled. The chirp rate, $\alpha$, that compensates the Doppler shift is determined by the relation \cite{Cheinet-ApplPhysB-2006}
\be
  \label{Phi_g}
  \Phi_g = (\bm{k}_{\rm eff} \cdot \bm{g} - \alpha) T^2 = 0.
\ee
This expression can be obtained from \Eq \refeqn{Phi_a} by setting the phases, $\phi_j = \phi(t_j) = -\alpha t_j^2/2$, where $t_j$ is the onset time of each pulse. The transition probability of the interferometer then oscillates sinusoidally as a function of $\Phi_g$, as shown in \Fig \ref{fig:Fringes-ChirpRate}. The central fringe, for which $\alpha = k_{\rm eff} g$, stays fixed for all values of $T$.

It should be noted that the phase shifts given by \Eqs \refeqn{Phi_a} and \refeqn{Phi_g} do not depend on the initial atomic velocity or on the mass of the particle---a direct consequence of the equivalence principle. The first precision cold atom gravimeter \cite{Peters-Metrologia-2001} achieved a resolution of 20 $\mu$Gal (1 $\mu$Gal = $10^{-9} \, g$) after one measurement cycle lasting 1.3 s. When compared to the best classical devices (such as the Scintrex FG5, which is based on optical interferometry with a falling corner cube), the two values of $g$ agreed to within 7 $\mu$Gal after accounting for systematic effects.

Following this first demonstration, atom gravimeters are currently under development at many institutions, some of which have already demonstrated improved performances. In particular, a record short-term sensitivity of 4.2 $\mu$Gal at 1 s was demonstrated in \Ref \cite{Hu-PRA-2013}, and a direct comparison between an atomic and a corner-cube gravimeter at their best level of performance, operating simultaneously in a low-noise environment, has recently confirmed the superior stability of the atomic device \cite{Gillot-Metrologia-2014}. Also, systematic effects have been thoroughly investigated, leading to an improved consolidated accuracy budget. An accuracy of a few $\mu$Gal has been claimed in \Ref \cite{Louchet-Chauvet-NJP-2011} and confirmed by the agreement found with the reference value obtained by averaging the measurements of a large ensemble of gravimeters at the last ``Key Comparisons of Absolute Gravimeters'' in 2009 and 2011 \cite{Jiang-Metrologia-2012, Francis-Metrologia-2013}, where so far the LNE-SYRTE gravimeter was the first and the only atom gravimeter to have participated.


The main limitation of this kind of gravimeter on Earth is due to spurious accelerations of the reference platform. One possibility for overcoming this problem is to measure the vibration of the platform using a sensitive mechanical accelerometer and correcting for phase fluctuations either in post-analysis or in real-time, as we will discuss in \Sec \ref{sec:ICE}. Another option is to perform simultaneous measurements with two different atomic samples with the same reference platform. This offers the possibility of rejecting any common-mode vibration noise on the measurements \cite{Sorrentino-ApplPhysLett-2012, Bonnin-PRA-2013}. Furthermore, if the two samples are spatially separated, simultaneous measurements would be sensitive to spatial gradients in $g$, and would also allow one to suppress a variety of systematic effects. We discuss such an apparatus in the next section.

\subsection{Gradiometers}
\label{sec:Gradiometers}

Measurements of the gradient of gravitational fields have important scientific and industrial applications ranging from the measurement of the Newtonian constant of gravity, $G$, and tests of general relativity, to covert navigation, underground structure detection, and geodesy. Initially at Stanford University, the development of atom-interferometric gravity gradiometers has been followed by other advances either for Space or fundamental physics measurements \cite{Bertoldi-EuroPhysJD-2006, Fixler-Science-2007, Lamporesi-PRL-2008}. A crucial aspect of every design is its intrinsic immunity to spurious accelerations.

\begin{figure}[!t]
  \centering
  \includegraphics[width=0.8\textwidth]{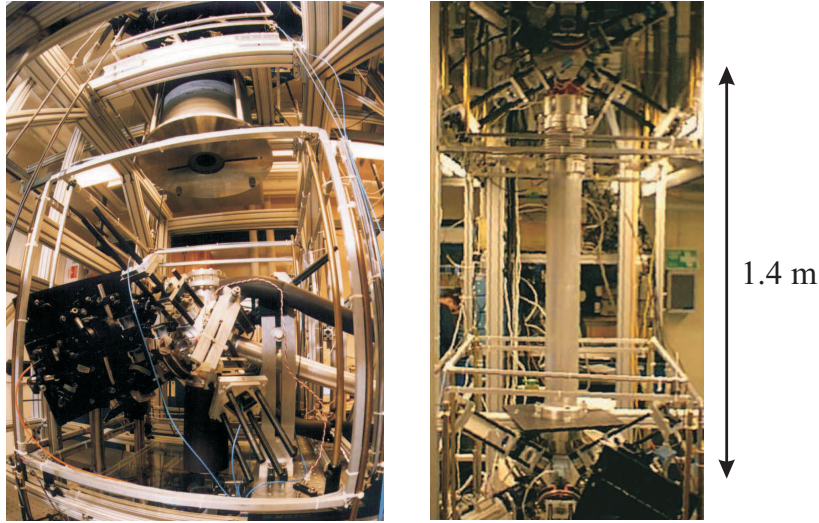}
  \caption{(Colour online) Gravity gradiometer developed in Stanford \cite{McGuirk-PRA-2002}. This system was also used for measurements of the Newtonian constant $G$ (see \Ref \cite{Fixler-Science-2007}) using a 540 kg mass shown at the top of the photo on the left. Photos courtesy of M. Kasevich.}
  \label{fig:Gradiometer}
\end{figure}

The gradiometer setup is illustrated in \Fig \ref{fig:Gradiometer}. It measures, simultaneously, the acceleration of two separate laser-cooled ensembles of atoms. The geometry is chosen so that the measurement axis passes through both atomic samples.  Since the acceleration measurements are made simultaneously at both positions, many systematic measurement errors, including the vibration of the experimental platform, are common to both measurements and can be subtracted. The relative acceleration of the two ensembles along the axis defined by the Raman beams is measured by subtracting the measured phase shifts $\Phi(\bm{r}_1)$ and $\Phi(\bm{r}_2)$ at the two locations $\bm{r}_1$ and $\bm{r}_2$. The gradient is extracted by dividing the relative acceleration by the separation of the ensembles. However, this method determines only one component of the gravity gradient tensor.

This type of instrument is fundamentally different from state-of-the-art classical sensors that are designed, for example, to measure $G$. First, the proof-masses are individual atoms rather than precisely machined macroscopic objects. This reduces systematic effects associated with the material properties of these objects. Second, the calibration for the two accelerometers is referenced to the wavelength of a single pair of frequency-stabilized laser beams, and is identical for both accelerometers. This provides long term accuracy. Finally, large separations ($\gg$ 1 m) between accelerometers are possible, enabling the development of high-sensitivity instruments. The apparatus shown in \Fig \ref{fig:Gradiometer}, with a separation of 1.4 m, has demonstrated a differential acceleration sensitivity of $4 \times 10^{-9} \, g/\sqrt{\rm Hz}$, corresponding to gravity-gradient sensitivity of 4 E$/\sqrt{\rm Hz}$ (1 E = $10^{-9}$ s$^{-2}$) \cite{McGuirk-PRA-2002}.

More recently, a compact gravimeter (consisting of just one atomic source) measured the vertical gravity gradient with a precision of 4 E \cite{Bidel-ApplPhysLett-2013}. This was done by placing the instrument on an elevator and measuring $g$ at various heights both above and below ground level.

\subsection{Gyrometers}

In the case of a spatial separation of the laser beams, and when the atoms have a velocity component perpendicular to $\bm{k}_{\rm eff}$, the interferometer is in a configuration similar to an optical Mach-Zehnder interferometer. Then, in addition to accelerations, the interferometer is also sensitive to rotations. This is the matter wave analog to the optical Sagnac effect. For a Sagnac loop enclosing an area $\bm{A}$, a rotation $\bm{\Omega}$ produces a phase shift (to first order in $\bm{\Omega}$) of
\be
  \varphi_{\rm rot} = \frac{4\pi}{\lambda_{\rm dB} v_l} \, \bm{\Omega} \cdot \bm{A}.
\ee
Here, $\lambda_{\rm dB}$ is the de Broglie wavelength and $v_l$ the atom's longitudinal velocity. The area $A$ of the interferometer depends on the distance between two light pulses, $L$, and on the transverse velocity $\bm{v}_t = \hbar \bm{k}_{\rm eff}/M$:
$$
  A = L^{2} \frac{v_t}{v_l}.
$$
For the Mach-Zehnder atom interferometer, the phase shift due to the rotation takes the same form as that of an acceleration, except the free evolution time becomes $T = L/v_l$ and the acceleration becomes $\bm{a}_{\rm cor} = -2 (\bm{\Omega} \times \bm{v})$ (the Coriolis acceleration)
\be
  \label{Phi_rot}
  \Phi_{\rm rot} = \bm{k}_{\rm eff} \cdot \bm{a}_{\rm cor} \left(\frac{L}{v_l}\right)^2 + (\phi_1 - \phi_2 + \phi_3).
\ee
By utilizing the small de Broglie wavelength of massive particles, atom interferometers can achieve a much higher sensitivity to rotations than optical interferometers with the same area. An atomic gyroscope \cite{Gustavson-PRL-1997, Gustavson-ClassQuantumGrav-2000}, using thermal caesium atomic beams (where the most-probable longitudinal velocity was $v_l \sim 300$ m/s) and with an overall interferometer length of 2 m has demonstrated a sensitivity of $6 \times 10^{-10}$ rad/s/$\sqrt{\rm Hz}$. The apparatus consists of a double interferometer using two counter-propagating sources of atoms, sharing the same lasers. The use of the two sources facilitates the discrimination between rotation and acceleration signals.

\subsection{Six-axis sensor}

\begin{figure}[!t]
  \centering
  \subfigure{\includegraphics[width=0.45\textwidth]{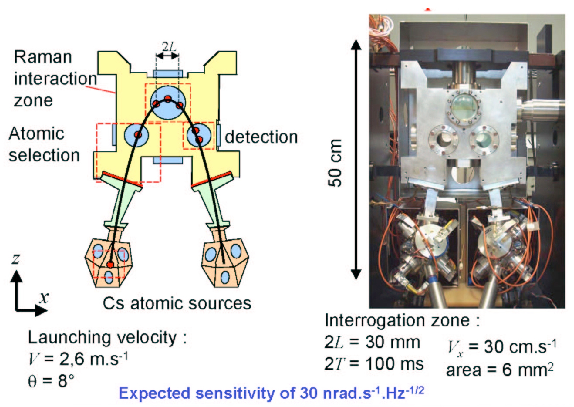}}
  \hspace{0.5cm}
  \subfigure{\includegraphics[width=0.45\textwidth]{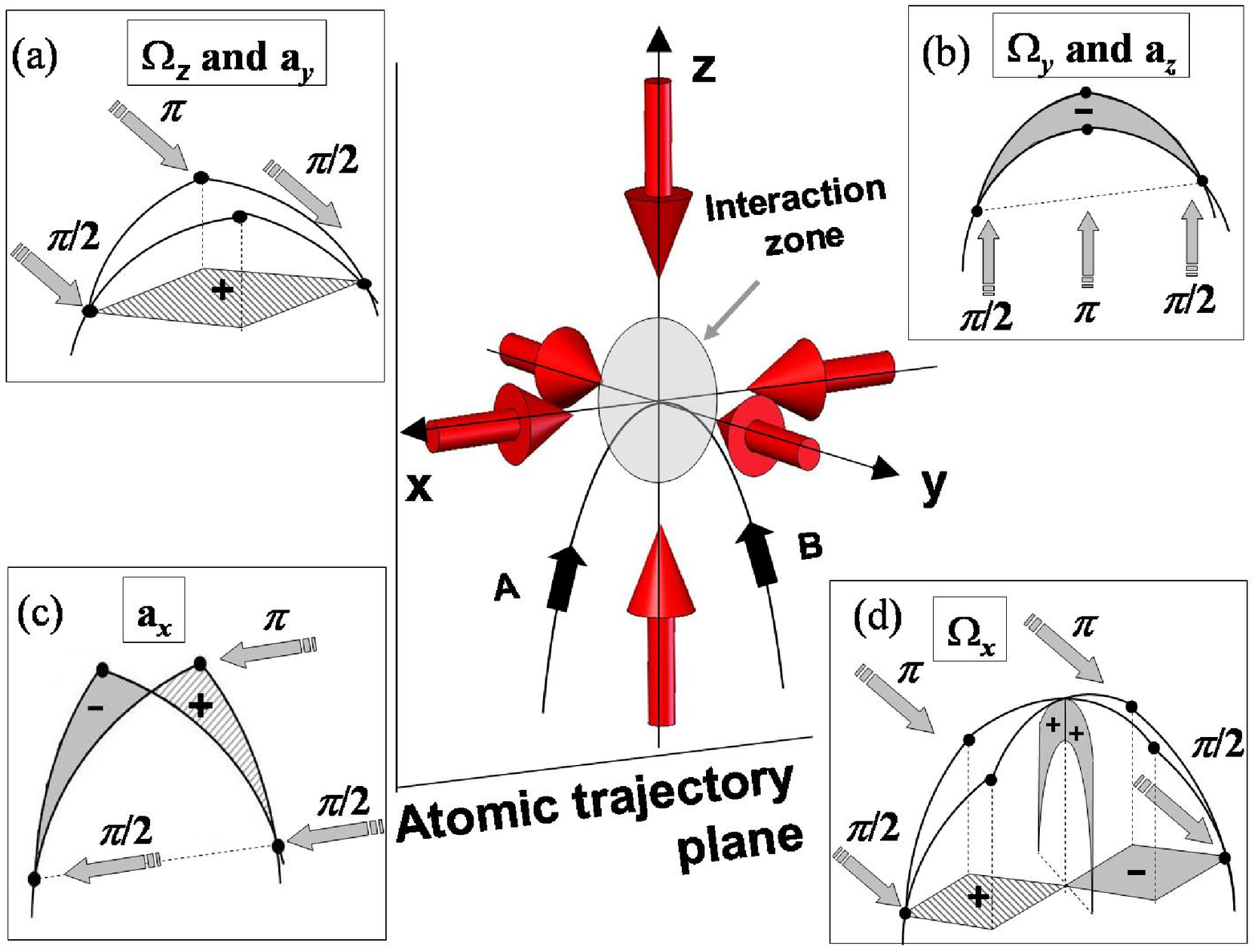}}
  \caption{(Colour online) Six-axis inertial sensor. On the left is a schematic of the experimental setup, and typical operating parameters. On the right we illustrate the principle of operation. Here, two atomic clouds, labeled ``A'' and ``B'', are launched on identical parabolic trajectories, but in opposite directions. The Raman lasers are pulsed on when the atoms are near the vertex of their trajectories. Four different interferometer configurations (a)--(d) give access to the 3 rotations and the 3 accelerations.}
  \label{fig:Six-Axis}
\end{figure}

The sensitivity axis of an interferometer is usually defined by the direction of the Raman interrogation laser with respect to the atomic trajectory. An experiment carried out in Paris \cite{Canuel-PRL-2006} demonstrated sensitivity to three mutually orthogonal accelerations and rotations by launching two atomic clouds in opposite parabolic trajectories. As illustrated in \Fig \ref{fig:Six-Axis}(a), with the usual $\pi/2 - \pi - \pi/2$ pulse sequence, a sensitivity to vertical rotation $\Omega_z$ and to horizontal acceleration $a_y$ can be achieved by placing the Raman lasers along the $y$-direction, perpendicular to the atomic trajectory. With the same sequence, using vertically-oriented lasers, the horizontal rotation $\Omega_y$ and vertical acceleration $a_z$ can be measured, as shown in \Fig \ref{fig:Six-Axis}(b). The phase shift in these two cases can be shown to be
\be
  \Phi_{\rm 3p} = \bm{k}_{\rm eff} \cdot \big[ \bm{a} -2 (\bm{\Omega} \times \bm{v}) \big] \, T^2.
\ee
It is also possible to access the other components of acceleration and rotation which lie along the $x$-axis (in the plane of the atomic trajectories). By utilizing these strongly curved launch trajectories, Raman lasers can be aligned along the $x$-direction---producing a sensitivity to the acceleration component $a_x$ but not to rotations [see \Fig \ref{fig:Six-Axis}(c)]. Access to the horizontal rotation $\Omega_x$ is achieved by changing the pulse sequence along the $y$-direction to a four-pulse ``butterfly'' configuration [see \Fig \ref{fig:Six-Axis}(d)]. This configuration was first proposed to measure gravity gradients \cite{McGuirk-PRA-2002}. It involves four pulses with areas $\pi/2 - \pi - \pi - \pi/2$, and separated by times $T/2 - T - T/2$, respectively. The projection of the interferometer area along the $x$-axis gives rise to a sensitivity to the $x$-component of rotation, $\Omega_x$, described by the phase shift
\be
  \Phi_{\rm 4p} = \frac{1}{2} \big[\bm{k}_{\rm eff} \times (\bm{g} + \bm{a}) \big] \cdot \bm{\Omega} \, T^3.
\ee
In contrast, the $z$-axis projection of the area cancels out, so the interferometer is insensitive to $\Omega_z$.

%% file: Varenna-MobileSensors.tex
\section{Compact and mobile inertial sensors}
\label{sec:MobileSensors}

Until now, we have discussed various applications of atom interferometry in terms of lab-based inertial sensors. These experiments are typically quite large, require a dedicated laboratory, and are designed to stay in one place. Furthermore, it is normal for sensors of this kind to operate well only in environments where the temperature, humidity, acoustic noise, \etc is tightly constrained. In this section, we describe three different projects that are designed to be compact, robust and mobile---making them distinctly different from most laboratory experiments. The development of this technology will help create a new generation of atomic sensors that can operate ``in the field'' under a broad range of environmental conditions.

\subsection{MiniAtom: a compact and portable gravimeter}
\label{sec:MiniAtom}

Here, we present the realization of a highly compact, absolute atomic gravimeter called ``MiniAtom'', which was developed jointly by labs at SYRTE (Observatoire de Paris) and LP2N (Institut d'Optique d'Aquitaine). The main purpose of this work is to demonstrate that atomic interferometers can overtake the current limitations of inertial sensors based on ``classical'' technologies for field and on-board applications in geophysics. We show that the complexity and volume of cold-atom experimental set-ups can be drastically reduced while maintaining performances close to state-of-the-art sensors---enabling such atomic sensors to perform precision measurements outside of the laboratory. As a feasible prototype, we chose to realize an absolute gravimeter to measure the acceleration of the Earth's gravity, which can be used to support geophysical surveys. This work has played an important role in the development of commercial cold atom gravimeters, one of which we will discuss in \Sec \ref{sec:muQuanS}.

The major design features---the reduction of the sensor head size and the significant simplification of the laser module---rely on the use of an innovative hollow pyramid as the retro-reflecting mirror of quantum inertial sensors and a laser system based on telecom technologies. This design allows us to perform all the steps of the atomic measurement (laser-cooling, selection, interferometry and detection) with just a single laser beam \cite{Bodart-ApplPhysLett-2010}. In contrast, other atomic gravimeters require up to 9 different optical beams (six beams for the MOT, one pusher beam, one for interferometry, and one for detection) coming from multiple frequency sources. As we will show, this key component is responsible for the simplifications of both the sensor head and the laser system.

\begin{figure}[!t]
  \centering
  \includegraphics[width=0.7\textwidth]{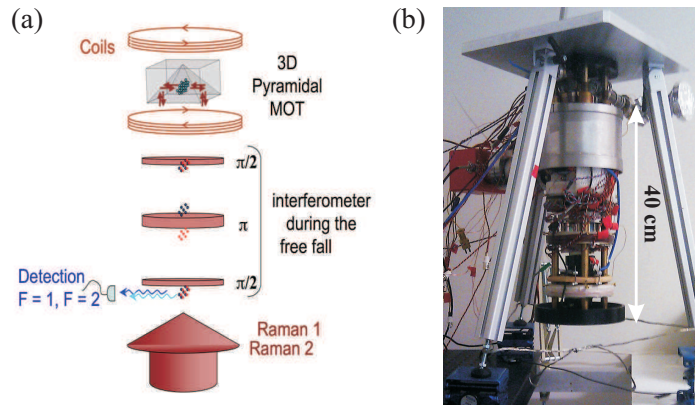}
  \caption{(Colour online) (a) Schematic of a compact gravimeter with only one laser beam and a pyramidal reflector. (b) The MiniAtom apparatus. The total height of the sensor head is 40 cm.}
  \label{fig:MiniAtom}
\end{figure}

The concept of a single beam interferometer performed with a pyramidal retro-reflector [illustrated in \Fig \ref{fig:MiniAtom}(a)] was validated on a previous experiment, as described in \Ref \cite{Bodart-ApplPhysLett-2010}. In that work, approximately $10^7$ $^{87}$Rb atoms were loaded from a vapor in $\sim 400$ ms. This is followed by 20 ms of molasses cooling which brings the atoms to a temperature of $\sim 2.5$ $\mu$K. A sequence of microwave and pusher-beam pulses selects the atoms in the state $\ket{F = 1, m_F = 0}$ at the beginning of their free-fall. Once the cloud has fallen clear of the pyramid, the two vertically-oriented, retro-reflected Raman beams are used to perform a velocity selection, followed by the usual $\pi/2 - \pi - \pi/2$ interferometer scheme. After the Raman pulses, the relative population between the two hyperfine ground states is obtained using fluorescence detection. With an interrogation time of $2T = 80$ ms, we demonstrated a relative sensitivity to $g$ of $1.7 \times 10^{-7}$ within one second of data acquisition, and we have shown promising long-term relative stability with a noise floor below $5 \times 10^{-9}$.

\begin{figure}[!t]
  \centering
  \includegraphics[width=0.5\textwidth]{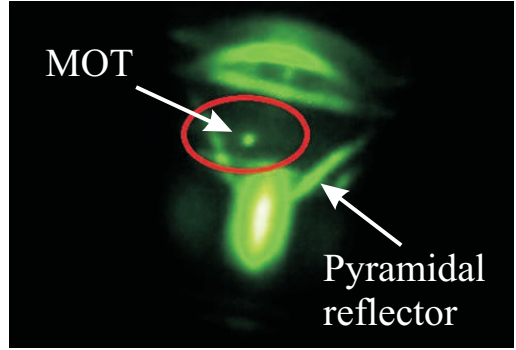}
  \caption{(Colour online) Image of the MiniAtom pyramidal MOT. The bright dot at the center is the fluorescence emitted by about $10^8$ atoms of $^{87}$Rb loaded from a background vapor.}
  \label{fig:MiniAtom-MOT}
\end{figure}

The sensor head [shown in \Fig \ref{fig:MiniAtom}(b)] consists of a 2 liter titanium vacuum chamber which is magnetically shielded by a single layer of mu-metal. The science chamber features several optical viewports to perform the atomic measurement sequence. Four indium-sealed rectangular windows are designed to measure the fluorescence of the atoms at the output of the interferometer. These viewports were made 10 cm long in order to adjust the trade-off between cycling-rate and sensitivity with respect to applications or environment. A maximum interrogation time of 100 ms for the interferometer is allowed, which is limited by the 15 cm height between the bottom viewport and the pyramidal reflector. To keep the design as simple as possible, we do not use any optics for imaging in the detection. Two sets of four 1 cm$^{2}$ area photodiodes allow for 3\% fluorescence collection efficiency for each state. The decrease in the number of optical beams has resulted in a drastic reduction of the volume of the sensor head---it fits in a cylinder 40 cm high and 20 cm in diameter, as shown in \Fig \ref{fig:MiniAtom}(b). In comparison, a separate transportable absolute gravimeter developed at SYRTE \cite{Louchet-Chauvet-NJP-2011} has a sensor head that is 80 cm high and 60 cm wide. Figure \ref{fig:MiniAtom-MOT} shows an image of the pyramidal MOT produced in the MiniAtom chamber.

The laser system was designed such that all the frequencies necessary for the gravimeter are carried along a single optical path with one linear polarization state. A liquid crystal variable retarder plate (LCVR) from Meadowlark is used to control the polarization state of the laser field reaching the science chamber at each step of the measurement. For the trapping and cooling stages, the LCVR creates a circular polarization from the incoming linear one so that after successive reflections on the faces of the pyramid, light is in the $\sigma^+/\sigma^-$ configuration. For the interferometer, the polarization is then changed to a linear state aligned at $45^{\circ}$ to the edges of the pyramid so that the two counter-propagating Raman beam polarizations are perpendicular. Just after the third Raman pulse, the polarization is switched back to circular to perform the state detection. An important feature of the reflector is that the faces of the pyramid have a special dielectric coating that prevents the two crossed polarizations to dephase from each other after successive reflections.

\begin{figure}[!t]
  \centering
  \includegraphics[width=0.5\textwidth]{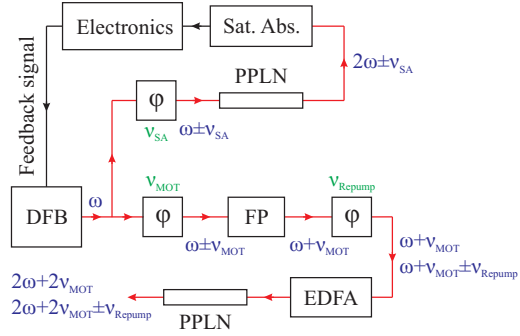}
  \caption{(Colour online) Architecture of the fiber-based laser system with only one laser diode. Quantities in blue represent the laser frequency at different parts of the optical chain, while quantities in green are modulation frequencies sent to the phase modulators. The modulations $\nu_{\rm MOT}$ and $\nu_{\rm repump}$ are set such that two output frequencies serve as the MOT and repump beams during the loading sequence, and they are adjusted to serve as the two Raman beams during the interferometry sequence. DFB = distributed feedback laser diode; $\varphi$ = electro-optic phase modulator; FP = Fabry-Perot interferometer; PPLN = periodically-poled lithium niobate waveguide; EDFA = erbium-doped fiber amplifier; Sat.~Abs.~= saturated absorption spectrometer.}
  \label{fig:MiniAtom-LaserSystem}
\end{figure}

For this project, we developed a compact laser architecture (see \Fig \ref{fig:MiniAtom-LaserSystem}) based on telecommunication technology with one key element: a periodically-poled lithium-niobate (PPLN) wave-guide (from NTT Electronics, Japan) which is used to frequency-double the 1560 nm laser source to 780 nm via second harmonic generation. This method of frequency doubling using a waveguide is particularly efficient, because the confinement of the optical mode within the guide leads to a conversion efficiency as high as 50\%. The telecom laser source is a cheap and convenient distributed feedback (DFB) laser diode.

A common laser architecture adopted in cold atom experiments is the master-slave configuration, where one fixed-frequency laser serves as a reference for multiple ``slave'' lasers whose frequencies are shifted relative to the ``master''. In this experiment, we use only one laser with a fixed optical frequency. The light from the DFB is split into two parts. A small amount of power is diverted to an electro-optic phase modulator (EOM) that shifts the laser frequency in such a way that, after doubling, the light is resonant with the $F = 3 \to F' = 4$ transition in $^{85}$Rb. It is then sent to a saturated absorption cell. The locking signal is deduced from synchronous detection at 5 MHz where the modulation is created using the same EOM. The second part of the DFB output is sent through another phase modulator driven by an independent yttrium iron garnet (YIG) oscillator whose detuning will be close to the $F = 2 \to F' = 3$ transition in $^{87}$Rb after frequency-doubling. The light is then filtered by a fiber-based Fabry-Perot cavity such that only one sideband remains. The frequency agility is supported by the very fast response of the EOM that enables the detuning for cooling and then for the Raman transition at $\sim 1$ GHz to the red of the transition. A third fiber-based EOM is used to create a second frequency that will be close to the $F = 1 \to F' = 2$ transition for repumping during the trapping stage or for the second Raman frequency during the interferometer. The beam carrying these two frequencies is then amplified by an erbium-doped fiber amplifier (EDFA) to produce enough power to obtain a sufficient frequency-doubling conversion efficiency. With this setup, we obtain an output power of 200 mW at 780 nm after fiber-coupling. This scheme allows us to change the frequency spacing in the optical domain by adjusting the rf modulation, which is accomplished almost instantaneously. As a result, the laser frequency can be rapidly controlled without changing the current of the laser diode.

Particular efforts have been made to integrate the frequency chain used to derive the 6.835 GHz reference for both the optical Raman transitions and the microwave pulse used for the quantum state selection. Although our frequency chain fits in a 2 liter volume, it features a phase noise that limits the sensitivity to gravity only at the level of $10^{-7}$ m/s$^{2}$ in one second. This is on the order of the best sensitivities achieved in the laboratory with the same interrogation time. Thus, this project has demonstrated an interesting trade-off between integration in a small package and a satisfying level of phase noise.

This prototype demonstrates that several mature pieces of technology can be gathered to produce precise measurements in a compact inertial sensor. Further work is being carried out to improve and simplify the filtering of ground vibrations. In addition, our sensor opens new doors toward the operation of an adjustable remote head gradiometer.

\subsection{Toward a commercial absolute quantum gravimeter}
\label{sec:muQuanS}

\begin{figure}[!t]
  \centering
  \includegraphics[width=0.8\textwidth]{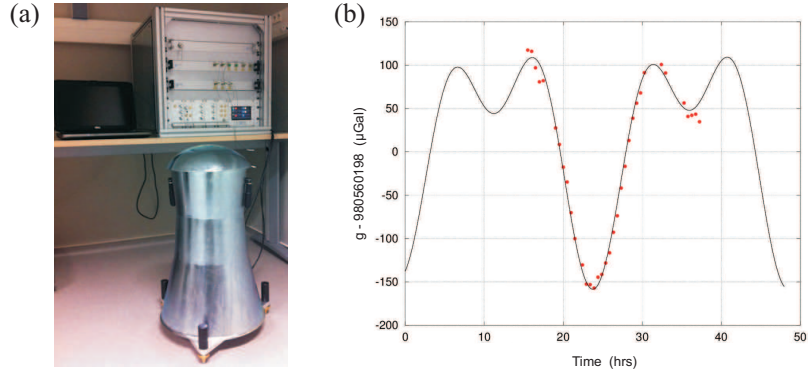}
  \caption{(Colour online) (a) Prototype of a commercial absolute quantum gravimeter. The laser and control electronics are shown in the 19 inch racks on the upper right. The gravimeter is surrounded by a layer of mu-metal to shield from external magnetic fields. (b) Preliminary measurements from the prototype taken over several hours (uncorrected for systematic effects). Here, the total interrogation time is $2T = 100$ ms, and the solid line is an overlayed tidal model that includes solid Earth and ocean-loading effects. The sensitivity is currently a few $\mu$Gal after 1000 seconds of integration. Photo and data courtesy of $\mu$QuanS.}
  \label{fig:muQuanS-gravimeter}
\end{figure}

As a result of the research involved with the MiniAtom project at two French laboratories (SYRTE in Paris and LP2N in Bordeaux), a commercial absolute quantum gravimeter is currently being developed for various applications in geophysics, including volcano monitoring, hydrology, and hydrocarbon and mineral exploration. The operational requirements for these applications are extremely stringent, but modern telecom laser technology presents very attractive features for the development of a high-performance absolute gravimeter compatible with field use.

The general architecture of the instrument is very similar to the one used in the MiniAtom experiment---it relies on the utilization of a pyramidal reflector, which enables all of the operations involved in the measurement sequence (cooling, interferometry, and detection) to be performed with a single laser beam. A strong technological effort was conducted in order to integrate the laser system required for the quantum manipulation of atoms and the driving electronics. The laser system is based on the utilization of a fiber-based telecom laser operating at 1560.48 nm, which is then amplified and frequency-doubled to the required wavelength of 780.24 nm. This compact design is extremely robust and reliable. A prototype of the gravimeter is shown in \Fig \ref{fig:muQuanS-gravimeter}, along with some preliminary gravity measurements taken over several hours.

\subsection{ICE: A mobile apparatus for testing the weak equivalence principle}
\label{sec:ICE}

\begin{figure}[!t]
  \centering
  \subfigure{\includegraphics[width=0.7\textwidth]{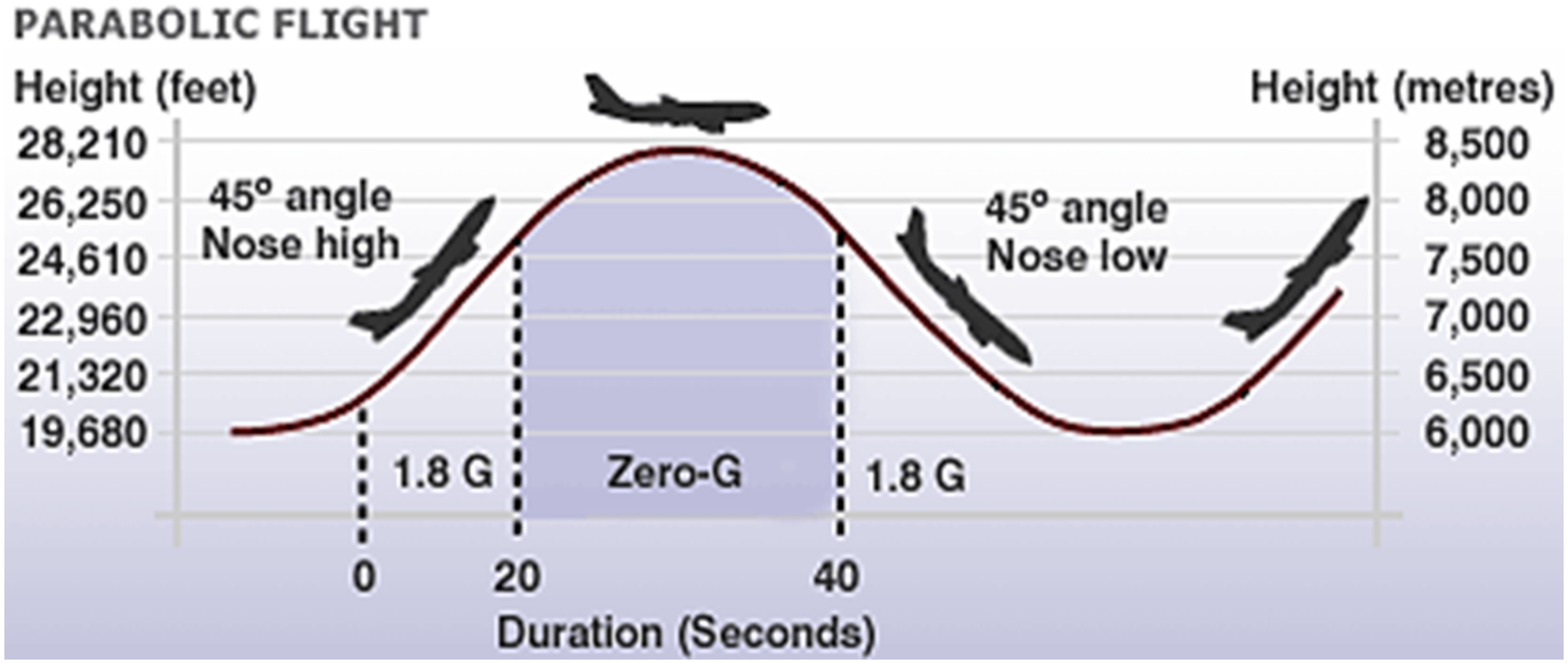}}
  \subfigure{\includegraphics[width=0.7\textwidth]{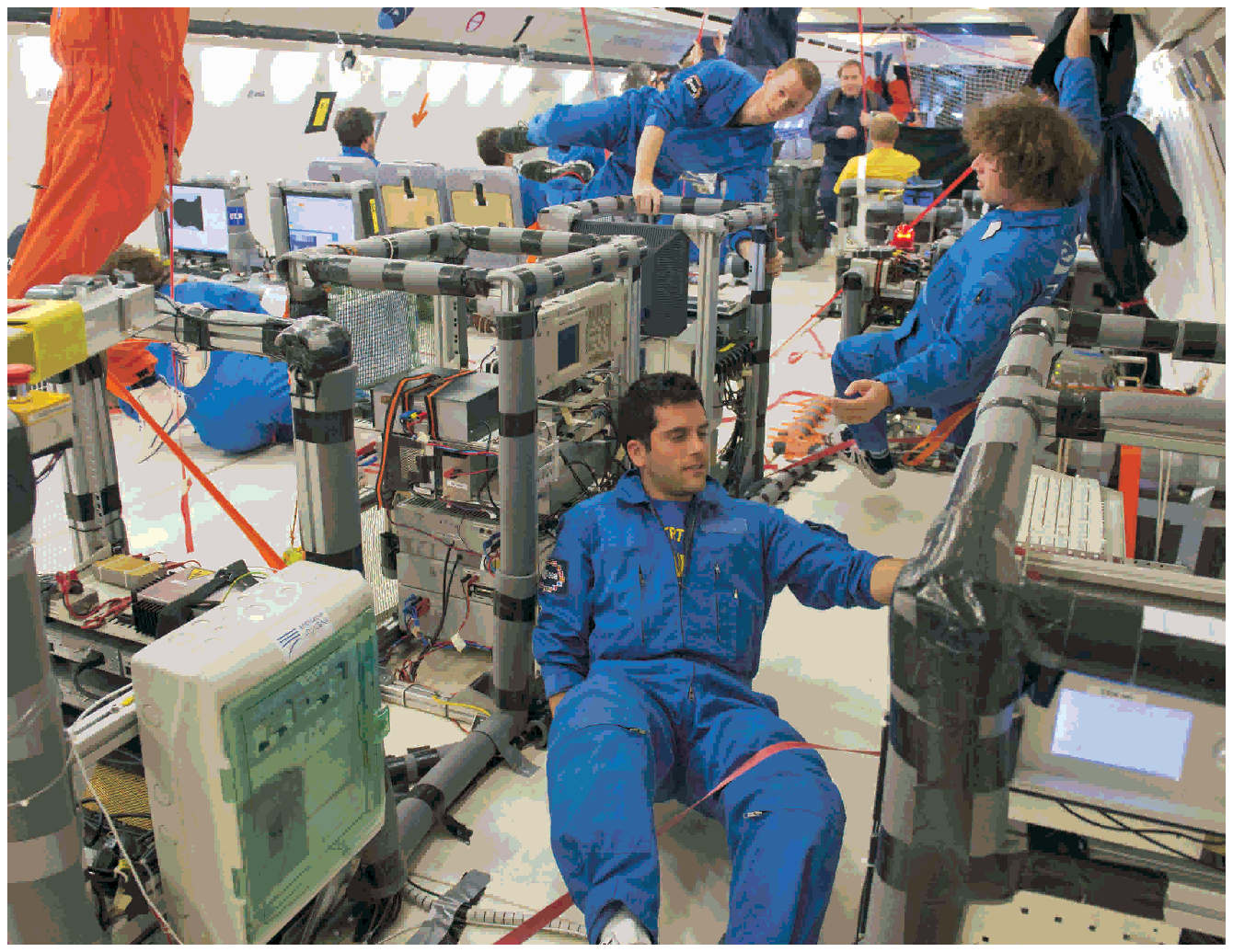}}
  \caption{(Colour online) Top figure: schematic of a parabolic flight in the zero-g A300 airbus, courtesy of Novespace. Bottom figure: the ICE team in micro-gravity. From top to bottom: B.~Barrett, B.~Battelier, and P.-A.~Gominet. Photo courtesy of the ESA and Novespace.}
  \label{fig:ParabolicFlight-ICESetup}
\end{figure}

The ICE experiment (an acronym for Interf\'{e}rometrie atomique \`{a} sources Coh\'{e}rentes pour l'Espace, or coherent atom interferometry for Space) is a compact and transportable dual-species atom interferometer. The main goal of ICE is to test the weak equivalence principle (WEP), also known as the universality of free-fall, which states that two massive bodies will undergo the same acceleration from the same point in space, regardless of their mass or internal structure. This principle is characterized by the E\"{o}tv\"{o}s parameter, $\eta$, which is the difference between the acceleration of two bodies, $a_1$ and $a_2$, divided by their average acceleration:
\be
  \eta = 2 \, \frac{a_1 - a_2}{a_1 + a_2}.
\ee
Historically, there have been a number of experiments to test the WEP using classical bodies. The most precise tests have previously been carried out using lunar laser ranging \cite{Williams-PRL-2004}, or using a rotating torsion balance \cite{Schlamminger-PRL-2008}, and have measured $\eta$ at the level of a few parts in $10^{13}$. Although these previous tests are very accurate, they were both done with classical objects. Various extensions to the standard model of particle physics have made predictions that would directly violate Einstein's equivalence principle \cite{Will-LivingRevRel-2006}, therefore it is interesting to test the WEP with ``quantum bodies''.

ICE aims to measure $\eta$ using a dual-species atom accelerometer that utilizes laser-cooled samples of $^{87}$Rb and $^{39}$K \cite{Varoquaux-NJP-2009, Geiger-NatureComm-2011}. By performing simultaneous measurements on the two spatially-overlapped atomic clouds, the acceleration of the two species can be measured and common-mode noise can be rejected. This concept is similar to the operation principle of gradiometers, as we discussed in \Sec \ref{sec:Gradiometers}.

The experiment is designed to perform this test in a micro-gravity environment (onboard the Novespace A300 ``zero-g'' aircraft) in order to extend the interrogation time, thereby increasing the sensitivity to acceleration. Similar research is being carried out in a lab-based experiment by a team in Paris that recently demonstrated a differential free-fall measurement at the level of $1.2 \times 10^{-7}\, g$ using a dual-species accelerometer with $^{85}$Rb and $^{87}$Rb \cite{Bonnin-PRA-2013}.

Other Earth-based atom interferometry experiments that exploit long interrogation times are taking place around the world. Two examples include the QUANTUS (Quantengase Unter Schwerelosigkeit --- Quantum gases under micro-gravity) experiment \cite{Vogel-ApplPhysB-2006, Muntinga-PRL-2013} at the ZARM drop tower in Bremen, Germany, and at Stanford University in a recently constructed 10 m vacuum chamber \cite{Dickerson-PRL-2013, Sugarbaker-PRL-2013}. However, the defining feature of interferometry experiments like ICE and QUANTUS is that the apparatus is designed to be in free-fall with the atoms.

\subsubsection{\textsf{Parabolic flights}}

On average, ICE takes part in two parabolic flight campaigns per year, which are organized by Novespace (based out of Bordeaux-M\'{e}rignac airport), and are funded by the European Space Agency (ESA) and the French Space agency (CNES). Each campaign consists of three flights where the zero-g A300 aircraft undergoes multiple parabolic trajectories, as shown in \Fig \ref{fig:ParabolicFlight-ICESetup}. Each flight typically contains 31 parabolas, and each of those consists of approximately 20 s of micro-gravity when the aircraft is in free-fall. This amounts to approximately 10 minutes of $0g$ per flight, or just over 30 minutes for the entire campaign.

During one parabola, the experiment has the potential of reaching a maximum interrogation time of $2T \sim 20$ s. In comparison, the QUANTUS experiment in the ZARM drop tower is currently limited to $2T = 4.7$ s, with plans to extend this to $2T = 9.4$ s when the tower is modified to accommodate a launched capsule \cite{Muntinga-PRL-2013}. Similarly, the 10 m fountain at Stanford has recently demonstrated $2T = 2.3$ s \cite{Dickerson-PRL-2013}.

One advantage of the A300 plane is that the experiment can be controlled in real-time during the flight, offering the possibility of changing experimental conditions ``on the fly''. The disadvantage is that there are many constraints to working on a plane---especially one that undergoes such extreme flight paths. For example, the experiment must be able to withstand the stress of frequent trips between the lab and the airport. During the flight, strong vibrations and changes in gravity call for stringent requirements on the mechanical structure. Power restrictions on the plane require that the experiment be turned off periodically during the flight, and overnight between flights. Finally, since the aircraft is not insulated, the temperature can vary by as much as $20^{\circ}$C throughout the day. These issues have presented many technical challenges to overcome when designing the experiment, but it has lead to the development of a very stable and robust setup that is capable of sensitive acceleration measurements in a noisy environment.

\begin{figure}[!t]
  \centering
  \includegraphics[width=0.8\textwidth]{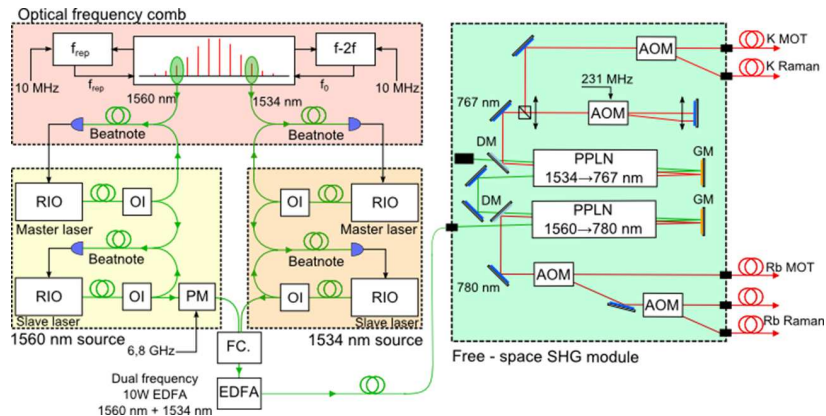}
  \caption{(Colour online) The ICE laser system. RIO = RIO laser diode; OI = optical isolator; PM = electro-optic phase modulator; PPLN = periodically-poled lithium niobate; FC = fiber combiner; AOM = acousto-optic modulator; EDFA = erbium-doped fiber amplifier; DM = dichroic mirror; GM = gold mirror.}
  \label{fig:ICE-LaserSystem}
\end{figure}

\subsubsection{\textsf{Experimental setup}}

We now give a brief description of the experimental setup and the laser system developed for the dual-species interferometer with rubidium and potassium. The setup is divided into six racks, as depicted in the bottom photo of \Fig \ref{fig:ParabolicFlight-ICESetup}, one rack each for the vacuum chamber, laser system, frequency comb, power supplies, rf frequency chain, and computer control system\footnote{We utilize the control software ``Cicero Word Generator'' to generate all of our experimental sequences, which is designed specifically for atomic physics experiments \cite{Keshet-RSI-2013}.}. These racks are designed to be fastened to the aircraft's interior, and to comply with Novespace regulations to withstand $9g$ of forward thrust in the event of an emergency landing.

\begin{figure}[!t]
  \centering
  \includegraphics[width=0.95\textwidth]{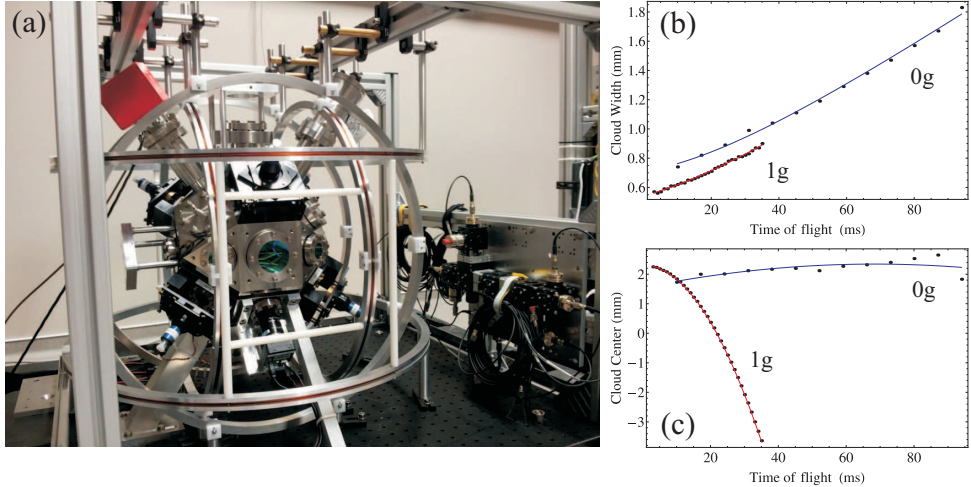}
  \caption{(Colour online) (a) Titanium vacuum chamber for the next generation of ICE experiments. (b,c) Recent rubidium time-of-flight data taken onboard the zero-g aircraft with the titanium vacuum system. Measurements of the $1/e$ cloud width along the axis of the Raman beams (b), and the cloud center along the vertical direction (c), were performed in $1g$ while the plane was grounded (red curves), and in $0g$ while in flight (blue curves). Fits to the cloud center give an acceleration $a = (-9.85 \pm 0.03)$ m/s$^2$ when on ground and $a = (-0.3 \pm 0.2)$ m/s$^2$ in micro-gravity. Similarly, fits to the cloud width yield temperatures of $\mathcal{T} = (2.4 \pm 0.2)$ $\mu$K and $(2.6 \pm 0.3)$ $\mu$K in $1g$ and $0g$, respectively.}
  \label{fig:Chamber-TOFResults}
\end{figure}

The laser system is based on optical fiber and telecom technology that is very robust and well-adapted for this type of environment. As light sources, we use Redfern Integrated Optics (RIO) external cavity diode lasers (ECDLs) at 1560 and 1534 nm. This light is frequency-doubled using second-harmonic generation (SHG) in a PPLN to 780 and 767 nm for $^{87}$Rb and $^{39}$K, respectively. These ECDLs are extremely compact, fiber-based lasers, with a gain chip and a planar waveguide circuit that includes a Bragg grating inside a butterfly package. They have a narrow linewidth ($\sim 15$ kHz in our case), ultra-low phase noise, and low sensitivity to bias current and temperature---making them highly suitable for use in noisy environments. We stabilize both rubidium and potassium diodes on a common frequency reference by using a fiber-based optical frequency comb \cite{Menoret-OptLett-2011}, which gives us precise knowledge of the optical frequencies for both atomic sources.

A schematic of the fiber-based components of the laser system is shown in \Fig \ref{fig:ICE-LaserSystem}. For each atomic species, we utilize a master-slave architecture, where the master laser diode is locked on the frequency comb, and the slave is locked to the master using an optical beat-note. The set-point of each slave laser can be adjusted over approximately 500 MHz at 1.5 $\mu$m (corresponding to $\sim 1$ GHz at 780 nm) within $\sim 2$ ms of settling time. The output of each slave laser is coupled into a dual-wavelength EDFA, where each light source can be amplified to $\sim 5$ W. For $^{87}$Rb, the slave light is coupled through an electro-optic phase modulator at 6.8 GHz before being amplified. This generates the sideband needed for laser-cooling and making Raman transitions in rubidium. The amplification stage is followed by a free-space SHG stage which generates approximately 1 W of 780 and 767 nm light. A second free-space module, composed of a series of shutters and acousto-optic modulators (AOMs), is used to split, pulse and frequency shift the light appropriately for cooling, interferometry and detection. Finally, this light is coupled into a series of single-mode, polarization-maintaining fibers and sent to the vacuum chamber.

The sensor head is composed of a non-magnetic titanium vacuum chamber\footnote{Previous experimental results \cite{Geiger-NatureComm-2011} were performed in a stainless steel chamber, where rubidium cloud temperatures were limited to $7 - 8$ $\mu$K. This was attributed to the presence of relatively large magnetic field gradients from the magnetized steel frame.}, as shown in \Fig \ref{fig:Chamber-TOFResults}(a). This chamber has 19 view ports for extended optical access, including four that are anti-reflection coated for 1.5 $\mu$m light (for a future dipole trap), and three mutually-perpendicular pairs of large-area view ports (for a future 3-axis inertial sensor). A custom 2-6 way fiber-splitter is used to combine the 780 and 767 nm light and divides it equally into six beams for laser-cooling purposes. With this system, we achieve rubidium temperatures of $\sim 2.5$ $\mu$K both in $1g$ and in $0g$, as shown by the time-of-flight measurements in \Fig \ref{fig:Chamber-TOFResults}(b). Here, we measured the cloud position at times as large as $\sim 100$ ms while in micro-gravity. This is not possible on ground because the atoms fall outside of the field of view of the camera.

\begin{figure}[!t]
  \centering
  \includegraphics[width=0.6\textwidth]{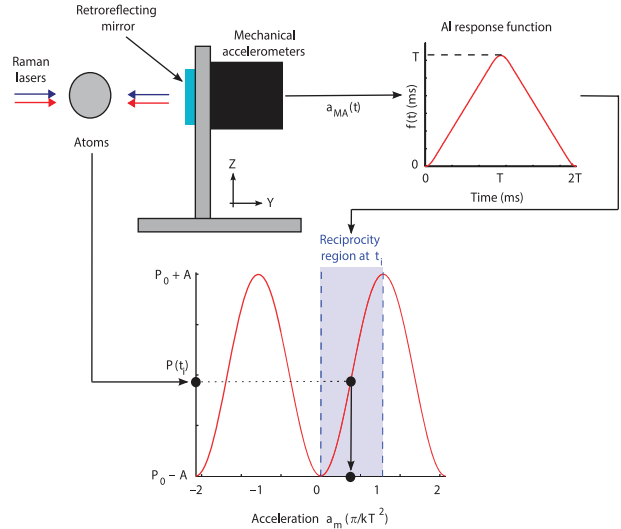}
  \caption{(Colour online) Schematic for reconstructing the interferometer fringe pattern using correlations with a mechanical accelerometer.}
  \label{fig:MA-Correlation}
\end{figure}

\subsubsection{\textsf{Airborne interferometer with $^{87}$Rb}}

\begin{figure}[!t]
  \centering
  \includegraphics[width=0.8\textwidth]{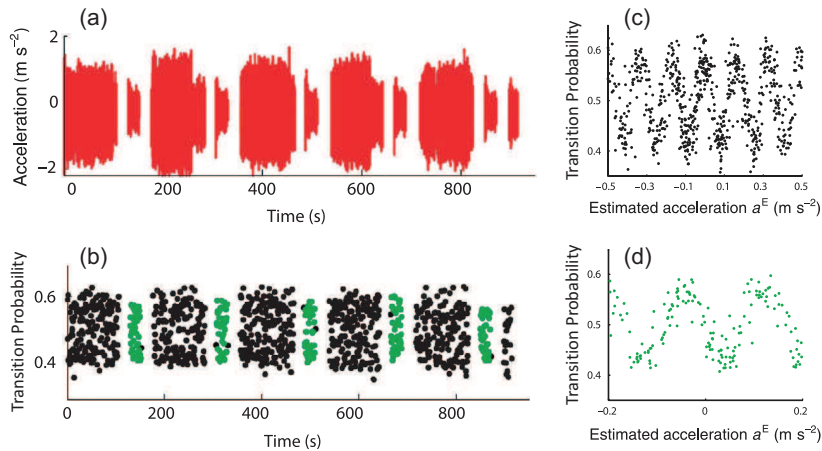}
  \caption{(Colour online) (a) Acceleration signal recorded by the MAs. The standard deviation $\sigma_a$ of the acceleration signal is about 0.5 m/s$^2$ during the $1g$ phase of the flight and 0.2 m/s$^2$ during $0g$. (b) Interferometer measurements of the transition probability corresponding to a total interrogation time of $2T = 3$ ms. The black and green points correspond to the $1g$ and $0g$ phases, respectively. (c,d) Atomic transition probability measurements as a function of the estimated mirror acceleration, $a_E = \Phi_E/k_{\rm eff} T^2$, during $1g$ (c) and $0g$ (d). The sinusoidal correlations show that the interferometer contains information on the acceleration of the plane.}
  \label{fig:Results-Rb}
\end{figure}

The first airborne matter-wave interferometer was achieved in the zero-g plane with rubidium \cite{Geiger-NatureComm-2011}, where we demonstrated sensitivity to the acceleration along the wings of the aircraft. The system combines a mechanical accelerometer (MA), which has a large dynamic range, and an atom interferometer, which has a high sensitivity. The MA is attached to the back of the retro-reflecting Raman mirror, which acts as the inertial reference frame for the interferometer. Since the Raman beams are aligned along the horizontal $y$-axis, the mean acceleration is zero in both the $1g$ and $0g$ phases of the flight. On the aircraft, the level of vibrations is extremely high, and the Raman mirror can move distances that correspond to phase shifts of much more than $\pi$ over the duration of the interferometer sequence. Under these conditions, the fringes are ``scanned'' by the vibrations, but the phase shift is random and unknown for each repetition of the experiment---which results in fringe smearing. However, by recording both the transition probability from the interferometer and the acceleration of the Raman mirror during the pulse sequence, $a_{\rm MA}(t)$, it is possible to reconstruct the fringes by utilizing the sensitivity function (see \Sec \ref{sec:SensitivityFunction}). The phase shift due to mirror vibrations during the $i^{\rm th}$ measurement is estimated using the relation
\be
  \label{PhiE}
  \Phi_E^{(i)} = k_{\rm eff} \int_{t_i}^{t_i + 2T} f(t - t_i) a_{\rm MA}(t) \diff t,
\ee
where $t_i$ is the start time corresponding to the $i^{\rm th}$ repetition of the pulse sequence, and $f(t)$ is the interferometer response function given by \Eq \refeqn{f(t)-3pulse}. This function is a triangle-like function with units of time that characterizes the sensitivity of the interferometer to phase shifts at any point during the pulse sequence.\footnote{The integrand $f(t - t_i) a_{\rm MA}(t)$ appearing in \Eq \refeqn{PhiE} can be thought of as a time-dependent velocity that must be integrated to obtain the effective displacement of the mirror at the end of the interferometer sequence, $\Delta y_i$. The phase shift is then $\Phi_E^{(i)} = k_{\rm eff} \Delta y_i$.} This phase is then correlated with the measured transition probability, as depicted in \Fig \ref{fig:MA-Correlation}. The first results of this implementation of the experiment are shown in \Fig \ref{fig:Results-Rb}.

This implementation of the mobile accelerometer demonstrated sensitivities at the level of $2 \times 10^{-4}$ m/s$^2/\sqrt{\rm Hz}$ while in micro-gravity. Furthermore, during the $1g$ phases of the flights, we detected inertial effects more than 300 times weaker than the vibration level of the plane.

\subsubsection{\textsf{Toward a mobile dual-species interferometer with $^{87}$Rb and $^{39}$K}}

\begin{figure}[!t]
  \centering
  \subfigure{\includegraphics[width=0.8\textwidth]{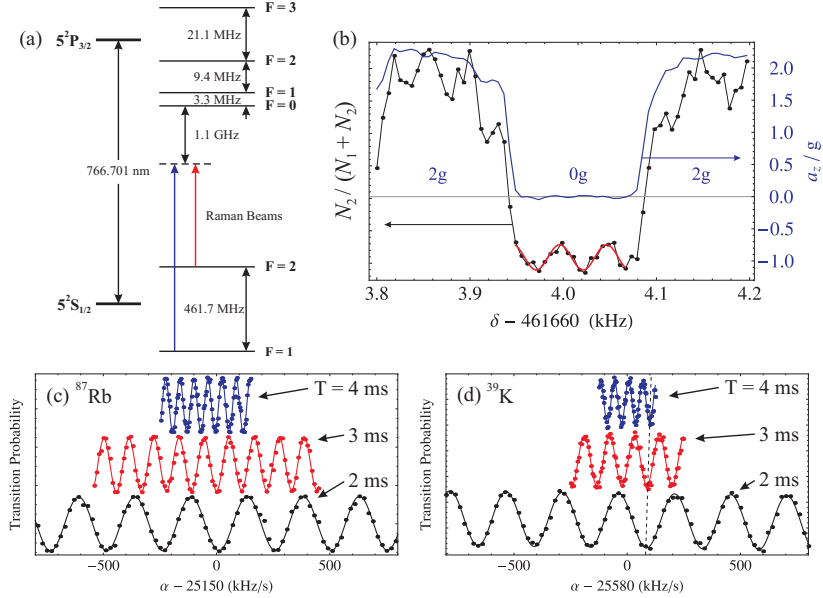}}
  \caption{(Colour online) (a) Energy level structure of $^{39}$K (not to scale). The spin-structure is identical to that of $^{87}$Rb, but the hyperfine states are separated by a factor of $10 - 20$ less in frequency. (b) Optical Ramsey fringes with $T_{\rm Ramsey} = 20$ ms using $^{39}$K in micro-gravity. Here, the black points (corresponding to the scale on the left) represent the transition probability as a function of the frequency difference, $\delta$, between Raman beams. The red curve is a sinusoidal fit to the data during $0g$. The blue curve, corresponding to the scale on the right, is a simultaneous measurement of the vertical acceleration during a single parabola. (c,d) Three-pulse interferometer fringes for a $^{87}$Rb and $^{39}$K atoms, respectively, as a function of the chirp rate, $\alpha$, that cancels the gravity-induced Doppler shift. Interrogation times of $2T = 4, 6$, and 8 ms are shown. These data were recorded in the same titanium vacuum system, with the same laser-cooling and Raman beam optics, but at separate times.}
  \label{fig:ICE-Rb+KResults}
\end{figure}

One of the main challenges in constructing a dual-species interferometer with $^{87}$Rb and $^{39}$K is working with potassium because of, for example, its compact energy level structure [see \Fig \ref{fig:ICE-Rb+KResults}(a)]. This makes potassium isotopes particularly difficult to cool to sub-Doppler temperatures without evaporation techniques \cite{Campbell-PRA-2010}. Similarly, the depumping time between ground states is on the order of $\sim 1$ $\mu$s for near-resonant excitation light due to the proximity of excited states---making state selection and detection of $^{39}$K more challenging than other alkali atoms. Nevertheless, we have made encouraging progress toward a mobile interferometer with these two isotopes.

By employing techniques similar to those discussed in \Refs \cite{Landini-PRA-2011, Gokhroo-JPhysB-2011}, we sub-Doppler cool our sample of $^{39}$K to temperatures around 25 $\mu$K\footnote{This temperature corresponds to a most probable speed of just $7.7\,v_{\rm rec} \sim 10$ cm/s, where $v_{\rm rec} \sim 1.3$ cm/s is the one-photon recoil velocity for $^{39}$K. Recent work \cite{Salomon-EPL-2013} has shown efficient cooling of $^{39}$K to temperatures as low as 6 $\mu$K, or $3.8\,v_{\rm rec}$, using a gray molasses on the D1 transition.}. We have also measured optical Ramsey fringes by inducing co-propagating Raman transitions with a $\pi/2 - \pi/2$ pulse sequence, separated by free-evolution times as large as $T_{\rm Ramsey} = 30$ ms. Although this configuration is essentially insensitive to the velocity of the atoms, it nonetheless shows that coherent two-photon transitions can easily be made with $^{39}$K, which is an important first step toward an interferometer.

Figure \ref{fig:ICE-Rb+KResults}(b) shows potassium Ramsey fringes at $T_{\rm Ramsey} = 20$ ms measured during a parabolic flight. Here, the ratio $N_2/(N_1 + N_2)$, between the total number of atoms and those in the $\ket{F = 2, m_F = 0}$ state is measured as a function of the two-photon detuning between the Raman beams, $\delta$. To the best of our knowledge, these are some of the first measurements of optical Ramsey fringes with $^{39}$K.

We have also achieved some of the first three-pulse interferometer fringes with potassium. Figures \ref{fig:ICE-Rb+KResults}(c) and (d) show fringes from $^{87}$Rb and $^{39}$K samples, respectively, for $T = 2$, 3 and 4 ms. These data were recorded at different times in the same laboratory setup. Here, the Raman beams were oriented along the vertical direction, and the frequency difference between the beams was chirped at various rates, $\alpha$, that allowed the fringes to be scanned while keeping the two-photon Raman transition on resonance. For $^{87}$Rb, the chirp rate that cancels the gravitationally-induced Doppler shift is $\alpha_{\rm Rb} = k_{\rm eff}^{\rm(Rb)} g/2\pi = 25.138$ MHz/s. Similarly, for $^{39}$K it is $\alpha_{\rm K} = 25.581$ MHz/s---which is slightly greater than that of rubidium owing to the different atomic transition frequencies. Notice that the fringe zeroes for $^{87}$Rb align near 25.150 MHz/s for all three values of $T$, which indicates that each data set gives a similar measurements of $g$\footnote{Systematic effects have not been accounted for in these preliminary results.}. However, for $^{39}$K, the fringe zeroes near 25.680 MHz/s appear to shift to the right for successive $T$---showing that there is a strong systematic effect on the measurement of $g$ as $T$ increases, and the atoms fall and expand. This is a result of the two-photon light shift in potassium, which cannot be suppressed as easily as a rubidium interferometer due to the fact that the one-photon detuning is greater than the hyperfine splitting, $|\Delta| > \omega_{\rm HF}$. A future test of the equivalence principle will require further investigation of this effect.

These results open the way toward the first mobile, dual-species interferometer, and precise tests of the WEP in the near future.

\subsection{Inertial navigation}

The navigation problem is easily stated: How do we determine an object's trajectory as a function of time? Nowadays, we take for granted that a hand-held global positioning system (GPS) receiver can be used to obtain meter-level position resolution. When GPS is unavailable (for example, when satellites are not in direct line-of-sight), position determination becomes much less accurate. In this case, stand-alone ``black-box'' inertial navigation systems, comprised of a combination of gyroscopes and accelerometers, are used to infer position changes by integrating the outputs of these sensors. State-of-the-art commercial navigation systems have position drift errors of kilometers over many hours of navigation time, significantly worse than the GPS solution. Yet many 21$^{\rm st}$ century applications require GPS levels of accuracy everywhere and at anytime. Examples of such applications include indoor navigation for emergency responders, navigation in cities and urban environments, and autonomous vehicle control. How can we close the gap between GPS system performance and inertial sensors? One way forward is improved instrumentation: better gyroscopes and accelerometers.

Inertial sensors based on light-pulse atom interferometry appear to be well suited to this challenge. The sensor registers the time evolution of the relative distance between the mean position of the atomic wavepackets and the sensor case (defined by the opto-mechanical hardware for the laser beams) using optical telemetry. Since distances are measured in terms of the wavelength of light, and since the atom is in a benign environment, the sensors are characterized by highly stable and low-noise operation.

However, there is an additional complication in the architecture of these sensors for high accuracy navigation applications: the so-called ``problem of the vertical''. Terrestrial navigation requires determining the sensor's position in Earth's gravitation field. Due to Einstein's equivalence principle, accelerometers cannot distinguish between the acceleration due to gravity and the sensor itself. So, in order to determine the sensor's trajectory in an Earth-fixed coordinate system, the local acceleration due to gravity needs to be subtracted from the accelerometer output. For example, existing navigation systems use a gravity map to make this compensation.  However, in present systems, this map does not have enough resolution or accuracy for meter-level position determinations (a $10^{-7}$ error in the knowledge of the local value of $g$ integrates to an error of $\sim 1$ m in 1 hour).

Two possible solutions to this issue are to obtain better maps of local gravitational acceleration with more precise surveys, or to perform simultaneous gravity field measurements. To do this, one could utilize the two outputs from a gravity gradiometer, such as those mentioned in \Sec \ref{sec:Gradiometers}. By integrating the gravity gradient over the inferred trajectory, one can determine gravity as a function of position. In principle, such an instrument can function on a moving platform, since acceleration noise that is common to the two outputs can be rejected. The central design challenge is the realization of a mobile instrument which has very good noise performance.

Another promising application of cold atom inertial sensors to navigation is to correct the intrinsic drift of a mechanical accelerometer in real time. The use of such a hybrid sensor---which effectively have zero bias---is promising because even the best mechanical accelerometers have a bias that fluctuates in time (with a standard deviation on the scale of $10^{-4}$ m/s$^2$). This phenomena can lead to errors in positions of several hundred meters after one hour of navigation \cite{Jekeli-JInstNav-2005}. The idea is to use the high accuracy of atom-based accelerometers to measure bias variations of the mechanical accelerometers and correct them. In this way, it is possible to benefit from the high bandwidth of mechanical accelerometers (quasi-continuous sampling of the acceleration signal) while suppressing the bias drift. Numerical simulations of this type of hybrid sensor (using $T = 4.5$ ms) have shown a reduction in the position error by a factor of $\sim 25$ after one hour of navigation \cite{Menoret-Thesis-2012}. This improvement is significant, even with such a small free-evolution time.

%% file: Varenna-Geophysics.tex
\section{Application to geophysics and gravitational wave detection}
\label{sec:Geophysics}

\subsection{Gravity and geophysics}

Historically, gravity has played a central role in studies of dynamic processes in the Earth's interior and is also important in the study of geophysical phenomena, geodesy and metrology. Gravity is the force responsible for the shape and structure of the Earth. The combined effect of gravitational attraction and centrifugal force acts to distribute dense material toward the innermost layers, and lighter material in the outer ones. High-precision measurements of the gravitational field and its variations (both spatial and temporal) give important information about the dynamical state of the Earth. However, the analysis of these variations in local gravity is quite challenging because the underlying theory is complex, and many perturbational corrections are necessary to isolate the small signals due to dynamic processes. With respect to determining the three-dimensional structure of Earth's interior, a disadvantage of a gravitational field (or any potential energy field), is that there is a large ambiguity in locating the source of gravitational anomalies.

The law of gravitational attraction was formulated by Isaac Newton (1642-1727) and was first published in 1687 \cite{Newton-Book-1687}---approximately three generations after Galileo had determined the magnitude of gravitational acceleration, and Kepler had discovered his empirical laws describing the orbits of planets. The gravitational force between any two particles with (point) masses $M$ at position $\bm{r}_0$, and $m$ at position $\bm{r}$, is an attraction along the line joining the particles:
\be
  \bm{F} = -G \frac{M m}{|\bm{r} - \bm{r}_0|^3}(\bm{r} - \bm{r}_0).
\ee
Here, $G$ is the universal gravitational constant: $G = 6.674 \times 10^{-11}$ N(m/kg)$^2$, which has the same value for all pairs of particles. \footnote{$G$ should not be confused with the gravitational acceleration, $g$, which is approximately given by $g = G M_{\oplus}/R_{\oplus}^2$, where $M_\oplus$ is the mass of the Earth, and $R_{\oplus}$ is Earth's effective radius.}

The value of Earth's gravitational acceleration was first determined by Galileo. Its magnitude is approximately  $g = 9.8$ m/s$^2 = 980$ Gal, but it varies over the surface of Earth between 9.78 and 9.82 m/s$^2$ depending on a number of factors such as latitude, elevation and local density.\footnote{In honor of Galileo, the unit often used in gravimetry is the Gal: 1 Gal = 1 cm/s$^2 = 10^{-2}$ m/s$^2 \sim 10^{-3} g$. Therefore, 1 mGal = $10^{-5}$ m/s$^2 \sim 10^{-6} g$ and 1 $\mu$Gal = $10^{-8}$ m/s$^2 \sim 10^{-9} g$.} Gravity anomalies are often expressed in mGal ($\sim 10^{-6} g$) or in $\mu$Gal ($\sim 10^{-9} g$), a level of precision that can be achieved by modern absolute gravimeters.

In general, geophysics is the quantitative study of the physics of the Earth. Nowadays, geophysicists are particularly interested in studying variations of Earth's local gravity field because they permit the detection of the anomalous distribution of masses, and the determination of geological structures such as faults, the crust-mantle boundary, and density anomalies in the mantle. Furthermore, it allows the study of dynamical processes like the movement of tectonic plate, mountain formation, convection in the Earth's mantle, and volcanic activity. All of these processes strongly affect the mass distribution within the Earth---generating large anomalies in the gravitational field. Gravity is therefore a basic tool for studies of structural geology. Some geological structures within the Earth's crust (such as faults, synclines, anticlines, or salt domes) are frequently associated with potential reservoirs for oil and gas. As a result, the study of the Earth's gravity field also plays an important role in the search for fossil fuels, as well as for geothermic activity.

\subsection{Gravity anomalies and how to use gravity data}
\label{sec:GravityAnomalies}

In general, a gravity anomaly is the difference between an observed value of local gravitational acceleration, $g_{\rm obs}$, and that predicted by a model. The combination of the gravity anomaly measurements and topographical data yield crucial information about the mechanical state of the Earth's crust and lithosphere. Both gravity and topography can be obtained by remote sensing and, in many cases, they form the basis of our knowledge of the dynamical state of planets, such as Mars, and natural satellites, such as Earth's Moon. Data reduction plays an important role in gravity studies since the signal of interest (caused by variations in density) is minuscule compared to the sum of the observed field and other effects, such as the influence of the position at which the measurement is made.

The following list describes various contributions to the gravitational field, with the name of the corresponding correction in parentheses:
\begin{enumerate}
  \item[] Observed gravity equals the attraction of the reference spheroid, \emph{plus}:
  \item Effects of elevation above sea level (free-air correction).
  \item Effect of ``normal'' attracting mass between observation point and sea level (Bouguer and terrain correction).
  \item Effect of masses that support topographic loads (isostatic correction).
  \item Time-dependent changes in Earth's shape (tidal correction).
  \item Changes in the rotation term due to motion of the observation point, for example, when measurements are made from a moving ship (E\"ot\"os correction).
  \item Effects of crust and mantle density anomalies (``geological'' or ``geodynamic'' process correction).
\end{enumerate}
We will now describe some of the most crucial corrections in more detail.

\subsubsection{\textsf{Free-air correction}}

The free-air correction to the measurement of gravitational acceleration adjusts the value of $g$ to correct for its variation due to elevation above sea level. It assumes there is no mass between the observer and sea level, hence the name ``free-air''' correction. For an altitude $h \ll R_{\oplus}$ above sea level, this correction amounts to the following shift in $g$:
\be
  \delta g_{\rm FA} = -2 \frac{h g}{R_{\oplus}}.
\ee
The shift is $\delta g_{\rm FA}/h \approx -3.14 \times 10^{-7}$ $g$/m at the equator, where $R_{\oplus} = 6.37 \times 10^6$ m is the Earth's effective radius (the ``geoid'' effect of Earth's ellipticity is often included in a separate model). Since this level of precision can be attained by field instruments, it shows that uncertainties in elevation can be a limiting factor in the accuracy that can be achieved.\footnote{For example, a realistic error in elevation of a few meters leads to an uncertainty in $g$ of $\sim 1$ mGal.}

\subsubsection{\textsf{The Bouguer anomaly}}

\begin{figure}[!t]
  \centering
  \includegraphics[width=0.8\textwidth]{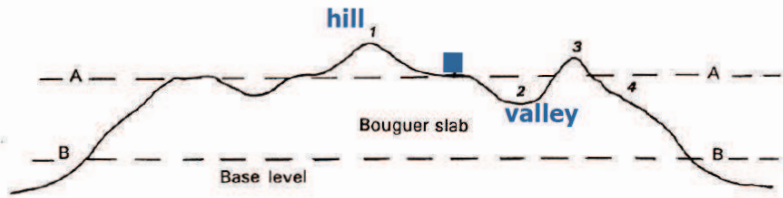}
  \caption{(Clour online) Bouguer and terrain corrections. The Bouguer correction refers to the gravity effect of the intervening plate between a station at elevation A and the base level B. The terrain correction takes into account the effects of topographic rises (like points 1 and 3) and depressions (like points 2 and 4). The observation point is indicated by the square between points 1 and 2.}
  \label{fig:Bouguer}
\end{figure}

The free-air correction does not correct for any attracting mass between the observation point and sea level. However, on land, at a certain elevation there will be attracting mass (even though it is often compensated---see \Sec \ref{sec:Isostatic}). Instead of estimating the true shape of, say, a mountain on which the measurement is made, one often resorts to what is known as the ``slab approximation'', where the rocks are assumed to be of infinite horizontal extent. The Bouguer correction is then given by
\be
  \delta g_{\rm B} = 2\pi G \bar{\rho} h,
\ee
where $\bar{\rho}$ is the mean density of crustal rock and $h$ is, again, the height above sea level. For $\bar{\rho} = 2700$ kg/m$^3$, we obtain a correction of $\delta g_{\rm B}/h \approx 1.15 \times 10^{-7}$ $g$ per meter of elevation (or $0.113$ mGal/m). If the slab approximation is not sufficient, for instance near the top of a mountain, one must apply an additional ``terrain'' correction. This is straightforward if one has access to topography/bathymetry data. Figure \ref{fig:Bouguer} illustrates a situation in which the Bouguer and terrain corrections would increase the accuracy of gravity anomaly measurements.

The Bouguer correction must be subtracted from the observed value of gravity, $g_{\rm obs}$, since one wants to remove the effects of the extra attraction, and it is typically applied together with the free-air correction. Ignoring the terrain correction, the Bouguer gravity anomaly is then given by:
\be
  \Delta g_{\rm B} \simeq g_{\rm obs} - g_{\rm pred} - \delta g_{\rm FA} - \delta g_{\rm B},
\ee
where $g_{\rm pred}$ is a model-based prediction of $g$ that includes effects such as the aspherical shape of the Earth. In principle, these two corrections account for the attraction of all rock between the observation point and sea level. Then, $\Delta g_{\rm B}$ represents the gravitational attraction of the material below sea level. Maps of the Bouguer anomaly [see \Fig \ref{fig:GravitySurvey}(b)] are typically used to study gravity on continents, whereas the free-air anomaly ($\Delta g_{\rm FA} = g_{\rm obs} - g_{\rm pred} - \delta g_{\rm FA}$) is more commonly used in oceanic regions.

\subsubsection{\textsf{Isostatic anomalies}}
\label{sec:Isostatic}

If the mass between the observation point and sea level is all that contributes to $g_{\rm obs}$, one would expect the free-air anomaly to be large and positive near topographical peaks (since this mass is unaccounted for) and the Bouguer anomaly to decrease to zero. This relationship between the two gravity anomalies and topography is what would be obtained in the case where the mass is completely supported by the strength of a tectonic plate (\ie no isostatic compensation). In early gravity surveys, however, it was found that the Bouguer gravity anomaly over mountain ranges was, somewhat surprisingly, large and negative. Apparently, a mass deficiency remained after the mass above sea level was compensated for. In other words, the Bouguer correction subtracted too much! This observation in the 19$^{\rm th}$ century lead Airy and Pratt to develop the concept of isostasy. In short, isostasy means that at depths larger than a certain compensation depth, the observed variations in height above sea level no longer contribute to lateral variations in pressure.

The basic equation that describes the relationship between the topographical height and the depth of the compensating body is
\be
  H = \frac{\rho_c h}{\bar{\rho} - \rho_c},
\ee
where $\rho_c$ is the density of the compensating body, $H$ is the depth it is buried and $H + h$ its total height. Assuming some constant density for crustal rock, one can compute a spatial grid of depths, $H(x,y)$, from measured topographical data, $h(x,y)$, and correct for the mass deficiency. This results in the isostatic anomaly---a small variation in $g$ due to the uncompensated density variations that result from local geological, or geodynamic processes.

\subsubsection{\textsf{Gravity surveying}}

\begin{figure}[!t]
  \centering
  \subfigure[]{\includegraphics[width=0.55\textwidth]{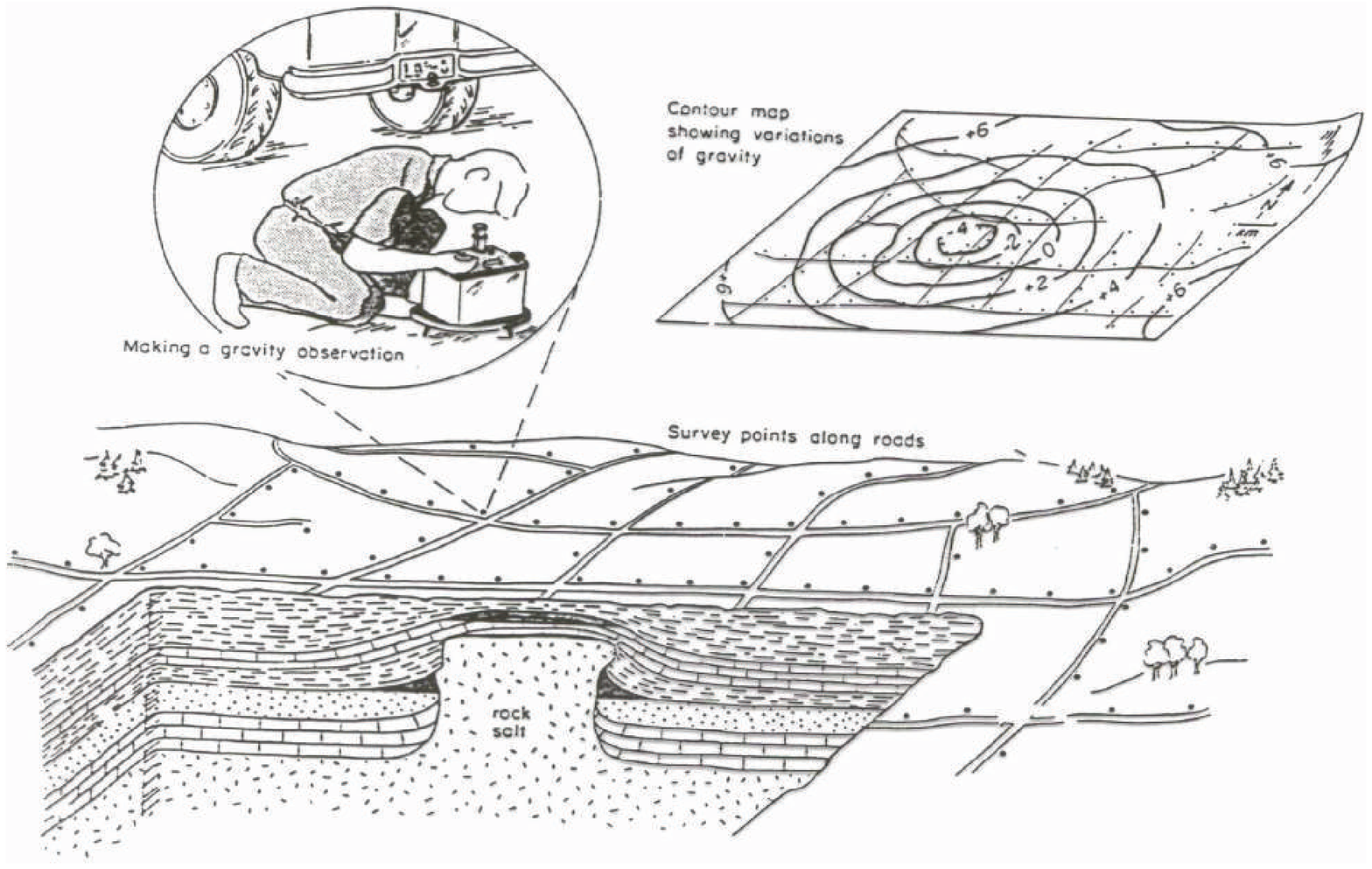}}
  \subfigure[]{\includegraphics[width=0.43\textwidth]{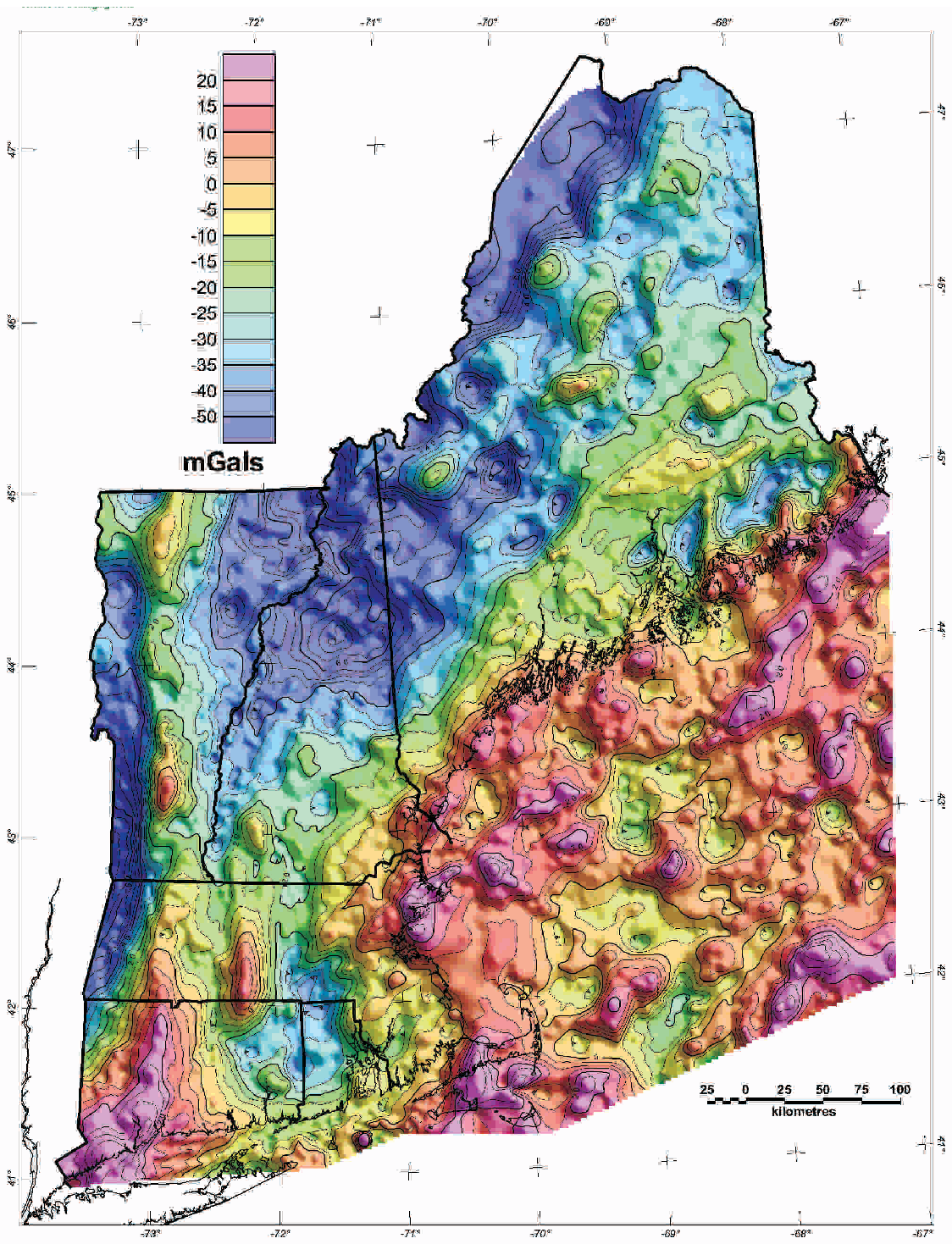}}
  \caption{(Colour online) (a) Cartoon of a gravity survey. By performing very precise measurements of $g$ and carefully applying the aforementioned corrections, a gravity survey can detect natural or man-made voids, variations in the depth of bedrock, and geological structures of engineering interest. (b) Free-air/Bouguer gravity anomaly map of New England and the Gulf of Maine in the United States. These data were compiled using 35,644 gravity stations on land and 27,251 sea-based measurements. Image courtesy of S.~L. Snyder and the US Geological Survey. Taken from \Ref \cite{USGS-Website}.}
  \label{fig:GravitySurvey}
\end{figure}

Figure \ref{fig:GravitySurvey}(a) depicts a gravity survey---a grid of precise, spatially-separated measurements of $g$ that are carefully corrected for local topographical variations. Data of this type can be used to detect man-made voids, variations in the depth of bedrock, geological structures or even buried resources. An example of a detailed gravity anomaly map of the New England area is shown in \Fig \ref{fig:GravitySurvey}(b).

\begin{figure}[!t]
  \centering
  \includegraphics[width=0.6\textwidth]{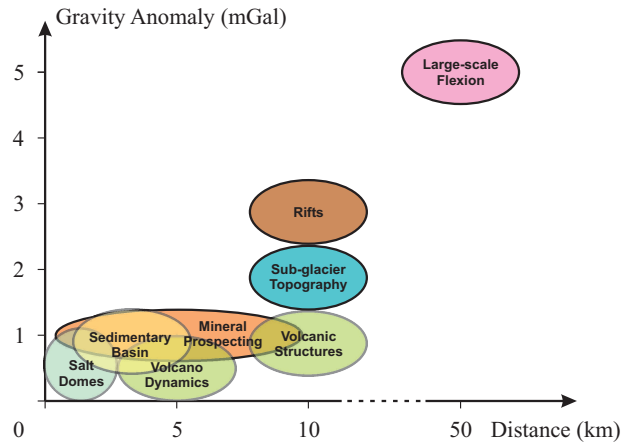}
  \caption{(Colour online) Range and sensitivity required for various applications in gravity surveying.}
  \label{fig:GravityAnomalySources}
\end{figure}

The information that can be extracted from a gravity survey is highly dependent on the sensitivity of the measurement device and the size of local gravity anomalies. The source of most gravity anomalies is a change in the lateral density of the sub-surface. Since the magnitude of the gravitational force due to a subterranean mass variation (from either a local concentration or void) is superimposed on the larger force due to the total mass of the Earth, these anomalies can be challenging to detect. For engineering and environmental applications, structures of interest are generally quite small (1-10 m in size) and gravity anomalies resulting from these are at the level of a few hundred $\mu$Gal. Therefore, high sensitivity gravity measurements are required at the level of or below 1 $\mu$Gal (\eg micro- and nano-gravimetry) in order to detect these structures. Figure \ref{fig:GravityAnomalySources} illustrates the range and sensitivity required for different gravity surveying applications.

When searching for underground structures, usually one first puts some constraints on the probable geometry, depth and mean density of the structure, in order to determine the approximate magnitude of the associated gravity anomaly. A general rule of thumb is that a body must be almost as big as it is deep. A gravity survey will measure the vertical component of the gravitational force at specific locations (\eg ground-based stations, or coordinates on water). Measurements can also be carried out in an aircraft, allowing consistent regional coverage in varied topography and offering rapid acquisition time compared to ground-based gravity surveys. \footnote{In an airborne environment, gravity gradiometers are preferred over gravimeters, because of their enhanced sensitivity, lower noise levels and higher spatial resolution. This technique has been successfully deployed on helicopters, single-engine and multi-engine fixed-wing aircraft \cite{Blomfield-Proc10SEGJ-2011}.} The total area, desired resolution, station spacing and topography of the region are a few of the factors to consider when deciding between a ground or airborne gravity survey. For instance, an airborne survey may provide better coverage and resolution in areas with steep topography or challenging terrain than a ground-based survey. However, a ground-based survey enables a more detailed characterization of an area, at the price of requiring a high station density. Typically, measurements separated by 1-3 m (with relative errors in spacing of $\sim 10\%$ and elevation of $\sim 1$ cm) are required to map anomalous masses with a maximum dimension of $\sim 10$ m.

At each location, gravity measurements will change with time due to ocean tides and sensor calibration drift. Ocean tides may cause changes of 0.24 mGal in the worst cases, but the effect has a period of about 12.5 hours and, generally, it can be calculated and removed. Sensor drifts will depend of the type of gravimeter. For instance, the drift of absolute gravimeters, such as those based on matter-wave interference discussed in \Sec \ref{sec:Gravimeters}, is far below the resolution limit of the instrument. Processing all of the gravimetric data is called \emph{gravity reduction}. It refers to the subtraction of all the corrections mentioned at the beginning of \Sec \ref{sec:GravityAnomalies} in order to obtain the residual gravity anomaly.

\begin{figure}[!t]
  \centering
  \includegraphics[width=0.7\textwidth]{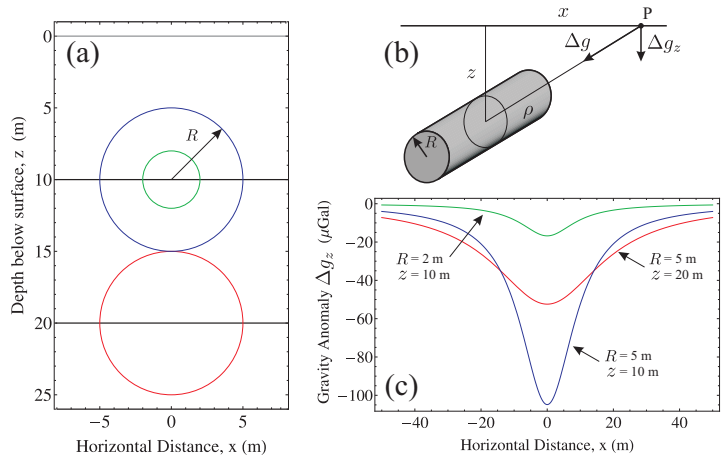}
  \caption{(Colour online) (a) Cross-section of three infinite cylinders at a depth, $z$, below ground, with radii $R = 5$ m (red \& blue) and $R = 2$ m (green). (b) Diagram of an infinite cylinder giving rise to the gravity anomaly, $\Delta g_z$, given by \Eq \refeqn{Delta g_z}. Here, $\rho$ is the mean density, $R$ is the radius of the cylinder, $z$ is the depth to the central axis, and $x$ is the horizontal distance from the observation point, $P$. (c) Plots of the gravity anomaly as a function of $x$ for the infinite cylinders shown in (a). The cavities are assumed to have a difference in mean density of $\Delta\rho = -10^3$ kg/m$^3$.}
  \label{fig:GravityAnomalyCylinders}
\end{figure}

The size and depth of gravity anomaly sources can be interpreted using a direct comparison with the signal that is produced by simple shapes. For example, a horizontal tunnel with a rectangular cross-section can be modeled by a cylinder or a line with an infinite length, depending on the distance at which the measurements are taken. In this case, the gravity anomaly can be computed analytically:
\be
  \label{Delta g_z}
  \Delta g_z
  = g_{\rm obs} - g_{\rm pred}
  \approx \frac{2 \pi G \Delta \rho R^2 z}{x^2 + z^2}.
\ee
Here, $R \ll z$ is the radius of the cylinder, $z$ is the depth of the structure, and $\Delta\rho$ is the difference in mean density between the inferred and the predicted mass distribution. Figure \ref{fig:GravityAnomalyCylinders} illustrates the effect of infinite cylinders with different depths and radii on the gravity anomaly. Indeed, unless one is very close to the structure, its exact shape is not important. More detailed information about the source can be obtained by using inversion algorithms, but gravity surveys will always be limited by ambiguity\footnote{A distribution of small masses at a shallow depth can produce the same effect as a large mass buried deeper.} and the assumption of homogeneity. Additional geological data about the surrounding region (measurements of local rock densities, seismic surveys, ground-penetrating radar and core-drilling information) are usually required in order to resolve this issue.

\subsection{Gravitational waves}

Einstein's general theory of relativity predicts the existence of gravitational waves (GWs)---disturbances of space-time that propagate at the speed of light and have two transverse quadrupolar polarizations \cite{Thorne-Book-1987}. Presently, the vast majority of astrophysical phenomena are observed with electro-magnetic waves, which originate from moving charges. In contrast, GWs are generated from the motion of massive (or energetic) bodies, and can therefore unveil an entirely new spectrum of information that is not possible to extract electro-magnetically. This will bring forth new possibilities to explore details about the early universe, and to observe new phenomena, such as binary systems of black-holes and neutron stars, that comprise a host of unexplored territory in fundamental physics.

In the mid-$20^{\rm th}$ century, it was confirmed that the theory of general relativity predicted GWs by the realization that energy could be extracted from these waves, which meant that, in principle, one could build a device to detect them \cite{Saulson-GenRelGrev-2011}. A more complete historical review of GWs and their detection can be found in \Ref \cite{Riles-ProgPartNucPhys-2013}. Gravitational waves manifest themselves as a periodic variation in the separation between two test masses---an effect on which all current GW detectors are based. A classic example is that of an astronaut's observations while orbiting the Earth in a windowless spacecraft. Observing the slow relative drift of two objects, initially placed at separate positions and at rest with respect to one another, allows the astronaut to detect the tidal influence of the Earth on local space-time.

\begin{figure}[!t]
  \centering
  \includegraphics[width=0.5\textwidth]{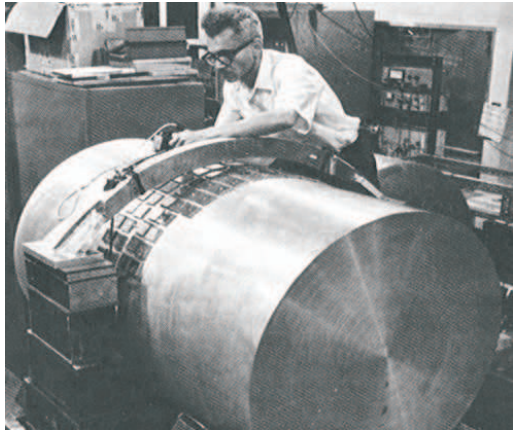}
  \caption{Joe Weber working on an early GW detection bar with piezo-electric strain sensors (circa 1965). Photo taken from \Ref \cite{Riles-ProgPartNucPhys-2013}, courtesy of the University of Maryland.}
  \label{fig:WeberBar}
\end{figure}

The earliest man-made GW detectors were based on the idea that two masses on a spring can be momentarily stretched apart and then compressed by a GW, with an enhancement of the system response if the characteristic frequency of the wave coincides with the resonance frequency of the mechanical system. Early approaches to detecting GWs involved searching for excitations of vibrational modes in the Earth's crust (sub-mHz and higher harmonics), however, large earthquakes made it unattractive for detection purposes. The first GW detectors were thus simple metal cylinders, pioneered by Joe Weber of the University of Maryland \cite{Weber-PR-1960} shown in \Fig \ref{fig:WeberBar}, where the energy converted to longitudinal mechanical oscillations was measured via piezoelectric transducers. However, these instruments failed to ever detect GWs \cite{Weber-PRL-1969}.

\begin{figure}[!t]
  \centering
  \includegraphics[width=0.6\textwidth]{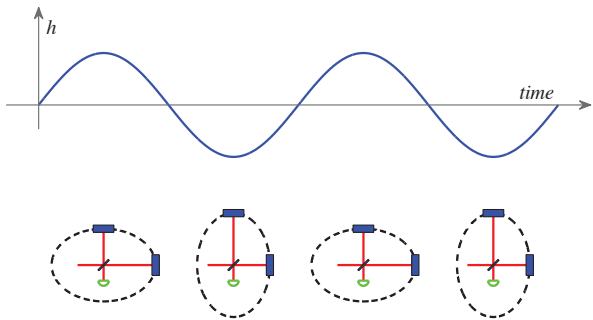}
  \caption{(Colour online) Illustration of the effect of a GW on the arms of a Michelson interferometer, where the readout photodiode is denoted by the green semi-circle. Taken from \Refs \cite{Riles-ProgPartNucPhys-2013, Abbott-RepProgPhys-2009}.}
  \label{fig:GW-Cartoon}
\end{figure}

Gravitational-wave interferometers, first introduced in 1962, take a different approach and it quickly became appreciated that laser interferometers had the potential to surpass bar detectors in sensitivity. As shown in the cartoon in \Fig \ref{fig:GW-Cartoon}, a linearly polarized GW acting on a simple right-angle Michelson interferometer will stretch one arm and contract the other, while simultaneously red- or blue-shifting the light in the respective arms. Since the red-shifted light takes longer to complete its round-trip in the arm than the blue-shifted light, the phase difference between the light returning from each arm will increases with time following the passage of the GW. Thus, the interferometer will have a finite and frequency-dependent response time.

\begin{figure}[!t]
  \centering
  \includegraphics[width=0.45\textwidth]{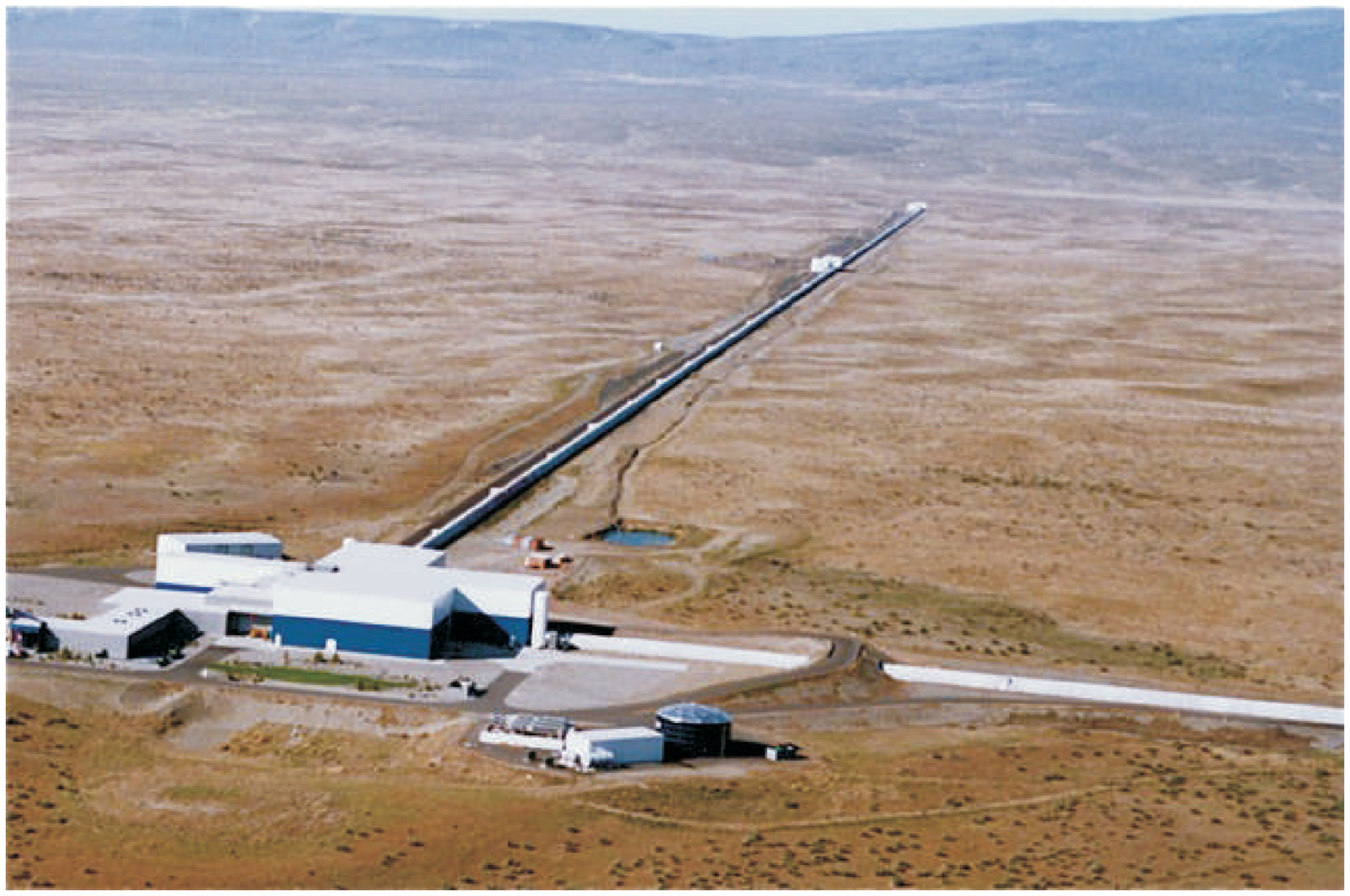}
  \hspace{0.1cm}
  \includegraphics[width=0.45\textwidth]{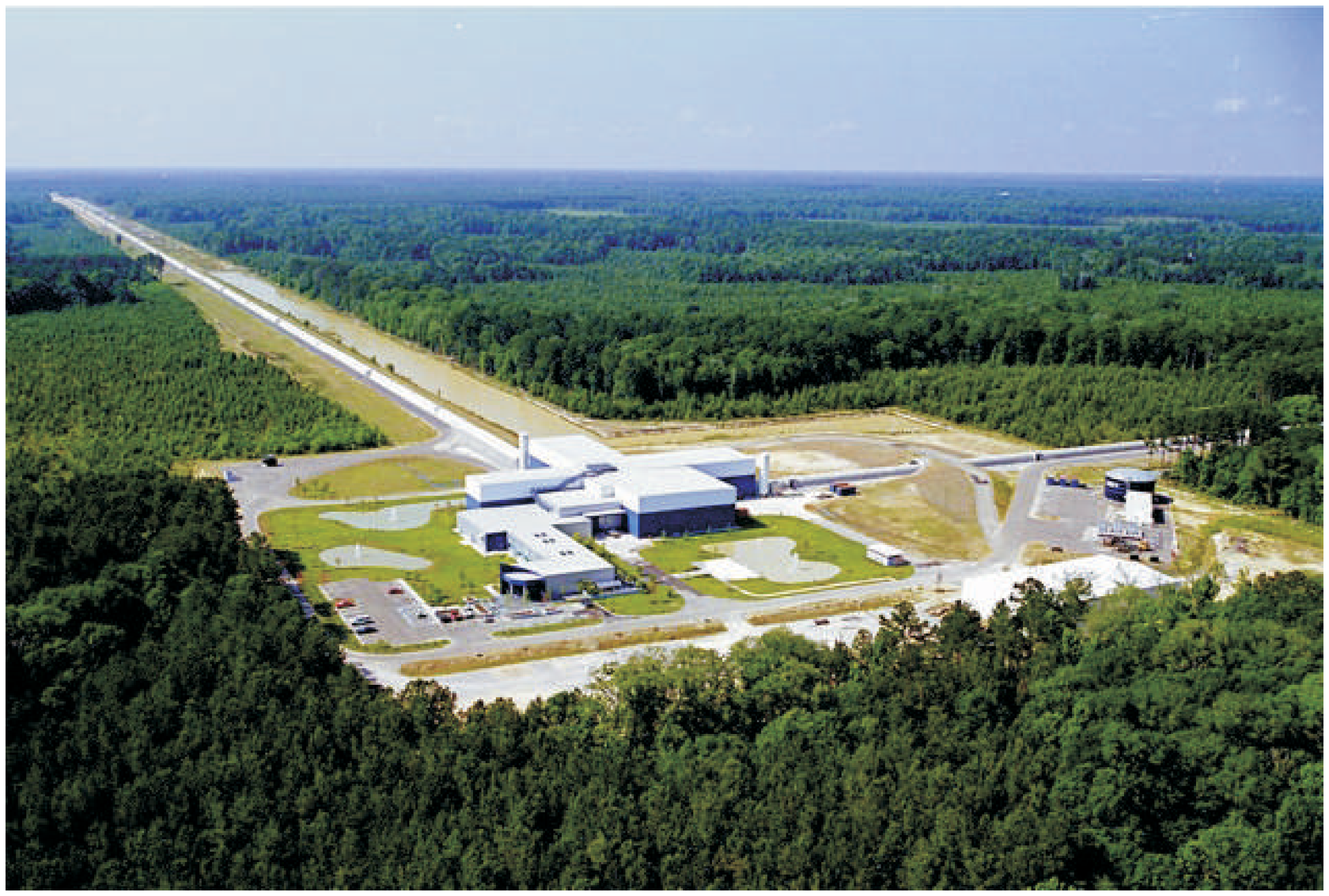}
  \caption{(Colour online) Aerial photographs of the LIGO observatories at Hanford, Washington (left) and Livingston, Louisiana (right). The laser sources and optics are contained in the blue and white buildings. From these buildings, evacuated beam tubes extend at right angles for 4 km in each direction, which are covered by concrete enclosures. Photos taken from \Ref \cite{Riles-ProgPartNucPhys-2013}, courtesy of the LIGO scientific collaboration.}
  \label{fig:LIGO-Observatories}
\end{figure}

In the case of the LIGO (Laser Interferometer Gravitational wave Observatory---shown in \Fig \ref{fig:LIGO-Observatories}) or VIRGO projects, one measures via light propagation time the influence of GWs on pairs of test masses (mirrors) separated by 4 km in a large-scale optical Michelson interferometer. These detectors include substantial improvements from the basic GW interferometer idea by, for example, (i) using Fabry-Perot cavities for the interferometer arms to increase the time of exposure of the laser light to the GW, (ii) the introduction of a ``recycling'' mirror between the laser and beam-splitter to increase the effective laser power, and (iii) the introduction of an additional mirror between the beam-splitter and photo-detector to allow tuning of the interferometer's frequency response \cite{Meers-PRD-1988}.

With these detectors, gravitational radiation is likely to be detected (with a frequency less than a few kHz) from sources with sizes comparable to the wavelength of the GW (\ie $\lambda \gtrsim 300$ km). Hence, the signal reflects coherent motion of extremely massive objects. The primary challenge in detecting GWs is that the magnitude of there effects in the vicinity of Earth is extremely small. The strength of GWs is characterized by the unitless amplitude $h$, which is given by \cite{Pitkin-LivRevRel-2011}
\be
  h = \frac{2 \Delta L}{L}.
\ee
Here, $\Delta L$ is the change in distance between the two test masses, and $L$ is their mean separation. To give an idea of the magnitude of $h$ that could be generated by man-made means, consider a rotating dumbbell consisting of two 1-ton masses separated by 2 meters and spinning at 1 kHz. For an observer 300 km away, one obtains an extraordinarily small $h \sim 10^{-38}$! Even for the most violent astrophysical events, such as coalescing binary neutron stars, $h$ is of order $10^{-21}$ or smaller \cite{Saulson-Book-1994}.

To lowest order, gravitational radiation is a quadrupolar phenomenon, which leads to polarizations that are crossed at $45^{\circ}$. In comparison, electric and magnetic dipole interactions are responsible for electromagnetic radiation, which have orthogonal polarizations (crossed at $90^{\circ}$). In passing through ordinary matter, gravitational radiation suffers no more than a tiny absorption or scattering (although, like light, it is subject to gravitational lensing by large masses). As a result, GWs can carry information about astrophysical phenomena from locations where electro-magnetic radiation is blocked or obscured, for example, from deep within stars or behind dust clouds. Even neutrinos have large scattering cross sections compared to GWs.

Gravitational wave sources can be classified in four broad categories \cite{Virgo-Report-2011}: (i) short-lived and well-defined, for which coalescence of a compact binary system is the canonical example; (ii) short-lived and \emph{a priori} poorly known, \eg supernovae; (iii) long-lived and well-defined, \eg continuous waves from spinning neutron stars; and (iv) long-lived and stochastic, \eg primordial GWs from the Big Bang. For existing terrestrial detectors, the most promising category of GW sources is the first. Detectable event rates for compact binary coalescence can be estimated with the greatest confidence, and their discovery by Advanced-LIGO and VIRGO detectors is highly likely.

\subsection{MIGA: An atom interferometric gravitational wave detector}

The MIGA (Matter wave -- laser Interferometry Gravitation Antenna) experiment \cite{MIGA-Website} will implement an atom-interferometry-based infrastructure for the investigation of the space strain tensor. The device will consist of an underground ``Very Long Baseline Atom Interferometer'' (VLBAI) to monitor the gravity field over a broad frequency band, ranging from 0.1 Hz to a few tens of Hz, that is unaccessible to the most sensitive Earth-based optical interferometers aiming at detecting GWs. The possible applications extend from fundamental physics (associated with GW detection), to monitoring geophysical signals, such as the evolution of Earth's gravitational field, tectonic plates or general underground mass movements, as well as studies of various effects in hydrology.

The baseline design consists of a matter wave-laser antenna, where two or more atom interferometers are coupled to a cavity-based optical interferometer through the radiation circulating in the resonator. The  optical field stored in the cavity, which has ultra-low phase noise, will be used to coherently split, reflect and recombine the matter waves. The combined atom-laser system will monitor the motion of the cavity and the forces acting on the atoms at the same time, and with a broad frequency resolution. The laser interferometer performs best at frequencies above 10 Hz, while the atom interferometers provide sensitivity at low frequency (10 Hz and below) due to their intrinsic high accuracy. The system will operate in a gravity-gradiometer configuration, with two (or more) atom interferometers residing in the one-arm optical gravitational detector, which is coupled to a highly precise laser link. This allows the variations of optical path between the two ensembles to be measured with extreme precision. These variations can be induced by the space strain due to a passing GW, or by fluctuating gravitational forces. During the measurement, the atoms are in free-fall, hence coupled to environmental vibrations only through gravity. Together with the use of the same laser light to operate the two interferometers, this strongly mitigates the effect of vibrations. The effect of the laser phase noise on the matter wave interferometer can be reduced using (i) a pre-stabilization cavity on the probe light source, (ii) a second interferometer baseline, or (iii) by adopting new interrogation schemes as proposed in \Ref \cite{Graham-PRL-2013}.

Intuitively the experiment can be seen as a periodic measurement of the propagation time of a laser between the two ensembles. A passing gravitational wave changes the distance between the two ensembles in a way proportional to their distance, and at the frequency of the GW. The effect on an interferometer with interrogation time $T$, and effective wave vector $k_{\rm eff}$, is a phase shift:
\be
  \label{GW-phase}
  \Delta \phi \simeq h(\omega) \cdot 2 k_{\rm eff} L \cdot \sin^2 \left ( \frac{\omega T}{2} \right ) \sin \big( \phi_{\rm GW} \big) \, ,
\ee
where $h(\omega)$ is the strain at frequency $\omega$, and $\phi_{\rm GW}$ the phase of the GW at the beginning of the interferometric sequence. To maximize the strain sensitivity, the baseline $L$ should be as large as experimentally achievable, which will not pose constraints on the matter wave sensors since only the cavity laser light travels between the two locations to establish the coherent optical link.

\begin{figure}[!t]
  \centering
  \includegraphics[width=0.8\textwidth]{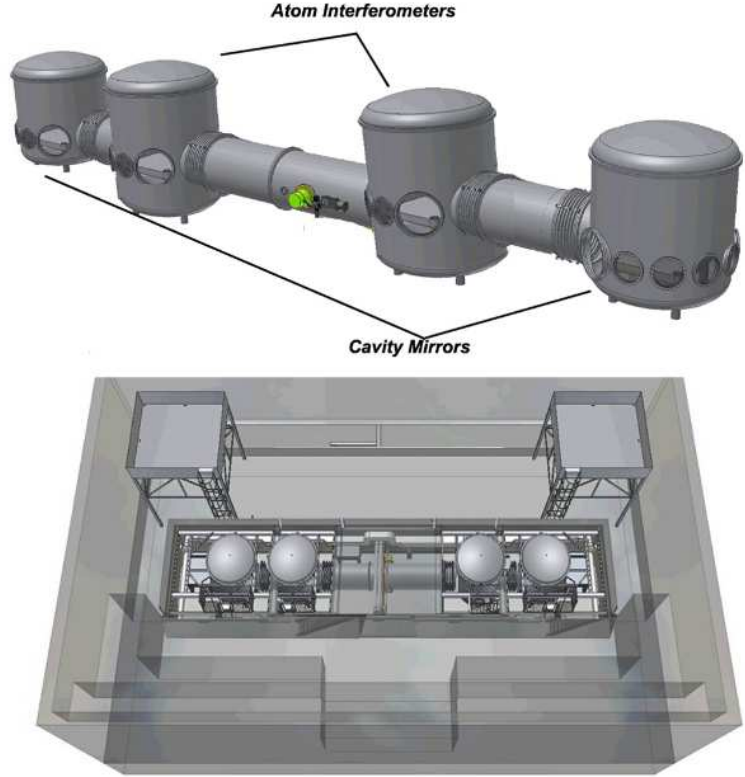}
  \caption{Design of the 10 m baseline MIGA--prototype at LP2N (Laboratoire Photonique Numerique et Nanosciences) in Bordeaux.}
  \label{fig:MIGA}
\end{figure}

The atom-interferometer-based GW detector we want to implement will be located at \textit{Laboratoire Souterrain {\`a} Bas Bruit} (LSBB), in Rustrel, France. This objective will be pursued in two separate phases. During the first phase, the key components of the antenna will be designed, built and tested. A dedicated test facility will be set up at LP2N in Bordeaux, where a reduced scale prototype of the final experiment will be operated. A scheme of the MIGA--prototype is shown in \Fig \ref{fig:MIGA}. It will consist of a 10 m linear optical cavity maintained in a high vacuum environment, with the two mirrors on ultra-stable vibration isolation systems, and two regions where the atom interferometers are operated. To obtain high sensitivity, a long interrogation time of 250 ms will be adopted for the interferometers by utilizing atoms launched along vertical parabolic trajectories. The laser link between the two cavity mirrors will be implemented with a telecom laser at 1560 nm maintained on resonance with the cavity. The interrogation of the rubidium atoms will require Bragg pulses at 780 nm, obtained via frequency doubling of the telecom laser. The use of a cavity-enhanced beam-splitter---a new concept in atom interferometry---will bring forward several advantages. First, it will enable coherent splitting and recombination of different atomic ensembles that are separated over hundreds of meters---thus enhancing the instruments sensitivity to gravity-gradients and space-time variations. Second, the cavity will filter the spatial laser mode, which is one of the main limiting factors in existing interferometers for long term measurements \cite{Louchet-Chauvet-NJP-2011}. Third, large-momentum-transfer (LMT) beam-splitters \cite{Chiow-PRL-2011, McDonald-PRA-2013} can be implemented as a result of the optical power that will accumulate in the resonator.

\begin{figure}
  \centering
  \includegraphics[width=0.6\textwidth]{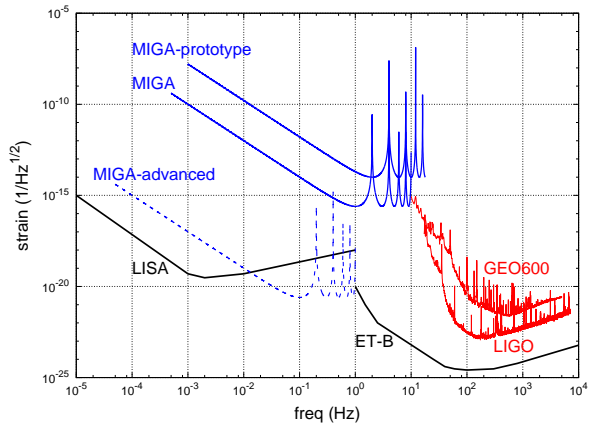}
  \caption{(Colour online) Comparison of strain sensitivity curves of different generations of atom-laser interferometers and their optical counterparts. The blue curves, for MIGA-prototype and MIGA, consider current technology ($T = 0.25$ s and 0.5 s, respectively; $4\hbar k$ beam-splitters; $L = 10$ m and 400 m, respectively; signal-to-noise ratio $= 10^{-3}$). The dashed MIGA--advanced curve is the projected sensitivity assuming a trapped sample ($T = 5$ s), LMT beam-splitters ($400\hbar k$), and improved sensitivity from a large number of atoms and squeezed states. In red, the sensitivity curves of LIGO and GEO600---two operational Earth-based optical GW detectors. In black is the projected sensitivity of future optical GW detectors: the space-based LISA, and the underground Einstein Telescope.}
  \label{fig:MIGA-Sensitivity}
\end{figure}

The second phase (2015--16) will consist of the construction of the antenna in the underground laboratory at LSBB. All the components previously validated in the test facility will be moved and installed at the final site, where the length, $L$, of the laser link will be 400 m. This will produce a 40-fold boost of the sensitivity when compared to the prototype, as indicated in \Fig \ref{fig:MIGA-Sensitivity}. The curve is also shifted to lower frequencies by a factor 2 because of a doubling of the interrogation time from $T = 0.25$ s to 0.5 s. The projected peak strain sensitivity for the antenna will be $\sim 10^{-16}$ Hz$^{-1/2}$ at 1 Hz. An important difference from the prototype will be the simultaneous use of more than two matter wave heads distributed along the optical cavity, which will give the possibility of measuring not only the gravity gradient, but also its curvature or higher spatial moments.

Several techniques will then be investigated to push the sensitivity of MIGA toward the dashed curve in \Fig \ref{fig:MIGA-Sensitivity}, like the use of quantum enhanced input states to boost the sensitivity beyond the atomic shot-noise-limit \cite{Appel-ProcNatlAcadSci-2009, Schleier-Smith-PRL-2010, Bohnet-arXiv-2013}, LMT beam-splitters, and long interrogation times using trapped interferometric schemes.

From \Eq \refeqn{GW-phase} the antenna is maximally sensitive at frequencies $\nu_n = (2n+1) / (2T)$, with the lowest maximally sensitive frequency of 2 Hz for the initial MIGA design, and potentially one order of magnitude less for an advanced version. The frequency band 0.1--100 Hz is expected to be the host of several astrophysical phenomena producing GWs---such as binary pulsars, white dwarfs, neutron stars or black holes---that should produce gravitational radiation with a sweeping frequency as the pair of super-massive objects merge \cite{Schultz-ClassQuantGrav-1999, Creighton-Book-2011}. At the same time, this range of frequency lies beyond the sensitive frequency band of all-optical GW detectors because of the effect of seismic and vibrational background noise. Atom interferometry could thus potentially extend the operational frequency range of existing optical interferometers.

%% file: Varenna-Space.tex
\section{Atom interferometry in Space}
\label{sec:Space}

Since the sensitivity of atom interferometers to inertial accelerations scales as the square of the time in free-fall, ground-based atom interferometric experiments are fundamentally limited. This is because an increasing time in free-fall implies an increasing path-length for the atom trajectory---making the control of systematic effects extremely challenging. An attractive alternative is the operation of an atom interferometer in a micro-gravity environment, as we discussed in \Sec \ref{sec:ICE}.

Performing Earth-based cold-atom experiments in $0g$ requires a very compact and robust design, which is built to withstand extreme environmental conditions, such as the vibrational noise onboard the zero-g aircraft of ICE, or the $\sim 50 g$ deceleration during capsule recapture in the ZARM drop tower of QUANTUS. The micro-gravity phase of these experiments lasts between just 5 and 20 seconds, and measurements can be carried out only during these times. Between repetitions of the $0g$ phase, the experiments can have long down times (2-3 drops per day for QUANTUS, two flight campaigns per year for ICE). Furthermore, the quality of micro-gravity in Earth-based systems is not perfect ($\sim 0.01\,g$ fluctuations during parabolic flights).

In contrast, performing an experiment onboard a satellite offers the possibility of extremely long interrogation times with essentially continuous micro-gravity operation. In principle, under these conditions $T$ would be limited only by the expansion time of the atoms out of the laser beams due to their temperature. \footnote{For example, with ultra-cold $^{87}$Rb atoms at 10 nK, and a beam diameter of 2.5 cm, the most probable time for an atom to traverse one beam radius is $\sim 9$ s in the absence of gravity.} Atom interferometers in Space promise sensitivities to differential accelerations on the order of $10^{-12}\,g$ with $T \sim 1$ s. This level of sensitivity is compatible with a test of the WEP on quantum objects of a few parts in $10^{15}$ \cite{Geiger-NatureComm-2011, Muntinga-PRL-2013, Tino-NucPhysB-2013, Dickerson-PRL-2013}. Several developing projects within the ESA, the French space agency (CNES) and the German Aerospace Center (DLR), are today investigating the potential of cold atom interferometry for precision measurements and fundamental tests in Space.

We now review two proposed Space missions to test the WEP with atom-based sensors. Other proposals involving atom-interferometric measurements in Space are discussed in \Refs \cite{Dimopoulos-PRD-2008, Sorrentino-MicrogSciTech-2010, Hogan-GenRelGrav-2011, Graham-PRL-2013}.

\subsection{Q-WEP and STE-QUEST: Testing the equivalence principle in Space}

The International Space Station (ISS) is a platform which is specifically designed for experiments in a continuously available micro-gravity environment. In 2011, the ESA launched the invitation to tender mission number AO/1-6763/11/NL/AF -- ``Atom Interferometry Test of the Weak Equivalence Principle in Space'', or Q-WEP. The main scientific objective of Q-WEP is to test the weak equivalence principle using an atom interferometer which is adapted for operation on the ISS. Additional applications, such as gravity gradiometry, have also been assessed as candidate experiments. The development of the first space-borne cold-atom sensor would make available new instruments for inertial measurements with an extremely high long-term stability, and a well-known calibration factor. Such technology is broadly applicable to a number of other interesting applications, such as inertial navigation and geodesy.

\begin{figure}
  \centering
  \includegraphics[width=0.6\textwidth]{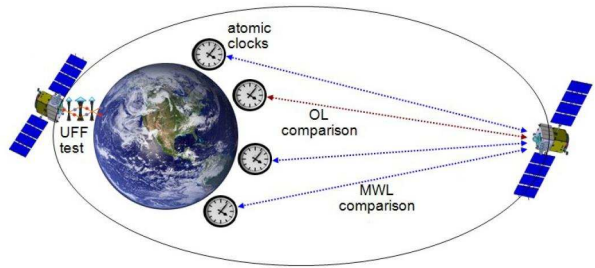}
  \caption{(Colour online) General concept of the STE-QUEST mission. During the perigee, the local acceleration of two rubidium isotopes is measured and compared.}
  \label{fig:STEQUEST-Concept}
\end{figure}

STE-QUEST (Space Time Explorer and Quantum Equivalence Space Test) was recommended by the Space Science Advisory Committee (SSAC) to be studied first internally, and afterwards with parallel industrial contracts. The primary goal of this mission is to perform a quantum test of the universality of free-fall by interferometrically tracking the propagation of two atomic species ($^{85}$Rb and $^{87}$Rb), with a projected accuracy of $10^{-15}$. The concept of the mission is illustrated in \Fig \ref{fig:STEQUEST-Concept}.

The baseline design of the atom interferometry payload, the experimental scheme, and the nature of performance limits are similar for both Q-WEP and STE-QUEST. In the following sections, a general description is given which applies to both missions, unless otherwise stated. The term ``spacecraft'' will refer to the ISS for Q-WEP, and to the satellite for STE-QUEST.

\subsubsection{\textsf{The measurement principle}}

The E\"otv\"os parameter $\eta$ can be defined in terms of the differential acceleration between the two test bodies and their mean acceleration: $\eta = \Delta a/\bar{a}$, where $\bar{a} = (a_1 + a_2)/2$, $\Delta a = a_1 - a_2$, and $a_1$ and $a_2$ are the accelerations of the two test objects. In the case of the atom interferometer, the two test bodies are samples of two ultra-cold atomic species. A direct readout of the differential acceleration is possible because the two species are prepared and interrogated simultaneously.

The basic operation of the interferometer is determined by the choice of atomic species. In both Q-WEP and STE-QUEST, the $^{85}$Rb--$^{87}$Rb pair is chosen because of the excellent common-mode noise suppression that is achievable with these two isotopes. However, this choice requires a rather complex cooling scheme, involving a dual-isotope MOT, and a crossed optical dipole trap (ODT). The ODT is necessary to obtain Bose-condensed $^{85}$Rb because a Feshbach resonance is needed to modify the scattering length from a negative to a positive value. This requires several $B$-field generating coils and magnetic shielding to suppress the effects of external fields. The experimental apparatus will consist of a two-dimensional (2D) MOT that loads a three-dimensional (3D) MOT on a chip inside the vacuum system. The cycle time for Q-WEP is expected to be 12 -- 18 s, depending on the duration of the atom interferometry sequence, and will be slightly longer for STE-QUEST.

A differential acceleration sensitivity of $\eta \sim 10^{-14}$ has been projected for Q-WEP using a free evolution time $T \sim 1$ s onboard the ISS. Similarly, $\eta \sim 10^{-15}$ has been projected for STE-QUEST using a free-evolution time $T \sim 5$ s on a dedicated satellite.

\subsubsection{\textsf{Preparation of the ultra-cold source}}

\begin{figure}[!t]
  \centering
  \includegraphics[width=0.8\textwidth]{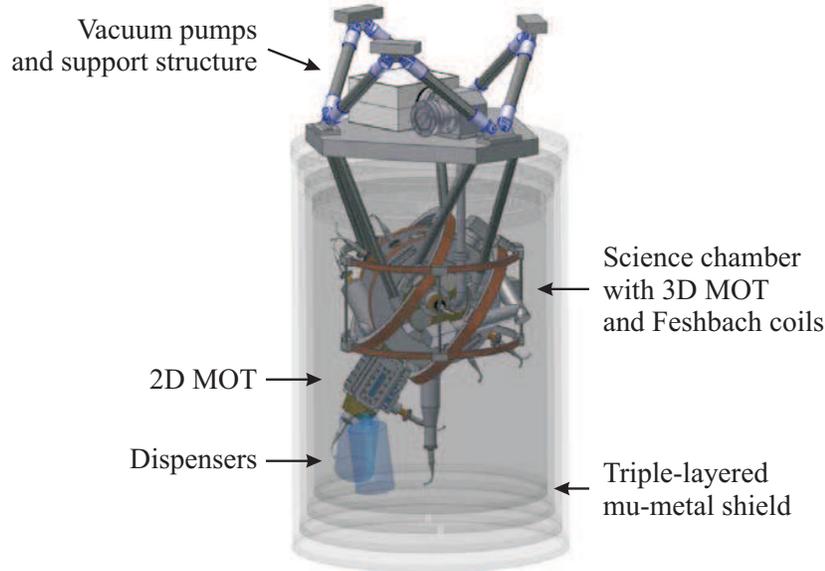}
  \caption{(Colour online) Design concept of the science chamber for STE-QUEST. A similar system is being constructed for Q-WEP. Image taken from \Ref \cite{STEQUEST-TechReport-2013}, courtesy of the German Aerospace Center (DLR).}
  \label{fig:STEQUEST-Setup}
\end{figure}

The ultra-cold dual-isotope sample is prepared in a vacuum system (shown in \Fig \ref{fig:STEQUEST-Setup} for STE-QUEST) that operates at extremely low pressures. To load the desired large number of atoms, a combination of a 2D$^+$-MOT and a 3D-MOT will be used. The 2D$^+$-MOT produces a slow, dual-isotope atomic beam toward the main science chamber, using the continuous flow of atoms provided by an atomic dispenser. The trap consists of four magnetic coils and six laser beams carrying four wavelengths (cooling + repumping transitions for both isotopes). The slow atoms from the 2D$^+$-MOT are captured and cooled on a chip by the combination of magnetic and light fields forming a 3D-MOT in a mirror configuration. The chip produces the magnetic gradient for 3D trapping, and provides a reflecting surface for two of the four laser beams. Three pairs of Helmoltz coils generate a uniform magnetic bias in the trapping region.

After loading $\sim 10^9$ atoms of both isotopes in the 3D-MOT, the magnetic and optical configuration is changed into optical molasses for cooling to sub-Doppler temperatures. This is followed by a transfer to a purely magnetic trap, which pre-evaporates the ensembles and ensures a high transfer efficiency to the crossed ODT. A strong magnetic field is necessary to drive $^{85}$Rb through the Feshbach resonances. This allows the two species to be efficiently condensed.

After their release from the ODT, the atoms are prepared in the magnetically insensitive $m_F = 0$ state of their lower hyperfine level via a microwave transition. This is followed by a transfer to the interferometer position, which must be at a sufficient distance from the chip. Precise positioning can be accomplished by means of coherent momentum transfer via stimulated Raman transitions.

\subsubsection{\textsf{Atom interferometry sequence}}

During this step, the atoms are simultaneously subjected to an interferometry pulse sequence where each pulse induces $2\hbar k_{\rm eff}$ of momentum to the atom (equivalent to a four-photon transition), as in the double-diffraction scheme presented in \Refs \cite{Leveque-PRL-2009, Malossi-PRA-2010}. Double-diffraction is advantageous for an interferometer sequence in Space, where the Doppler shift of the atoms is zero. Due to its symmetric momentum transfer, the center-of-mass of the atom remains in the same location throughout the entire sequence---making it an ideal choice for equivalence principle tests. This feature also makes a double-diffraction-based interferometer insensitive to a number of systematic effects and sources of noise, because they are common to both arms.

\begin{figure}[!t]
  \centering
  \includegraphics[width=0.6\textwidth]{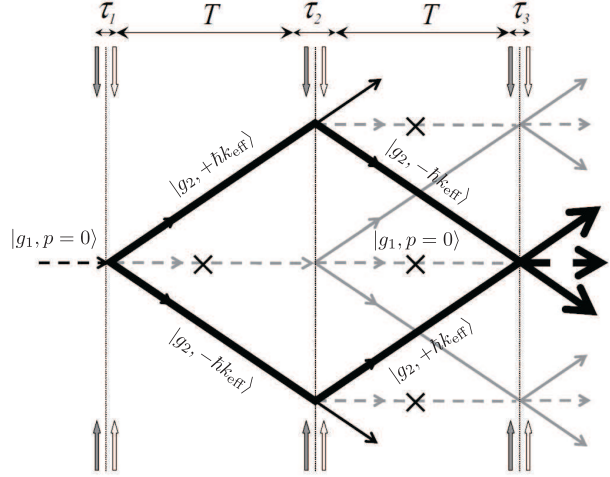}
  \caption{Illustration of the atomic trajectories involved in double-diffraction, taken from \Ref \cite{Malossi-PRA-2010}. The bold lines show the states contributing to the interference at $t = 2T + \tau_1 + \tau_2 + \tau_3$. The crosses indicate where the blow-away beam is applied to remove left-over population in the lower state, $\ket{g_1,p=0}$.}
  \label{fig:DoubleDiffraction}
\end{figure}

Figure \ref{fig:DoubleDiffraction} depicts the trajectories of the double-diffraction scheme. During the sequence, the first pulse acts as a splitter by inducing Raman transitions from $\ket{g_1,p=0}$ to an equal superposition of states $\ket{g_2,+\hbar k_{\rm eff}}$ and $\ket{g_2,-\hbar k_{\rm eff}}$. The second pulse acts as a mirror by coupling each path to its opposite momentum state, $\ket{g_2,+\hbar k_{\rm eff}} \to \ket{g_2,-\hbar k_{\rm eff}}$ for example. Atoms on both the momentum states $\pm\hbar k_{\rm eff}$ are in the same internal state, $\ket{g_2}$. After the first and the second pulse, a blow-away pulse removes residual atoms in the internal state $\ket{g_1}$. Finally, a third pulse combines the two pathways, and the population in the two internal states can be read out.

\subsubsection{\textsf{Differential noise rejection}}

\begin{figure}[!t]
  \centering
  \includegraphics[width=0.95\textwidth]{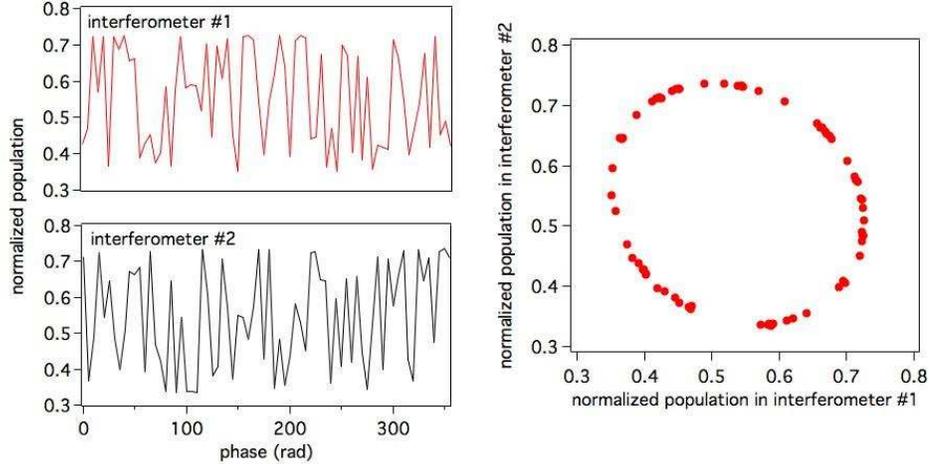}
  \caption{(Colour online) On the left is the signal from a pair of simultaneous interferometers, where vibrationally-induced phase noise is larger than one period. On the right is Lissajous plot resulting from the composition of the two signals. The differential acceleration is determined from the ellipse rotation angle, as discussed in \Ref \cite{Foster-OptLett-2002, Sorrentino-ApplPhysLett-2012}.}
  \label{fig:Fringes-Ellipse}
\end{figure}

With such a high sensitivity to acceleration, phase noise induced by vibrations of the reference mirror can span several interferometer fringe periods. A sensitive measurement of the relative phase of the two interferometers, which is proportional to the differential acceleration of the two atomic ensembles, can be obtained with the ellipse fitting method described in \Ref \cite{Foster-OptLett-2002}. This is an effective technique to suppress common-mode phase noise. Figure \ref{fig:Fringes-Ellipse} illustrates this method, where the interference signal of one interferometer is plotted versus the interference signal of the other---resulting in an ellipse. The relative phase shift can be obtained from the eccentricity and rotation angle of the ellipse. A high common-mode rejection ratio requires a precise matching of the scale factor of the two atom interferometers, as has been shown in recent work by Sorrentino \emph{et al.} \cite{Sorrentino-ApplPhysLett-2012}.

\subsubsection{\textsf{Mission duration}}

For Q-WEP, assuming a single-shot sensitivity to differential acceleration in the range of $4.4 \times 10^{-11}$ m/s$^2$, and a cycle duration of 18 s, an integration time of $10^7$ s (corresponding to $5.6 \times 10^5$ experimental cycles) will be required to reach the target of $\sim 6 \times 10^{-14}$ m/s$^{2}$ accuracy on differential acceleration, \ie one part in $10^{14}$ of the E\"{o}tv\"{o}s parameter. Assuming a mission duty cycle of 40\%, the WEP test will require about ten months, including an additional five months for the secondary objectives, and six months for commissioning and calibration, the entire mission duration is expected to be about 21 months.

For the STE-QUEST mission, the single-shot sensitivity to differential acceleration is projected to be $\sim 3 \times 10^{-12}$ m/s$^2$. Since the atom interferometry measurement is only performed at the perigee passage, where the gravitational acceleration is large enough, a few thousands orbits are required to reach the targeted sensitivity of $\sim 10^{-15}$ m/s$^2$ in $\eta$. Thus, the mission is expected to last 5 years.

\subsection{Advantages of Space-based atom interferometry}

As mentioned above, both STE-QUEST and Q-WEP project free-evolution times of $T \sim 1 - 5$ s in order to reach their desired accuracy. In principle, these values are also possible in Earth-based systems, such as the 10 m vacuum system at Stanford \cite{Dickerson-PRL-2013}. However, in this case the atoms cannot simply be released out of the trap as in micro-gravity---they need to be launched in order to extend $T$ to this scale. Due to the launch and Earth's gravitational acceleration, the center of mass of the atomic trajectories is displaced by almost 10 m with respect to the retro-reflection mirror during the interferometer sequence. The launching process needs to be precisely controlled, because even small differential displacement or velocities of the atoms will impose non-negligible systematic errors. Additionally the apparatus needs to be carefully shielded against external magnetic fields over the launch height.

On the other hand, in micro-gravity a launch is not necessary. Both the atoms and the retro-reflection mirror are in free-fall after trap-release. Then, in the ideal case, there is no relative velocity between the atoms and the mirror, alleviating the need for a phase-continuous linear-frequency chirp that is required in ground-based gravimeters. More importantly, this also implies a compact setup since the vacuum chamber only needs to contain the splitting of atomic trajectories on the order of several centimeters. This reduces the volume over which external fields need to be suppressed.

To reach the atomic temperatures necessary for a high signal-to-noise ratio, and to reduce systematic errors, an optical dipole trap is planned for Q-WEP and STE-QUEST. Micro-gravity ensures precise localization of two different atomic species in the same dipole trap, since there is no longer a differential sag from the gravitational potential.

Relative to terrestrial matter-wave tests of the equivalence principle, Q-WEP and STE-QUEST also have the distinct advantage of a high rotation frequency of the spacecraft relative to the Earth. The makes possible the measurement of relativistic terms in the gravitational potential that scale as $g(v/c)^2$.

%% file: Varenna-Conclusion.tex
\section{Conclusion}
\label{sec:Conclusion}

In summary, we have reviewed a number of past, present and future projects in atom interferometry, with a particular focus on applications of the technology that has been developed over the last 20 years. Although this field has shown rapid growth, we have only begun to see cold atom experiments leave the laboratory to be tested as portable field instruments during the past few years. Mobile and remote inertial sensing with cold atoms promises a new era of measurements for both applied and fundamental science. Although there are already many industrial applications of these highly sensitive devices---such as oil and mineral prospecting, remote object detection, and tidal chart correction---they may one day reach their limits in Space, where current research aims to perform precise tests of the weak equivalence principle and to detect gravitational waves. In addition to advancing the development of cold-atom-based technology, these projects may reveal new and interesting physics related to, for example, our understanding of the early universe.